\documentclass[pdflatex,sn-nature]{sn-jnl}

\usepackage{graphicx}%
\usepackage{multirow}%
\usepackage{amsmath,amssymb,amsfonts}%
\usepackage{amsthm}%
\usepackage{mathrsfs}%
\usepackage[title]{appendix}%
\usepackage{xcolor}%
\usepackage{textcomp}%
\usepackage{manyfoot}%
\usepackage{booktabs}%
\usepackage{algorithm}%
\usepackage{algorithmicx}%
\usepackage{algpseudocode}%
\usepackage{listings}%
\usepackage[mathlines]{lineno}  

\theoremstyle{thmstyleone}%
\theoremstyle{thmstyletwo}%

\theoremstyle{thmstylethree}%

\unnumbered

\begin{document}
\pagestyle{empty}

\title[Imaging aerosolized viruses with an X-ray free-electron laser using single-particle rotational invariants]{Imaging aerosolized viruses with an X-ray free-electron laser using single-particle rotational invariants}


\author[1] {\fnm{Tim} B. \sur{Berberich}}
\author[1] {\fnm{Johan} \sur{Bielecki}}
\author[2] {\fnm{Jonas} A. \sur{Sellberg}}
\author[3] {\fnm{Benedikt} J. \sur{Daurer}}
\author[4] {\fnm{Carl} \sur{Nettelblad}}
\author[5,6,1,7] {\fnm{Paul Lourdu}  \sur{Xavier}}
\author[8] {\fnm{Ivan} A. \sur{Vartanyants}} 
\author[9] {\fnm{Garth} J. \sur{Williams}}
\author[1] {\fnm{Luca} \sur{Gelisio}}
\author[1] {\fnm{Richard}  \sur{Bean}}
\author[1,10,11] {\fnm{Serguei} L. \sur{Molodtsov}}
\author[12] {\fnm{Petra}  \sur{Fromme}}
\author[7] {\fnm{Andrew}  \sur{Aquila}}
\author[1,13,3] {\fnm{Adrian}  P. \sur{Mancuso}}
\author*[1]{\fnm{Ruslan} P. \sur{Kurta}}\email{ruslan.kurta@xfel.eu}

\affil[1]{ \orgname{European XFEL}, \orgaddress{\street{Holzkoppel 4}, \city{Schenefeld}, \postcode{22869}, \country{Germany}}}

\affil[2]{\orgdiv{Department of Applied Physics}, \orgname{AlbaNova University Center, KTH Royal Institute of Technology},  \orgaddress{ \city{Stockholm}, \postcode{SE-10691}, \country{Sweden}}}

\affil[3]{\orgname{Diamond Light Source, Harwell Science and Innovation Campus},  \orgaddress{ \city{Didcot, Oxfordshire}, \postcode{OX11 ODE}, \country{United Kingdom}}}

\affil[4]{\orgdiv{Division of Scientific Computing, Department of Information Technology}, \orgname{Science for Life Laboratory, Uppsala University},  \orgaddress{ \city{Uppsala}, \postcode{SE-75237}, \country{Sweden}}}

\affil[5]{\orgdiv{Center for Free-Electron Laser Science CFEL}, \orgname{Deutsches Elektronen Synchrotron DESY}, \orgaddress{\street{Notkestrasse 85}, \city{Hamburg}, \postcode{22607}, \country{Germany}}}

\affil[6]{\orgname{Max-Planck Institute for the Structure and Dynamics of Matter}, \orgaddress{\street{Luruper Chaussee 149}, \city{Hamburg}, \postcode{22761}, \country{Germany}}}

\affil[7]{\orgdiv{Linac Coherent Light Source}, \orgname{SLAC National Accelerator Laboratory},  \orgaddress{\street{2575 Sand Hill Road}, \city{Menlo Park}, \postcode{CA 94025}, \country{USA}}}

\affil[8]{\orgname{Deutsches Elektronen Synchrotron DESY}, \orgaddress{\street{Notkestrasse 85}, \city{Hamburg}, \postcode{22607}, \country{Germany}}}

\affil[9]{\orgdiv{NSLS-II}, \orgname{Brookhaven National Laboratory},  \orgaddress{\city{Upton, New York}, \postcode{11973}, \country{USA}}}

\affil[10]{\orgdiv{Institute of Experimental Physics}, \orgname{TU Bergakademie Freiberg},  \orgaddress{\street{Leipziger Str. 23}, \city{Freiberg}, \postcode{09599}, \country{Germany}}}

\affil[11] {\orgdiv{Center for Efficient High Temperature Processes and Materials Conversion (ZeHS)}, \orgname{TU Bergakademie Freiberg},  \orgaddress{\street{ Winklerstrasse 5}, \city{Freiberg}, \postcode{09599}, \country{Germany}}}

\affil[12]{\orgdiv{Biodesign Center for Applied Structural Discovery and School of Molecular Sciences}, \orgname{Arizona State University},  \orgaddress{\street{797 E Tyler Street}, \city{Tempe}, \postcode{AZ 85281}, \country{USA}}}

\affil[13]{\orgdiv{Department of Chemistry and Physics}, \orgname{La Trobe University},  \orgaddress{\city{Melbourne}, \postcode{VIC 3086}, \country{Australia}}}


\abstract{X-ray free-electron lasers (XFELs) enable diffraction-before-destruction measurements of individual nanosized bioparticles, making it possible to study the structure and dynamics of non-crystalline targets under near-biologically relevant conditions. In this work, we employ rotational invariants for model-guided and \textit{ab initio} three-dimensional (3D) structure determination of aerosolized bacteriophages PR772 measured with an XFEL. The rotational invariants derived from diffraction patterns collected during multiple independent XFEL experiments facilitate the characterization of similarities and structural variations within the measured ensembles of PR772 particles. Despite modest experimental resolution, we can identify various structural features of the viruses, including the asymmetric nature of capsid distortions from the perfect icosahedral shape, density variations in the encapsulated content, and an extension at one of the capsid vertices. Rotational invariants combine structural sensitivity with applicability to forward-scattering modeling and inverse problem solving, making them powerful tools for probing the structure and temporal evolution of nano- and bioparticles using an XFEL, particularly enhancing the fidelity of structural analysis at limited experimental resolution.}

\maketitle
\pagestyle{empty}
\section{Introduction}\label{sec1}

Viruses are tiny bioparticles that can profoundly impact cellular life on Earth, as evidenced, for example, by the outbreak of COVID-19 \cite{Henderson2020, Reinke2024}.
Most viruses are in the size range of 20 to 300 nm and consist of a genome enclosed in a protective capsid shell \cite{Rossmann2013}.
The structure of a virus undergoes complex transformations during its life cycle, involving attachment to and penetration into the host cell, uncoating, replication, assembly and release \cite{Rossmann2013, Bruinsma2021}. 
Viruses also exhibit equilibrium dynamics, which can be described, for instance, in terms of normal modes \cite{Hadden2018, Bruinsma2021}.
Both steady-state and non-equilibrium structures can be influenced by environmental conditions such as temperature, light, relative humidity, and the local chemical environment \cite{Hadden2018, Bruinsma2021, Turgeon2014, Verreault2015, Dubuis2020, Lin2020, French2023}.
A detailed understanding of how virus structure evolves in different environments enables the identification of virus transmission routes, the development of optimal disinfection strategies, and the creation of effective antiviral agents \cite{Turgeon2016, Lin2020, Ouyang2023}.
 
The structural properties of viruses can be accessed using various techniques, including X-ray diffraction \cite{Rossmann2013, Reinke2024}, cryo-electron microscopy (cryo-EM) \cite{Rossmann2013, Kaelber2017, Henderson2020}, and atomic force microscopy \cite{Ivanovska2011, Mateu2012}.
X-ray free-electron lasers \cite{Tschentscher2017, SynchrandFELs2016}, which produce intense and ultrashort X-ray pulses, enable high-throughput diffraction-before-destruction measurements of bioparticle solutions and aerosols \cite{Neutze2000, Chapman2006, Seibert2011, Reddy2017, Li2020}, particularly at high repetition rates \cite{Decking2020}.
Single-particle imaging (SPI) with an XFEL offers an evolving portfolio of methods and instrumentation for nanoscale structural investigations \cite{Seibert2011, Ekeberg2015, Kurta2017, Rose2018, Assalauova2020, Kierspel2023}.
With XFEL pulse durations now reaching the attosecond regime, it is possible to capture particles’ structures and dynamics before the onset of radiation damage.

At the same time, the lack of translational invariance in non-crystalline bioparticles increases the complexity of structure determination, affecting both instrumental and methodological aspects \cite{Mancuso2010, Hantke2014, Kimura2014, Aquila2015, Hosseinizadeh2017, Daurer2017, Nakano2018, Chapman2019, Bielecki2020, Ekeberg2024}. 
The application of rotational invariants — which can be accessed \textit{a posteriori} from X-ray diffraction measurements — offers a promising approach for SPI with XFELs \cite{Donatelli2015, Berberich2024} and has the potential to extend imaging capabilities to multiple-particle scattering measurements in aerosols or liquid solutions \cite{Kam1977, Pande2018, Kurta2019, Kurta2023}.
The multi-tiered iterative phasing (MTIP) algorithm \cite{Donatelli2015} enables iterative \textit{ab initio} single-particle 3D structure determination from the measured invariants without the need to impose symmetry constraints. 
The rotational invariants are also instrumental for forward modeling approaches, such as bead modeling of particle structures, that are commonly used in biological small-angle X-ray scattering (SAXS) investigations \cite{Vela2020}.

Conventional X-ray imaging and scattering approaches for structural investigations of nanoobjects usually rely on directly accessible observables, namely the scattered intensities $I(\mathbf{q})$ measured at the momentum-transfer vectors $\mathbf{q}$.
In the kinematical, far-field scattering regime, the corresponding scattered amplitudes $A(\mathbf{q})$ (with $I(\mathbf{q})=|A(\mathbf{q})|^2$)  
are directly related to the real-space electron densities $\rho(\mathbf{r})$ through the inverse Fourier transform (FT), $\rho(\mathbf{r})=\textrm{FT}^{-1}[A(\mathbf{q})]$, where $\mathbf{r}$ is the real-space coordinate.
This simple relationship allows the application of \textit{ab initio} iterative phasing approaches to recover the phases of the complex-valued $A(\mathbf{q})$ lost during measurements of the real-valued $I(\mathbf{q})$,
and thus to reconstruct the real-space structure $\rho(\mathbf{r})$ \cite{Chapman2006}. 
Alternatively, the orientationally averaged SAXS intensities $\langle I(\mathbf{q}) \rangle$ can be employed to fit the best 3D structural model $\rho(\mathbf{r})$ \cite{Vela2020}.
In contrast, in this work we analyze products of the form $\langle I(\mathbf{q})I(\mathbf{q'}) \rangle$.
This higher-order statistical correlation approach provides access to extensive structural information that can be expressed as rotationally invariant quantities \cite{Donatelli2015, Berberich2024}. By construction, these invariants remove the rotational degrees of freedom and encode only the intrinsic structural information about the particle, which can be used concurrently in forward modeling and inverse problem solving.

In this work, we perform diffraction-before-destruction measurements of aerosolized bacteriophages PR772 \cite{Lute2004} with an XFEL and employ single-particle rotational invariants for \textit{ab initio} and model-guided 3D structure determination.
We use advanced methods for extracting and regularizing invariants implemented in the in-house software suite xFrame \cite{Berberich2024} for \textit{ab initio} structure reconstructions, and also perform extensive bead-modeling simulations to provide new insights and increase the fidelity of the structural analysis. 
This combined approach allows us to detect and characterize various structural features of aerosolized viruses, including density nonuniformities and deviations of the capsid from a perfect icosahedral shape.

\section{Results}\label{sec2}

\subsection{Single-particle rotational invariants}\label{sec2:sub1}

The 3D distribution of scattered X-ray intensity from a single particle, such as a virus, $I(\mathbf{q})$, can be represented as a function of the scattering vector $\mathbf{q}$ in spherical coordinates $(q,\theta,\phi)$ in the form of a spherical harmonics expansion: 
\begin{equation}
  \label{Eq:SphHarmExp}
  I(q,\theta,\phi) = \sum_{l=0}^\infty \sum_{m=-l}^{l}I^l_m(q)Y^l_m(\theta,\phi),
\end{equation}
where $q=\lvert \mathbf{q}\rvert$ is the momentum transfer magnitude, $\theta$ and $\phi$ are the polar and azimuthal angles, respectively,  $Y^l_m(\theta,\phi)$ are spherical harmonics of degree $l$ and order $m$, and $I^l_m(q)$ denotes the expansion coefficients. 
The single-particle rotational invariants of degree $l$ can be defined in the following form \cite{Saladin09, Berberich2024}:
\begin{equation}
\label{Eq:InvarBl}
  B_l(q,q')=\sum_{m=-l}^{l}I^l_m(q)I^{l\ast}_m(q').
\end{equation}
The invariants $B_l(q,q')$ represent a complex fingerprint of the complete 3D structure of the virus and can be experimentally accessed using angular X-ray cross-correlation analysis (XCCA) of the measured 2D diffraction patterns $I(q,\theta(q),\phi)$ \cite{Kam1977}. Here, the dependence $\theta(q)$ indicates that each measured 2D diffraction pattern $I(q,\theta(q),\phi)$
represents a curved section of the 3D single-particle intensity distribution $I(q,\theta,\phi)$ (Eq.~\eqref{Eq:SphHarmExp}), defined by the Ewald sphere (see Fig.~\ref{fig:fig1}). Partial information contained in individual experimental 2D diffraction patterns necessitates orientational averaging in the present formalism, achieved by averaging over a set of $M$ diffraction patterns measured from particles in random orientations.
In this work, to minimize unwanted background effects, we employ the average difference-pattern two-point angular cross-correlation function (CCF) of the form
\begin{equation}
 C_{\textrm{diff}}(q,q',\Delta)=\langle\langle I_{\textrm{diff},i}[q,\theta(q),\phi]I_{\textrm{diff},i}[q',\theta'(q'),\phi+\Delta]\rangle_{\phi}\rangle_{i},
\label{eq:Cdiffmain}
\end{equation}
where $0\leq\Delta,\phi<2\pi$ (see Fig.~\ref{fig:fig1}b). Here, $\langle \cdot\rangle_{\phi}$ defines the azimuthal averaging over the angle $\phi$, while $\langle \cdot\rangle_{i}$ denotes statistical averaging over $M_{\textrm{diff}}=\left\lfloor \frac{M}{2} \right\rfloor $ difference diffraction patterns, $I_{\textrm{diff}, i}(q,\theta(q),\phi)=I_j(q,\theta(q),\phi)-I_k(q,\theta(q),\phi)$, each composed of a unique pair $(j,k)$ of images, and $\lfloor \cdot\rfloor$ is the floor operation. Assuming a uniform rotational probability distribution of the intensity slices $I(q,\theta(q),\phi)$ (or, equivalently, of the particle orientations), the rotational invariants $B_l(q,q')$ for $l>0$ are related to the Fourier coefficients (FCs) of the difference-image CCF
$C_{\textrm{diff},n}(q,q')=1/(2\pi)\int_{0}^{2\pi} \mathrm{d}\Delta\ C_{\textrm{diff}}(q,q',\Delta)\exp{(-in\Delta)}$, as given by \cite{Berberich2024}
\begin{equation}
  C_{n}(q,q')=\sum_{l\geq |n|} B_l(q,q') \overline{P}^{|n|}_l(q) \overline{P}^{|n|}_l(q').
    \label{eq:CnBl}
\end{equation}
Here, $C_{n}(q,q')=C_{\textrm{diff},n}(q,q')/2$ and $\overline{P}^{|n|}_l(q) = \sqrt{(l-n)!/[4\pi(l+n)!]} P^{|n|}_l[\cos{\theta(q)}]$, where $P^{|n|}_l[\cos{\theta(q)}]$ is the associated Legendre polynomial of degree $l$ and order $|n|$.
In practical applications, it is customary to identify the maximum degree $L$ such that $B_l(q,q')$ has vanishing values for $l>L$ (see Methods). This allows one to rewrite Eq.~\eqref{eq:CnBl} as a system of linear equations, which can be expressed in matrix form as $\mathbf{C}_n={\mathbf{P}}_{n}^l \mathbf{B}_l$, where $n,l\leq L$, and ${\mathbf{P}}_{n}^l$ is the $(L+1)\times (L+1)$ upper-triangular matrix defined by ${\mathbf{P}}_{n}^{l}=\overline{P}^{|n|}_l(q) \overline{P}^{|n|} _l(q')$. The invariants $B_l(q,q')$ for $0<l\leq L$ are then obtained by directly solving this linear system of equations using the back substitution method.
The zero-degree invariant is determined independently through azimuthal and statistical averaging over $M$ individual diffraction patterns, given by $B_0(q,q')=4\pi I_{\textrm{SAXS}}(q)I_{\textrm{SAXS}}(q')$, where $I_{\textrm{SAXS}}(q)=\langle\langle I_j(q,\theta(q),\phi)\rangle_{\phi}\rangle_{j}=I^0_0(q)$ is the SAXS intensity. As noted, conventional biological SAXS analysis is fully characterized by the zero-degree rotational invariant. A complete set of non-vanishing $B_l(q,q')$ provides significantly more constraints for structure determination, which can be utilized in forward modeling and \textit{ab initio} iterative phasing \cite{Donatelli2015, Kommera2021, Berberich2024}. In fact, since $C_{n}(q,q')$ is a linear combination of rotational invariants (Eq.~\eqref{eq:CnBl}), it is itself invariant under rotations and can be directly used for structural analysis alongside $B_l(q,q')$. To showcase their combined application, we employ rotational invariants $\mathbf{C}_n$, directly accessible from experimental data, for model-guided structural analysis, and we use rotational invariants $\mathbf{B}_l$, determined via Eq.~\eqref{eq:CnBl}, as the basis for \textit{ab initio} structure reconstructions.

\subsection{Angular cross-correlation analysis of experimental X-ray diffraction from viruses}\label{sec2:sub2}

\begin{figure}[b]
\centering
\includegraphics[width=0.8\linewidth]{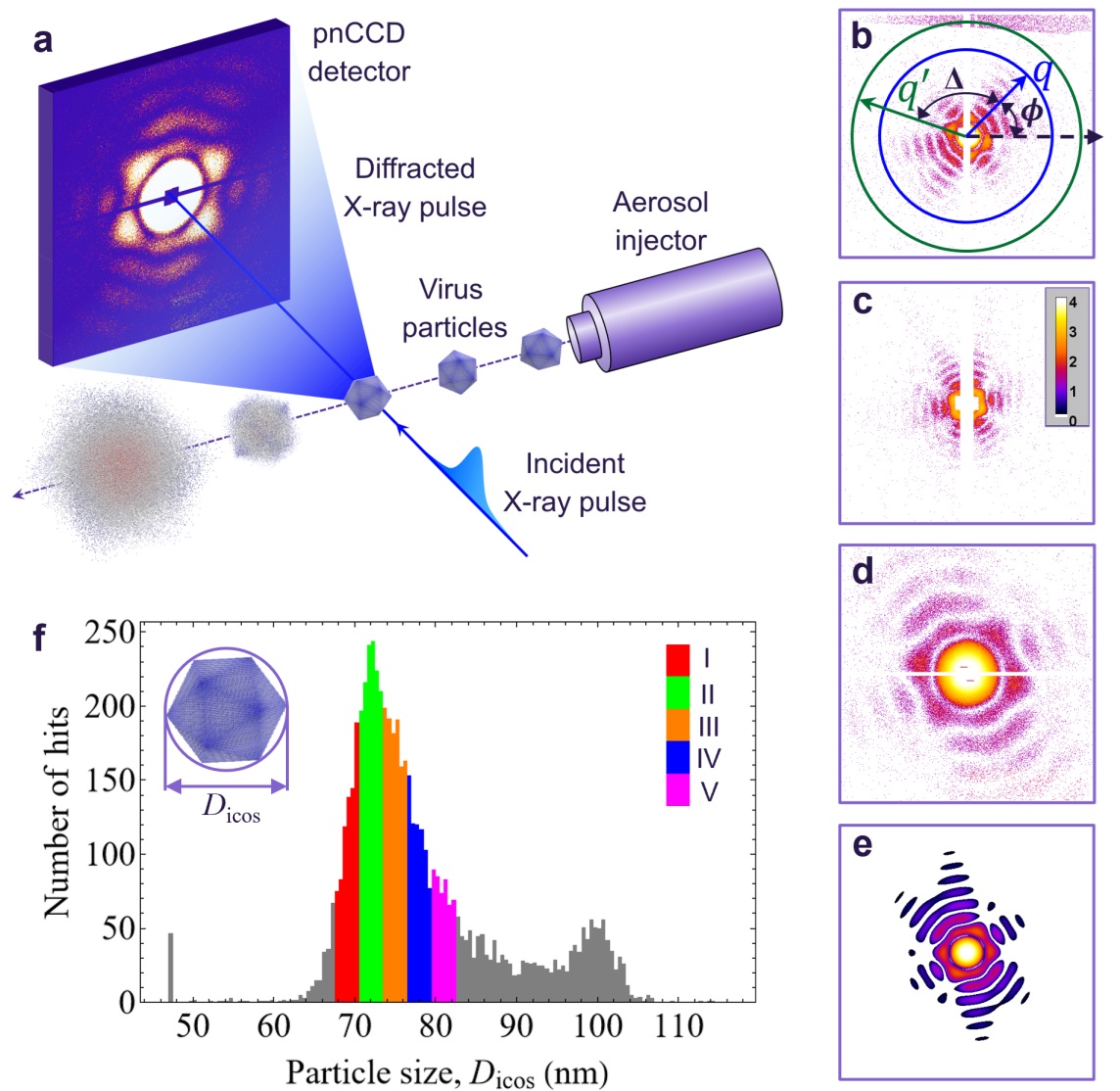}
\caption{\textbf{Single-particle diffraction experiments at the Linac Coherent Light Source.} \textbf{a} An aerosol injector delivers PR772 viruses in random orientations to the interaction region, where they are hit by intense, femtosecond coherent X-ray pulses. The resulting far-field 2D diffraction patterns are recorded by the pnCCD pixel detector before the viruses disintegrate due to radiation damage. \textbf{b-d} Representative single-particle diffraction patterns measured from PR772 bacteriophage in the (b) amo06516, (c) amo11416 and (d) amo86615 experiments. The intensity is shown in analog-to-digital units (ADUs) on a logarithmic scale; noisy pixels with values $<60$~ADU were set to zero and are displayed in white (see Methods). Vectors and angles used in the definition of the CCF (Eq.~\eqref{eq:Cdiffmain}) are shown in (b).
\textbf{e} Diffraction pattern simulated for a bead model of a distorted core-shell icosahedral particle of 70 nm in size. 
\textbf{f} Distribution of the virus sizes $D_{\textrm{icos}}$ determined for a selected set of 3069 single-particle diffraction patterns from the amo06516 experiment. 
Portions of the histogram I to V denote five subsets of data, defined by a 3~nm bin width, used in the \textit{ab initio} reconstructions.}
\label{fig:fig1}
\end{figure}

We recorded the diffraction patterns $I(q,\theta(q),\phi)$ from bacteriophage PR772 at the AMO station of the Linac Coherent Light Source \cite{Emma2010} in three independent X-ray diffraction experiments—amo06516, amo86615, and amo11416—as reported in \cite{Reddy2017} and \cite{Li2020}. 
The PR772 virus particles were aerosolized using aerodynamic lens-stack injectors and probed with femtosecond XFEL pulses of $1.6$ and $1.7$ keV energy in forward scattering geometry (see Fig.~\ref{fig:fig1}a and Methods). 
In this work, for each experiment, a set of high-intensity single-particle hits was selected and classified according to the scattering particle size, 
resulting in a total of 3069 (amo06516), 5030 (amo86615), and 221 (amo11416) selected patterns for their respective experiments, with the determined average particle size being approximately 70 nm (see Fig.~\ref{fig:fig1}f and Methods for details on diffraction data reduction and filtering).
Figs.~\ref{fig:fig1}b-d display representative diffraction patterns from the selected datasets, alongside a simulated pattern (Fig.~\ref{fig:fig1}e) from a core-shell icosahedral particle shown in Fig.~\ref{fig:fig5}f.

The extracted experimental datasets, consisting of single-particle diffraction patterns, were further divided into several subsets defined by 2~nm-wide bins (amo86615) and 3~nm-wide bins (amo06516 and amo11416) in the size distribution histogram (see Fig.~\ref{fig:fig1}f and Methods). These subsets were used independently for single-particle structure determination to explore similarities and variability within the studied virus ensembles. We employed the Fourier components $C_{n}(q,q')$ (see Eq.~\eqref{eq:CnBl}) for direct comparison of the results from XCCA of distinct experimental datasets, represented as real-valued quantities $\textrm{Re}[C_{n}(q,q')]$, and as phases $|\textrm{arg}[C_{n}(q, q')]|$ wrapped to the interval $[0,\pi]$ (see Figs.~\ref{fig:fig2}a-c and Supplementary Figs.~3 and 11 to 17 for details).
The invariants $C_{n}(q,q')$ determined for distinct subsets within individual experiments exhibit a high degree of similarity, which is reflected in the overall morphology of the feature distribution on the 2D maps of $C_{n}(q,q')$ for different orders $n$ (Supplementary Figs.~11 to 16). Moreover, the similarity of $C_{n}(q,q')$ is clearly preserved across the three experiments (Figs.~\ref{fig:fig2}a-c), even despite the very low number of diffraction patterns available in the amo11416 experiment (see Table~1). 
This suggests that ensembles of viruses with similar structures were measured across the experiments considered here, and also underscores the robustness of the invariants with respect to variation in experimental conditions and the quantity and quality of the measured diffraction data. 
At the same time, differences in $C_{n}(q,q')$ that indicate the presence of structural heterogeneity within the studied ensembles of viruses can also be identified, as will be evident from the results of \textit{ab initio} structure reconstructions and forward modeling.

\subsection{\textit{Ab initio} 3D structure reconstructions of viruses}\label{sec2:sub3}

We use the MTIP algorithm \cite{Donatelli2015}, implemented in the software package xFrame \cite{Berberich2024}, to reconstruct the 3D spip tructure of the PR772 bacteriophage \textit{ab initio} from the experimentally determined rotational invariants $B_l(q,q')$. 
Conventional coherent X-ray diffractive imaging (CXDI) approaches enable real-space 3D structure determination through iterative phasing of the measured reciprocal-space 3D intensity distribution $I(q,\theta,\phi)$ \cite{Ekeberg2015, Rose2018, Assalauova2020}. In the present context, this requires determination of all non-vanishing spherical harmonic coefficients $I^l_m(q)$ (see Eq.~\eqref{Eq:SphHarmExp}). However, the information contained in the experimentally accessible $B_l(q,q')$ allows for the determination of $I^l_m(q)$ only up to a unitary matrix for each expansion order \cite{Kam1977}. Recovering these unknown unitary matrices is analogous to solving the orientation determination problem in SPI, which requires the mutual alignment of the measured 2D patterns $I(q,\theta(q),\phi)$ to recover the full 3D intensity distribution $I(q,\theta,\phi)$ \cite{Loh2009}. The MTIP algorithm addresses these problems concurrently in one iterative loop: it finds the unitary matrices that recover the 3D intensity distribution $I(q,\theta,\phi)$ and simultaneously performs phasing of $I(q,\theta,\phi)$ to reconstruct the real-space single-particle structure \cite{Donatelli2015, Berberich2024}. Thus, the 3D real-space electron density distribution $\rho(\mathbf{r})$ is recovered by iteratively applying Fourier transforms between real and reciprocal space, while enforcing a finite support constraint in real space and using the invariants $B_l(q,q')$ as a reciprocal space constraint.
Here, we also applied additional data-processing procedures to regularize the invariants $B_l(q,q')$, which allowed us to optimize the transformation of the raw experimental invariants into a form suitable for MTIP and to extend the usable data range in reciprocal space (see Methods).

The 3D reconstructions obtained using xFrame from the selected data subsets of the amo06516 and amo86615 experiments are presented in Figs.~\ref{fig:fig3} and \ref{fig:fig4}, respectively, while the overall low signal-to-noise ratio and substantial missing data at low $\mathbf{q}$ in the amo11416 experiment precluded successful reconstructions.
The experimental 3D isosurface plots and 2D density cuts in Fig.~\ref{fig:fig3} display icosahedrally shaped virus particles with notable deviations from icosahedral symmetry.
In contrast, the three density cuts through the structure obtained from the simulated data for perfect solid icosahedral particles (Fig.~\ref{fig:fig3}S) using a similar reconstruction pipeline appear almost identical, as expected. The concentric oscillatory density variations within the particle, visible in these density cuts, are ringing artifacts caused by the Gibbs phenomenon \cite{Wang2023}. Meanwhile, the symmetry-equivalent cuts through the individual experimental reconstructions reveal not only deviations in the capsid shape from perfect icosahedral symmetry but also a significantly non-uniform distribution of density inside the viruses in the amo06516 experiment. Additionally, the structures reconstructed from subsets $\mathrm{II}$ and $\mathrm{III}$ of amo06516 show a faint density extension emerging from one of the 12 icosahedral vertices (see Fig.~\ref{fig:fig3} and Supplementary Fig.~7). 
Such structural features overall resemble transformations during viral genome delivery reported for the PRD1 virus, which belongs to the same family (Tectiviridae) as PR772. 
The cryo-EM study of PRD1 has revealed complex remodeling of the internal lipid membrane of the virus, leading to a nonuniform redistribution of material inside the capsid and ultimately resulting in a nanotube protruding from one of its vertices \cite{Peralta2013}. 
Cryo-EM images of PR772 reported in \cite{Li2020} indeed captured some virus particles with a nanotube, suggesting that such particles could also have been present in the amo06516  measurements.

The reconstructions from the amo86615 experiment (Fig.~\ref{fig:fig4}) also show deviations of the virus capsid from the ideal icosahedral shape, in agreement with the results from the amo06516 experiment (Fig.~\ref{fig:fig3}).
Note the consistency in the asymmetry observed in reconstructions corresponding to different particle sizes (II-V). Identical cuts through different reconstructions (e.g., those shown in the middle column of Fig.~\ref{fig:fig4}) appear more similar than two symmetry-equivalent cuts through any individual structure. 
At the same time, no additional structure at any of the capsid vertices has been detected, and the internal virus density distribution appears more uniform in the amo86615 than in the amo06516 reconstructions. The apparent density fluctuations inside the amo86615 virus reconstructions arise primarily from the Gibbs phenomenon, as is evident from the comparison with model reconstructions shown in Fig.~\ref{fig:fig4}S.  

The amo06516 experimental data are characterized by higher heterogeneity compared to the amo86615 data, which manifests in the particle size distribution histogram shown in Fig.~\ref{fig:fig1}f as an extended shoulder toward larger particle dimensions.
Considering that a similar sample preparation protocol was used in the amo06516 and amo86615 experiments (see Methods), it cannot be excluded that the observed differences in the reconstructions were induced by the rapidly changing environmental conditions during the sample injection, which were individually tuned in each experiment. 
Nonetheless, the \textit{ab initio} reconstructions allow us to identify different structural features and variations among distinct virus batches, despite the moderate resolution of the obtained structures, which ranges from 6 to 13~nm for amo06516 and 7 to 13~nm for amo86615, depending on the used subset of data and metric (see Methods and Supplementary Figs.~9 and 10). 
At the same time, the limited resolution of the \textit{ab initio} reconstructions—typical for current biological SPI studies using an XFEL— necessitates complementary analyses to enhance the fidelity of the results, which can also be achieved by using the rotational invariants, as shown in the next subsection.

\subsection{Model-guided structural analysis}\label{sec2:sub4}

One of the advantages of employing the rotational invariants is that they can be used directly in model-guided structural analysis.
Here, we apply a bead-modeling approach (see Methods), commonly used in biological SAXS studies, and test distinct model structures to verify and complement the results of the \textit{ab initio} reconstructions. In this analysis, we primarily focus on
the phases $|\textrm{arg}[C_{n}(q, q')]|$, which allow us to track the fine-structure features in the morphology of the 2D correlation maps induced by specific real-space structural modifications. 

We begin with a bead model of a perfect solid icosahedral particle of uniform density $\rho_{\textit{s}}$ (Fig.~\ref{fig:fig5}a), where the set of rotational invariants is dominated by $B_6(q,q')$ of degree $l=6$ (see Supplementary Note~8 and Supplementary Fig.~18f), leading to identical 2D morphologies of $C_{n}(q, q')$ for orders $n=2,4$ and 6 (Supplementary Figs.~18d,e), in agreement with Eq.~\eqref{eq:CnBl}. Such a morphology of the invariants is indicative of perfect icosahedral symmetry, as seen in the results of simulations for the empty icosahedral PR772 capsid (Supplementary Figs.~18a-c). However, the model of a perfect icosahedron is not in agreement with the experimental results (compare Supplementary Figs.~18d,e with Supplementary Figs.~11 to 17), where the lowest order harmonics ($n=2,4,6$) exhibit clearly distinct morphologies. The agreement between the simulated $n=2$ invariant and its experimental counterpart can be significantly enhanced by introducing a distortion to the ideal icosahedral shape (Supplementary Fig.~19), such as a radial uniaxial compression of the capsid by $7.5\%$ (compare the experimental 2D phase map $|\textrm{arg}[C_{2}(q, q')]|$ in Fig.~\ref{fig:fig5} with the phase maps in Figs.~\ref{fig:fig5}a and \ref{fig:fig5}b). 
This morphology of the $n=2$ invariants may arise from various types of radial distortions of specific magnitudes (see Supplementary Figs.~20a-d) and can serve as a characteristic indicator of deviations from perfect icosahedral symmetry. 

Our simulations show that models with more complex shape distortions are necessary to achieve better agreement with experimental invariants of higher orders $n$. We found that composite distortions with radial and tangential components (Supplementary Figs.~20e-h), particularly those producing asymmetric shapes, lead to morphologies of the $n=4$ invariants that match the experimental observations (compare the experimental phase map $|\textrm{arg}[C_{4}(q, q')]|$ in Fig.~\ref{fig:fig5} with the phase maps in Figs.~\ref{fig:fig5}c and \ref{fig:fig5}d; the red rectangle in the experimental map highlights the speckle configuration arising from the models with complex distortions).

\begin{figure*}
\centering
\includegraphics[width=\textwidth]{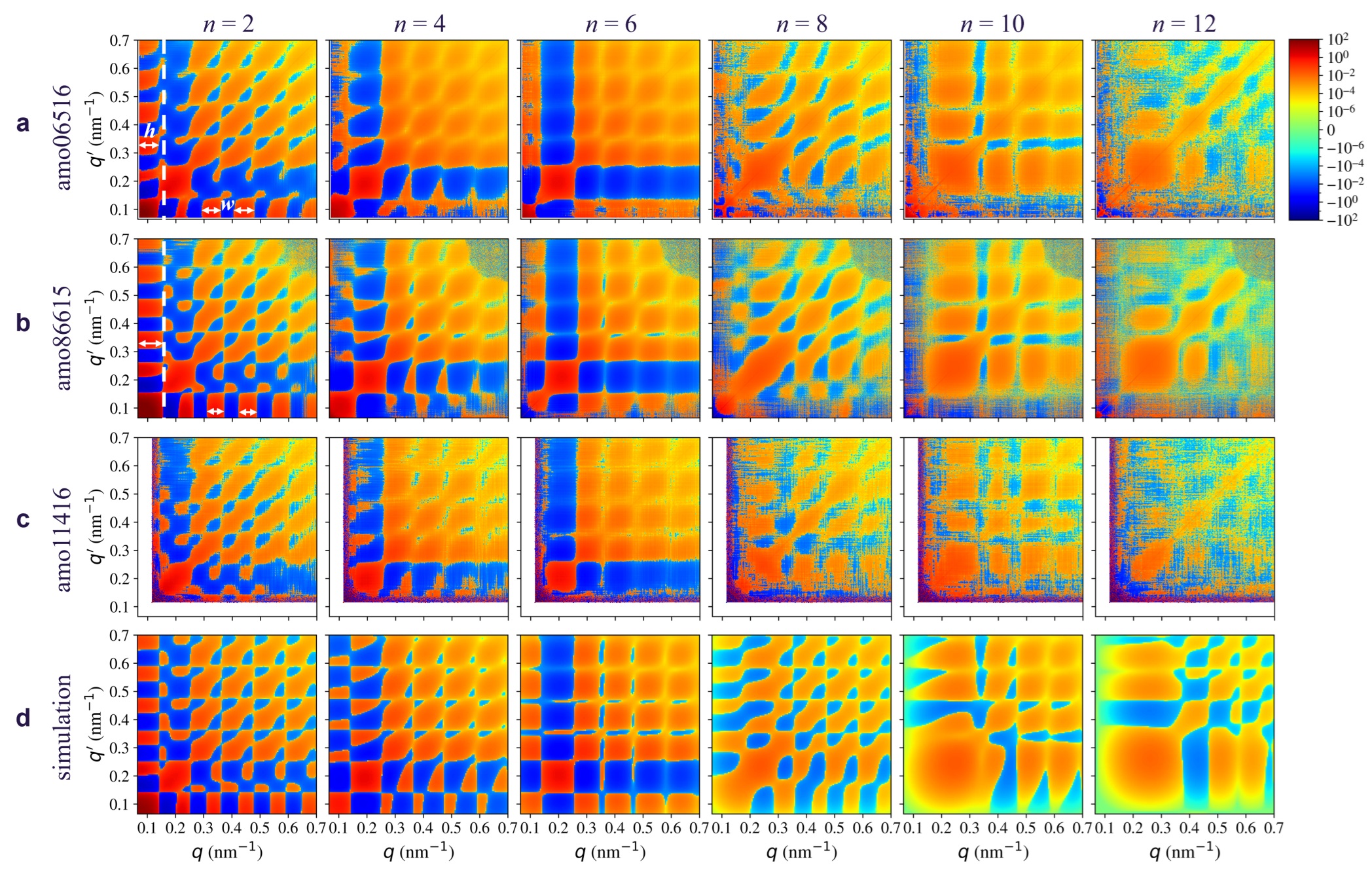}
\caption{\textbf{Experimental and simulated rotational invariants.} \textbf{a-d} Real parts of the Fourier components $\textrm{Re}[C_{n}(q, q')]$ determined for (a) subset $\mathrm{II}$ of the amo06516 experiment, (b) subset $\mathrm{III}$ of the amo86615 experiment, (c) subset $\mathrm{II}$ of the amo11416 experiment, and (d) simulated structure shown in Fig.~\ref{fig:fig5}f. Different columns correspond to the specified orders $n$ of $C_{n}$. The values of $C_{n}$ are given in arbitrary units and plotted on a symmetrical logarithmic scale. The white dashed line at $n=2$, along with the arrows indicating the height $h$ and width $w$ of the selected speckles in the amo06516 and amo86615 experiments, highlight features related to variations in the density of the internal content of the virus.}
\label{fig:fig2}
\end{figure*}

\begin{figure}
\centering
\includegraphics[width=0.65\linewidth]{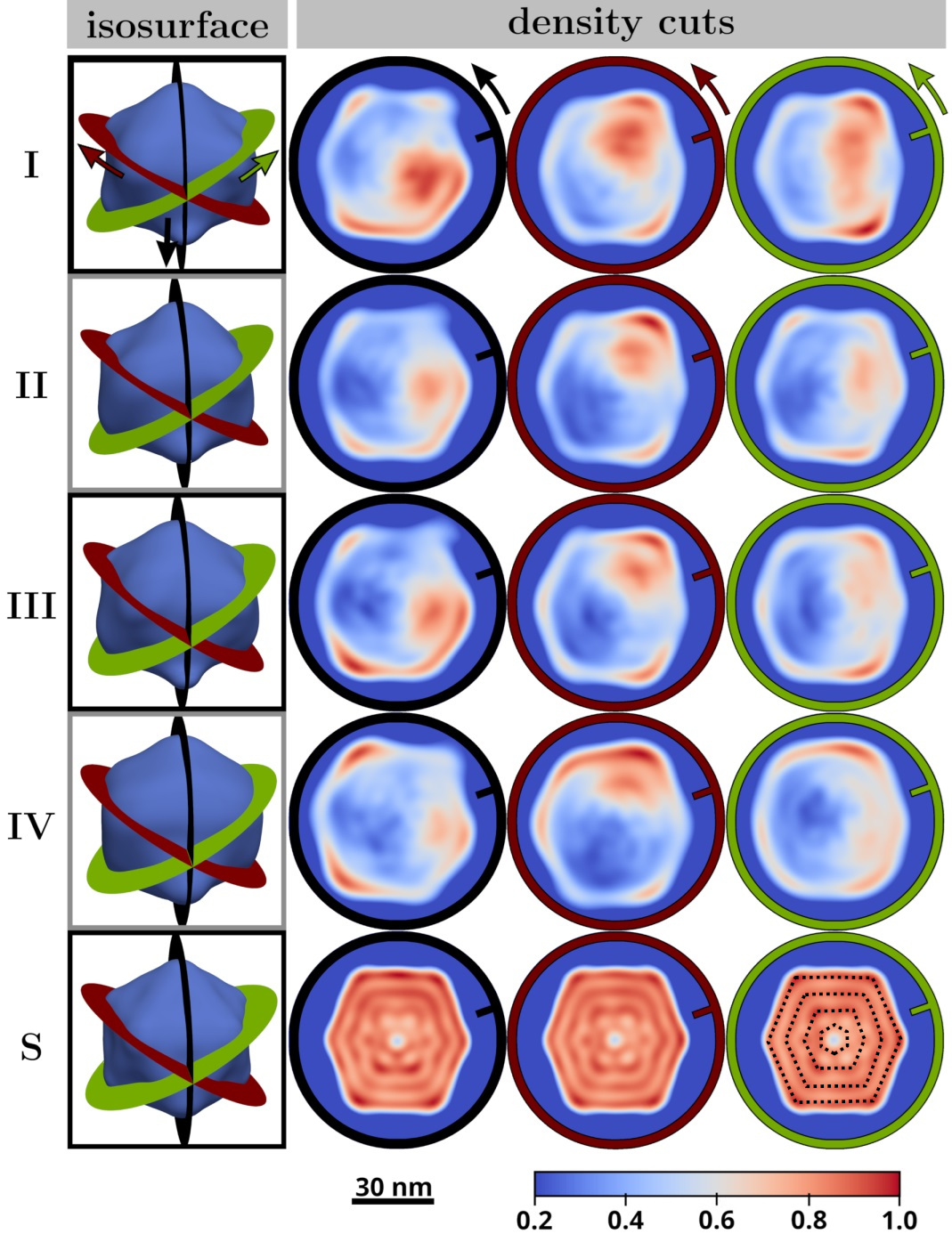}
\caption{\textbf{\textit{Ab initio} PR772 reconstructions from the amo06516 experiment.}
Rows $\mathrm{I}$-$\mathrm{IV}$ show reconstructions from the corresponding four subsets of data ($\mathrm{I}$-$\mathrm{IV}$) denoted in Fig.~\ref{fig:fig1}f. The bottom row (S) presents a reconstruction based on simulated X-ray scattering from ideal solid icosahedral particles with an average size of 70~nm and a polydispersity of 2~nm (see Methods).
Each reconstruction displays a 3D isosurface plot at the $30\%$ density level (leftmost column), as well as three color-coded (red, green, and black) density sections through the 3D structure.
These density sections are chosen to be equivalent under perfect icosahedral symmetry.
Their orientations (the same in all reconstructions) within the 3D isosurfaces are marked by the colored arrows in row $\mathrm{I}$.
The position of the intersection line of the three cutting planes is indicated in each density cut by a short line segment pointing from the circle rim to its center. 
The black dashed lines in the rightmost density section of (S) indicate the ringing artifacts due to the Gibbs phenomenon.
The bounding squares in the isosurface plots show the same field of view, while the color bar and the $30\,\textrm{nm}$ scale bar are valid for all density cuts.} 
\label{fig:fig3}
\end{figure}

\begin{figure}
\centering
\includegraphics[width=0.6\linewidth]{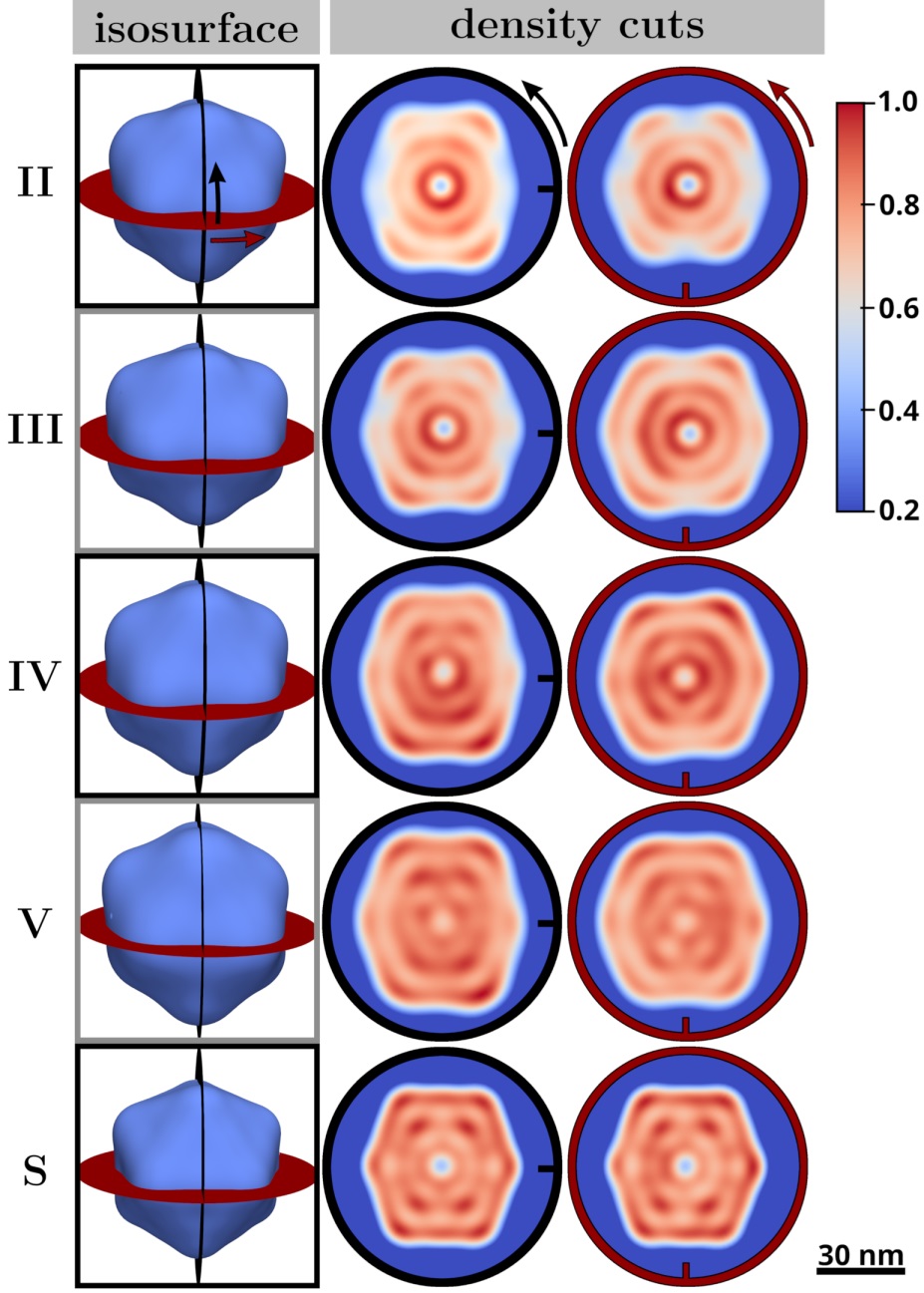}
\caption{\textbf{\textit{Ab initio} PR772 reconstructions from the amo86615 experiment.}
Rows $\mathrm{II}$-$\mathrm{V}$ show reconstructions from the four subsets of data ($\mathrm{II}$-$\mathrm{V}$) denoted in Supplementary Fig.~2e. The bottom row ($\mathrm{S}$) presents a reconstruction based on simulated X-ray scattering from perfect solid icosahedral particles, constrained by the experimental resolution of amo86615. The same data representation as in Fig.~\ref{fig:fig3} is used, featuring two instead of three symmetry-equivalent cutting planes.}
\label{fig:fig4}
\end{figure}

Some of the structural fingerprints in $|\textrm{arg}[C_{n}(q, q')]|$ can be interpreted by introducing a core-shell model, which aligns with the expected structure of the PR772 virus \cite{Reddy2019}, consisting of a lipid vesicle filled with genetic material (core) and enclosed by the icosahedral capsid (shell). The introduction of a spherical core with a reduced density $\rho_{\textit{c}}$ into the models of distorted icosahedra particularly alters the morphology of the $n=6$ invariants in the region of the 2D map highlighted by the red dashed square in the experimental phase map $|\textrm{arg}[C_{6}(q, q')]|$ in Fig.~\ref{fig:fig5} (also compare Supplementary Figs.~20 and 21). 
The width of the narrow speckles in this region (denoted by red arrows in Figs.~\ref{fig:fig5}e and \ref{fig:fig5}f) decreases as the size of the core approaches biologically relevant dimensions, thereby improving the agreement with the experimental map $|\textrm{arg}[C_{6}(q,q')]|$ in Fig.~\ref{fig:fig5} (see Supplementary Fig.~22 for additional simulation results).
This region of $C_{6}(q,q')$ also appears to be quite sensitive to the internal density of the virus, indicating a reduced density of the core content compared to the capsid density, $\rho_{\textrm{c}}<\rho_{\textrm{s}}$ (see Figs.~\ref{fig:fig5}g, \ref{fig:fig5}h, and Supplementary Fig.~23).
Variation of the core density leads to changes in the morphology (width $w$ and height $h$) of specific speckles in $C_{2}(q,q')$ for $n=2$ (Supplementary Fig.~23). Similar changes in $C_{2}(q,q')$ can be observed when comparing the invariants from the amo06516 and amo86615 experiments (see Figs.~\ref{fig:fig2}a,b and Supplementary Fig.~17), indicating that the virus particles in the amo06516 experiment are characterized by a reduced internal density compared to those in the amo86615 experiment.
It is also evident, that the aforementioned region in $|\textrm{arg}[C_{6}(q,q')]|$ is almost featureless in the amo06516 experiment (Supplementary Fig.~14), in contrast to the amo86615 (Supplementary Fig.~15) and amo11416 experiments (Supplementary Fig.~16).
This can be explained by the higher heterogeneity in the internal content of the viruses in the amo06516 experiment compared to the other two experiments, given the dependence of the morphology of $|\textrm{arg}[C_{6}(q,q')]|$ on core density variations, which can lead to the reduction and smearing of the narrow speckles in the specified part of the 2D map. 

As seen, the results of the model-guided analysis not only confirm our findings from the \textit{ab initio} reconstructions but also provide valuable insights into the 3D structure of the viruses, thereby increasing the overall fidelity of the analysis.
While the phases $|\textrm{arg}[C_{n}(q,q')]|$ provide a relatively simple yet powerful means for model-guided structural fingerprint analysis in correlation space, the amplitudes $|C_{n}(q,q')|$ are also an indispensable part of this analysis. We exploited the relationships between the SAXS intensity (directly related to the zero-order invariant) and higher-order invariants $C_n(q,q')$ for $n \geq 2$, to investigate the potential effect of solvent or debris covering the aerosolized virus particles (see Supplementary Note~9). Quantitative analysis based on several ``coated particle'' models suggests that such effects play a minor role in the considered experiments (see Supplementary Figs.~24 to 26). Additionally, these simulations demonstrate that analyzing particle sphericity—relevant for studying virus solvation or drying \cite{French2023}, maturation-induced remodeling \cite{Podgorski2025}, and osmotic swelling \cite{Zandi2020, Harder2023}—can be conveniently implemented using the rotational invariants.

The rotational invariants $\textrm{Re}[C_n(q,q')]$ determined for the asymmetrically distorted icosahedral core-shell structure shown in Fig.~\ref{fig:fig5}f are presented in Fig.~\ref{fig:fig2}d. The simulation results reproduce the experimental $C_n(q,q')$ for orders $n=2,4$ and 6 very well (see Figs.~\ref{fig:fig2}a-c). The imperfect agreement for higher orders $n \geq 8$ can be explained by the limited capability of the bead models used. Further improvements may be achieved by employing advanced forward-modeling approaches that go beyond the uniform-density and two-density approximations applied here, potentially including rigid-body models \cite{Vela2020} or all-atom molecular dynamics simulations \cite{Hadden2018, Miao2010, Jana2023}.

\section{Discussion}\label{sec3}

\textit{Ab initio} and model-guided 3D structure determination from multiple diffraction experiments with an XFEL allow us to characterize nanoscale structural features of aerosolized PR772 viruses, including distortions in the capsid shape, nonuniformities in the encapsulated virus content, and an extension at one of the capsid vertices. 

Remarkably, the determined rotational invariants reveal deviations of the PR772 capsid from ideal icosahedral symmetry, consistently observed across all three experiments considered here, suggesting that such distortions reflect a common structural property of the viruses in the studied aerosols.
Independent realizations of the amo06516, amo86615, and amo11416 experiments with variable parameters (photon energy, sample-detector distance, etc) allow us to exclude the experimental geometry and X-ray properties as potential sources of spurious structural distortions. 
Several studies have shown that aerosolization and environmental conditions (temperature, relative humidity) 
can affect the structural integrity and infectivity of PR772 on minute-to-hour timescales \cite{Turgeon2014, Verreault2015, Dubuis2020, French2023}.
 While in our experiments the diffraction patterns were measured within a few seconds after aerosolization, we cannot exclude the possibility of fast structural response of the virus to rapidly changing environmental conditions in our experiments, including solvent evaporation and the pressure and temperature changes that virus particles experience during aerosolization into a vacuum. Analysis of scaling relationships between SAXS and higher-order invariants (see Supplementary Note~9) does not reveal evidence of significant solvent coverage of the virus capsid in the studied experimental datasets. 

The femtosecond duration of XFEL pulses used in our diffraction experiments allows, in principle, for capturing instantaneous snapshots of PR772 virus dynamics. Simulations of icosahedral capsids indicate that both symmetric and asymmetric dynamics are functionally important \cite{Jana2023}. For instance, all-atom molecular dynamics simulations of the smaller T = 4 icosahedral capsid of the Hepatitis B virus (HBV) reveal intrinsically asymmetric global dynamics, exhibiting subtle ellipsoidal distortion even under equilibrium conditions \cite{Hadden2018}. Therefore, it is plausible to interpret the asymmetries in virus shapes observed in our experiments as resulting from asymmetric capsid dynamics in aerosols. While the rotational invariants indicate considerable reproducibility of capsid distortions in our experiments, the influence of environmental conditions on the virus dynamics represents an important topic for future XFEL studies \cite{Lin2020}.

Note that the high-resolution structures of PR772 and RDV capsids, as reported in previous cryo-EM \cite{Reddy2019} and X-ray crystallography \cite{Nakagawa2003} studies, respectively, possess perfect icosahedral symmetry. 
The application of symmetry constraints in determining virus capsid structures is a common practice, primarily aimed at reducing the complexity of structural analysis. In X-ray crystallography measurements, symmetry can be intrinsically imposed by the crystal itself, leading to apparently symmetric average structures. At the same time, the results of our XFEL experiments on ultrafast scattering from individual free-flying viruses, obtained without applying symmetry constraints, consistently show  deviations from perfect icosahedral shape.
Several other experimental XFEL studies have also reported distortions of icosahedral shape for PR772 and RDV viruses \cite{Kurta2017, Rose2018, Assalauova2020}, although providing limited detail on the nature of those distortions.
The potential of the correlation approach to extract single-particle rotational invariants from multiple-particle X-ray measurements in liquid solution \cite{Pande2018} under near-biologically relevant conditions offers hope for further insights into the origins of the broken virus symmetries.

Density nonuniformities inside the virus and the faint density extension from one of the vertices observed in the amo06516 reconstructions overall align with the structural transformations necessary for genome transmission reported for the PRD1 virus, a close relative of PR772 from the \textit{Tectiviridae} family \cite{Peralta2013}.
At the same time, because the sample-preparation workflow was essentially the same across all our experiments, these structural features may have been introduced during aerosolization. The injection parameters were optimized individually in each experiment to maximize the hit rate, which could have affected virus integrity and resulted in increased sample heterogeneity in the amo06516 experiment. 

All-atom molecular dynamics simulations indicate that viral capsids not only respond to, but also can influence, the properties of the surrounding environment \cite{Hadden2018}. Thus, the ability to access the structure of viruses — unaffected by imposed or naturally arising symmetry constraints — can provide new insights into their functionality. 
The femtosecond XFEL pulses can capture virus conformations during both equilibrium dynamics and non-equilibrium transformations, when the virus may exhibit symmetry breaking \cite{Miao2010, Castellanos2012, Hadden2018, Jose2022, Jana2023}.
With their remarkable sensitivity to the structural properties, the rotational invariants are perfectly suited for studying structure and dynamics of bioparticles using XFEL. The availability of complementary \textit{ab initio} and forward-modeling approaches based on single-particle rotational invariants \cite{Liu2012, Donatelli2015, Kurta2017, Pande2018, Zhao2024, Berberich2024} can, in particular, enhance the fidelity of structural analysis at limited experimental resolution.

\section{Methods}\label{sec4}

\subsection{Determination of $\mathbf{C}_n$ using difference-pattern CCF} \label{subsec4A}
The difference pattern CCF $C_{\textrm{diff}}(q,q',\Delta)$ defined in Eq.~\eqref{eq:Cdiffmain} allows to enhance the signal-to-noise ratio of the experimental rotational invariants, as compared to the ``classical'' CCF \cite{Kam1977},
$
 C(q,q',\Delta)=\langle\langle I_{j}[q,\theta(q),\phi]I_{j}[q',\theta'(q'),\phi+\Delta]\rangle_{\phi}\rangle_{j},
$
the latter being determined as an average over $M$ individually measured diffraction patterns. It can be readily shown (see Supplementary Note~1), that the FCs of $C(q,q',\Delta)$, defined as 
$C_n(q,q')=1/(2\pi)\int_{0}^{2\pi}C(q,q',\Delta)\exp{(-in\Delta)}\mathrm{d}\Delta$, are directly related to the FCs of $C_{\textrm{diff}}(q,q',\Delta)$ as $C_{n}(q,q')=C_{\textrm{diff},n}(q,q')/2$, for $n\neq 0$.
Using the latter relationship we arrive at Eq.~\eqref{eq:CnBl} \cite{Berberich2024}.
\subsection{Treatment of the masked areas on the difference diffraction patterns \label{subsec4B}}

Certain parts of the difference diffraction patterns do not contain useful information (e.g., due to gaps between tiles of a modular detector, dead pixels, etc.) and therefore need to be excluded from the analysis (masked). Considering each difference pattern on a polar grid, where the angular grid points, $\Delta_t$ and $\phi_t$, are given by $\Delta_t=\phi_t=t2\pi/N_{\phi}$, with $N_\phi$ as the number of angular grid points, and $q_{p}$ are the radial grid points with a total of $N_q$ radial sampling points, Eq.~\eqref{eq:Cdiffmain} 
can be rewritten in the following form:
\begin{align}
  \label{eq:CdiffAverPrac2}
  C_{\textrm{diff}}(q_p,q'_p,\Delta_t) = \frac{\langle C_{\textrm{diff\_masked},i}(q_p,q'_p,\Delta_t)\rangle_i}{C_{\textrm{mask}}(q_p,q'_p,\Delta_t)}.
\end{align}
Here, $C_{\textrm{diff\_masked},i}(q_p,q'_p,\Delta_t)=\langle I_{\textrm{diff},i}(q_p,\phi_f) W(q_p,\phi_f) I_{\textrm{diff},i}(q'_p,\Delta_t + \phi_f)W(q'_p,\Delta_t + \phi_f)\rangle_{\phi_f}$ is the CCF of the $i$-th masked difference pattern, $W(q_p,\phi_f)$ is the binary mask, assumed to be the same for each difference image, and $C_{\textrm{mask}}(q_p,q'_p,\Delta_t)=\langle W(q_p,\phi_f) W(q'_p,\Delta_t + \phi_f) \rangle_{\phi_f}$ is the CCF of the mask. The mask $W(q_p,\phi_f)$ has the value of $0$ for all sampling points $(q_p,\phi_f)$ for which image data should be excluded (masked) from the analysis, and the value of $1$ otherwise. Eq.~\eqref{eq:CdiffAverPrac2} has been applied in this work to the experimental data, to properly account for the masked (missing) data for each coordinate triplet $(q_p,q'_p,\Delta_t)$ (see  Supplementary Note~2 for details).

\subsection{Calculation of the CCF using the Discrete Fourier Transform\label{subsec4C}}

Direct calculation of the CCF in the form of Eq.~\eqref{eq:Cdiffmain} or \eqref{eq:CdiffAverPrac2} may quickly become computationally expensive for a large number $M_\textrm{diff}$ of difference patterns considered in the average, or for high angular sampling $N_\phi$. Therefore, it is customary to speed-up such computations by exploiting the known relationships between the angular FCs of the CCF and the angular FCs of the 2D images from which the CCF is determined \cite{Altarelli2010}.
For instance, one can compute the CCF of the mask $C_{\textrm{mask}}(q_p,q'_p,\Delta_t)$ in three steps, starting from a pure mask $W(q_p,\phi_f)$ as,
\begin{subequations}\label{eq:MaskFT}
\begin{align}
&w_{n}(q_p) =\frac{1}{N_\phi}\sum_{f=0}^{N_\phi-1}W(q_p,\phi_f)\exp{(-in\phi_f)},\label{eq:MaskFT1}\\
&C_{\textrm{mask},n}(q_p, q'_p) =\frac{1}{N_\phi}\sum_{t=0}^{N_\phi-1}C_{\textrm{mask}}(q_p,q'_p,\Delta_t)\exp{(-in\Delta_t)}=w_{n}(q_p)w^{\ast}_{n}(q'_p),\label{eq:MaskFT2}\\
&C_{\textrm{mask}}(q_p,q'_p,\Delta_t) =\sum_{t=0}^{N_\phi-1}C_{\textrm{mask},n}(q_p,q'_p)\exp{(in\Delta_t)},\label{eq:MaskFT3}
\end{align}
\end{subequations}
where $w_{n}(q_p)$ are the angular FCs of the mask, and $C_{\textrm{mask},n}(q_p, q'_p)$ are the angular FCs of the mask CCF. This method allows the CCFs to be computed much more efficiently using the angular Discrete Fourier Transform (DFT). We applied this approach in the calculations of all CCFs in this work.

\subsection{Regularization of rotational invariants $\mathbf{B}_l$\label{subsec4D}}

If we interpret the invariants $B_l(q,q')$ for each fixed $l$ as a matrix $\mathbf{B}_l$ in $q,q'$ of size $(N_q, N_q)$, we know \cite{Kam1977, Saladin09, Donatelli2015} that there exists a $(N_q,2l+1)$ matrix $\mathbf{V}_l$ such that, in the absence of experimental noise and numerical errors,
\begin{equation}
  \mathbf{B}_l= \mathbf{V}_l\mathbf{V}_l^{\dagger}.
  \label{eq:bl2vl_2}
\end{equation}
This implies that $\mathrm{rank}(\mathbf{B}_l)=\mathrm{rank}(\mathbf{V}_l)\le 2l+1$, for the typical case of $N_q \ge 2l+1$. 
In practice, the rank condition is never exactly satisfied, since already numerical rounding errors in extracting $\mathbf{B}_l$ from Eq.~\eqref{eq:CnBl} ensure that it has full rank.
The problem is therefore to find $\mathbf{V}_l$ such that $\mathbf{V}_l\mathbf{V}_l^{\dagger}$ optimally approximates $\mathbf{B}_l$ in some metric.
Typically $V_l$ is constructed using the eigenvectors associated with the $2l+1$ highest eigenvalues of $\mathbf{B}_l$ \cite{Donatelli2015,Berberich2024}, which minimizes the Frobenius norm $||\mathbf{B}_l - \mathbf{V}_l\mathbf{V}_l^{\dagger}||_F$.
However, this norm neglects the intrinsic steep decrease of $\mathbf{B}_l$ as a function of $q,q'$ (by several orders of magnitude in the present work), which leads to $\mathbf{V}_l$ that approximate $\mathbf{B}_l$ very well in the low $q,q'$ region but not at all in the mid to high range (see Supplementary Fig.~4).
To mitigate this effect, we implement the following algorithm to determine $\mathbf{V}_l$:
\begin{enumerate}
\item Use a symmetric variant of Ruiz equilibration \cite{Ruiz2001}, to find a column vector $\mathbf{d}$ such that $(\mathbf{\widetilde{B}}_l)_{i,j}= \mathbf{d}_i(\mathbf{B}_l)_{i,j}\mathbf{d}_j$ and the maximum of each individual row and each individual column of $\mathbf{\widetilde{B}}_l$ is equal to 1. This removes the fast decay in $q,q'$.
\item Solve the low-rank approximation problem under the Frobenius norm $\|\cdot\|_F$,
\begin{equation}
  \label{eq:vl_opt}
  \mathbf{\widetilde{V}}_l = \underset{\text{rank}(\mathbf{\widetilde{V}}_l)=2l+1}{\text{argmin}} \|\mathbf{\widetilde{B}}_l - \mathbf{\widetilde{V}}_l\mathbf{\widetilde{V}}_l^{\dagger}\|_F,
\end{equation}
for $\mathbf{\widetilde{B}}_l$ instead of $\mathbf{B}_l$. That is, use the $2l+1$ highest positive eigenvalues $\widetilde{\lambda}_n$ and the associated eigenvectors $\widetilde{e}_n$ of $\mathbf{\widetilde{B}}_l$ to obtain $(\mathbf{\widetilde{V}}_l)_{i,n}= \sqrt{\widetilde{\lambda}_n}(\mathbf{\widetilde{e}}_n)_i$.
\item Finally, determine $\mathbf{V}_l$ by removing the equilibration through $(\mathbf{V}_l)_{i,n} = \frac{1}{\mathbf{d}_i}(\mathbf{\widetilde{V}}_l)_{i,n}$. 

Note that by construction we have:
  \begin{align*}
    \label{eq:rengularize_V}
    \mathbf{V}_l\mathbf{V}_l^{\dagger} = \frac{1}{\mathbf{d}_i\mathbf{d}_j} \sum_{n=0}^{2l+1}(\mathbf{\widetilde{V}}_l)_{i,n}(\mathbf{\widetilde{V}}_l)_{n,j}^* \approx \frac{1}{\mathbf{d}_i\mathbf{d}_j} (\mathbf{\widetilde{B}}_l)_{i,j} = \mathbf{B}_l.
  \end{align*}  
\end{enumerate}

As can be seen in Supplementary Fig.~4, this computation scheme leads to better preservation of sign (phase) boundaries in the step from $\mathbf{B}_l$ to $\mathbf{V}_l\mathbf{V}_l^\dagger$ (implemented in xFrame), which are known to hold valuable structural information \cite{Kurta2017, Kurta2023}. 
We also tested an additional regularization approach, based on partitioning the matrices $\mathbf{B}_l$ and $\mathbf{V}_l$ (see Supplementary Note~3). The latter, however, leads to rather minor improvements, if applied in combination with the Ruiz regularization approach.

\subsection{Sample preparation and injection \label{subsec4S}}

The sample preparation and injection procedures,  described in detail in \cite{Reddy2017} for the amo86615 experiment and in \cite{Li2020} for the amo06516 and amo11416 experiments, are briefly overviewed here. Sample preparation followed a similar workflow for each experiment. The bacteriophage PR772 was cultured on agar using the overlay method. Purified PR772 viruses were maintained in storage buffer (TRIS 50 mM, NaCl 100 mM, MgSO4 1 mM, EDTA 1 mM, pH 8.0) and transferred into a volatile ammonium acetate buffer (250 mM, pH 7.5) prior to X-ray diffraction measurements. The sample solution was aerosolized with a gas dynamic virtual nozzle at a flow rate of 1-2 $\mu$l $\textrm{min}^{-1}$. The resulting aerosol passed through a differentially pumped skimmer and relaxation chamber for removal of excess carrier gas and pressure reduction. A focused stream of virus particles was then delivered into the X-ray interaction region via an aerodynamic lens stack. The sample-solution and carrier-gas flows were tuned to optimize the single-particle hit rate.

\subsection{Experimental data reduction and filtering \label{subsec4E}}

Diffraction data measured in the amo06516, amo86615, and amo11416 experiments include different numbers of patterns identified as single PR772 hits \cite{Reddy2017, Li2020}. 
The pixel masks and background corrections have been determined and applied individually in each experiment. The centers of diffraction patterns were refined individually, and the images were converted to discrete polar coordinates representation using second-order spline interpolation.
Diffraction patterns were then  corrected for geometric experimental factors, such as polarization of incident X-rays and different solid angle covered by individual detector pixels.
These sets of single hits were further classified based on the analysis of the $q$-dependent, azimuthally averaged intensity profiles, $I(q)\equiv\langle I(q,\phi) \rangle_{\phi}=1/(2\pi)\int_{\phi=0}^{2\pi} I(q,\phi) \mathrm{d}\phi$, where $q$ and $\phi$ denote the polar coordinates in the detector plane (Fig.~\ref{fig:fig1}b). These one-dimensional (1D) profiles were fitted with a form factor of a spherical particle, $I_{\textrm{spher}}(q)=A[(\sin(qR_{\rm s})-qR_{\rm s}\cos(qR_{\rm s}))/q^3]^2$ (see Supplementary Figs.~2a-c). Here $A$ is the scaling parameter, and $R_{\rm s}$ is the radius of the volume-equivalent spherical particle. The size of an ideal icosahedral particle $D_{\text{icos}}$, which we define here as the maximum pair distance in the particle, or $D_{\text{icos}}=2R_\text{cirscr}$ (see the inset in Fig.~\ref{fig:fig1}f), where $R_\text{cirscr}$ is the radius of a circumscribed sphere, can then be approximately determined as $D_{\rm icos}\approx2.36R_{\rm s}$ \cite{Kurta2017}. The fitting range was adjusted individually for each experiment depending on the data quality and missing data; that is, $q=(0.07,0.2)\;\text{nm}^{-1}$ for amo86615, $q=(0.09,0.34)\;\text{nm}^{-1}$ for amo06516, and  $q=(0.18,0.44)\;\text{nm}^{-1}$ for the amo11416 experiment. 
Finally, the root-mean-square error (RMSE) of a fit, as well as the average intensity of an image, $I_{\text{aver}}=1/N_{\text{pix}}\sum_{i=1}^{N_{\text{pix}}}I_i$, where the summation of intensities $I_i$ is performed over $N_{\text{pix}}$ informative (not masked) image pixels, were used to remove images with poor fitting results, as well as weak hits. All images with $\text{RMSE}\leq 0.7$ (arb. units) and $80 \leq I_{\text{aver}}\leq 500$ ADU in amo86615 experiment, as well as  $\text{RMSE}\leq 10.0$ (arb. units) and $5 \leq I_{\text{aver}}\leq 70$ ADU in the amo06516 experiment, were selected for further analysis (Table~\ref{TabS:TableNumbersPatterns}). In the case of the amo11416 experiment, a substantial number of images was recorded only by one of the detector halves. Such images were excluded from analysis in the present work. Considering a limited number of the remaining single-particle hits, no additional filtering using $\text{RMSE}$ or $I_{\text{aver}}$ has been applied to the amo11416 dataset. 
After such data processing and filtering, the histogram in Fig.~\ref{fig:fig1}f show virus size distributions determined for the remaining 3069 single hits in the amo06516 experiment. In all three experiments the main peak in the size distributions is located near 70~nm (see Supplementary Figs.~2d-f), that corresponds to the expected size of the PR772 virus. At the same time, the particle size distribution in the amo06516 experiment appears to be weakly bimodal. An additional smaller peak appears at about 100~nm, with a continuous distribution of sizes bridging the two peaks, suggesting a larger heterogeneity of the ensemble of virus particles studied in the amo06516 experiment, as compared to the other two experiments. The number of diffraction patterns in the subsets I-V noted in Supplementary Figs.~2d-f, corresponding to different particle sizes and experiments, notably varies (see Table~\ref{TabS:TableNumbersPatterns}), impacting the signal-to-noise ratio (SNR) of the obtained correlation functions, especially in the case of the amo11416 experiment.
Before determining the CCFs, all selected diffraction patterns were modified by setting the pixels with intensity values smaller than the threshold of $I_{\textrm{min}}=60\;\textrm{ADU}$ to zero (see Supplementary Fig.~1), and were normalized by the corresponding intensities $\langle I(q) \rangle_{q}$, averaged in the range of $q=(0.2, 0.4)\;\textrm{nm}^{-1}$.
\begin{table}
\centering
\caption{Single-particle hit counts in the selected subsets of data from the three experiments.}
\begin{tabular}{cccc}
Subset & amo86615 & amo06516 & amo11416 \\
\midrule
I & 58 & 497 & 20 \\
II & 762 & 842 & 55 \\
III & 1928 & 653 & 57\\
IV & 1404 & 387 & 35\\
V & 511 & 247 & 18\\
\bottomrule
\label{TabS:TableNumbersPatterns}
\end{tabular}
\end{table}

\subsection{Determination of experimental rotational invariants \label{subsec4F}}

The CCF $C_{\textrm{mask}}(q_p,q'_p,\Delta_t)$ of the mask $W(q_p,\phi_f)$ is first determined using Eqs.~\eqref{eq:MaskFT}. Similarly, the CCFs $C_{\textrm{diff\_masked},i}(q_p,q'_p,\Delta_t)$ are computed for all masked difference images $I(q_p,\phi_f)$ in a given subset of data, and the average difference-pattern CCF $C_{\textrm{diff}}(q_p,q'_p,\Delta_t)$ is determined via Eq.~\eqref{eq:CdiffAverPrac2}. The FCs of the average difference-pattern CCF,  $C_{\textrm{diff},n}(q_p,q'_p)$, are determined using the DFT transform of $C_{\textrm{diff}}(q_p,q'_p,\Delta_t)$, similar to Eq.~\eqref{eq:MaskFT2}, and normalized by a factor of 2 to get the desired $C_n(q_p,q'_p)$ for $n>0$. The maximum order $n=n_{\textrm{max}}$ of nonvanishing $C_n(q_p,q'_p)$ is identified to set $L=n_{\textrm{max}}$. The invariants $\mathbf{B}_l$ for $0<l\le L$ are then derived from $\mathbf{C}_n$ by solving Eq.~\eqref{eq:CnBl}, while the 0-th degree invariant $\mathbf{B}_0$ is determined independently, as specified in the main text. The invariants $\mathbf{B}_l$ are subsequently regularized according to Eq.~\eqref{eq:bl2vl_2} and used in \textit{ab initio} reconstructions with xFrame.

\subsection{3D structure reconstructions using xFrame \label{subsec4G}}

For each subset of data from the amo86615 and amo06516 experiments, a total of 114 reconstructions were obtained, which were subsequently centered, rotationally aligned and averaged to produce the final reconstructed densities shown in Figs.~\ref{fig:fig3} and \ref{fig:fig4}. The same procedure was also applied to the simulated datasets.

As a starting density in all reconstructions, a spherically symmetric bump function multiplied with white noise was applied,
\begin{equation}
  \label{eq:bump_function}
  \rho_0(\mathbf{r})= X(\mathbf{r})
  \begin{cases}
    e^{\frac{1}{2} r_{\mathrm{max}}^2/(r_{\mathrm{max}}^2-|\mathbf{r}|^2)}& |\mathbf{r}|\leq r_{\mathrm{max}} \\
    0 & \text{otw.}
  \end{cases} \,,
\end{equation}
where $r_{\mathrm{max}} = 43$\,nm and $X(\mathbf{r})$ takes uniformly distributed random values in the interval $[1,2]$.
The initial real-space support was given by a sphere with a radius of 45\,nm. 
During phase retrieval, the support was updated using the shrinkwrap algorithm \cite{Marchesini03}, while additionally restricting the updated support to be the largest connected component of the shrinkwrap output. 
We also imposed reality and positivity constraints by setting the imaginary parts and negative real values of the electron density to zero. 

The phasing process consisted of a main loop and a refinement loop.
The main loop comprised 30 iterations of $60\times$HIO steps, followed by a shrinkwrap support update and $40\times$ER steps.
During the main loop, the HIO parameter $\beta$ \cite{Donatelli2015,Berberich2024} was kept constant at $\beta=1$.
The threshold parameter of the shrinkwrap support update \cite{Marchesini03} was set to 15\%, while the standard deviation of the Gaussian blur varied linearly from $\sigma=30$\,nm to $\sigma = 15$\,nm over 30 loop iterations.
During the refinement loop, we used 2 iterations of $10\times$HIO steps followed by $10\times$ER steps, while decreasing the HIO parameter $\beta$ on each $j$-th step from 0.6 to 0.45 according to the formula $\beta(j) = ae^{-\alpha j} + b$,
with $a=0.87$, $\alpha = 0.01$, and $b=-0.27$.

The obtained reconstructions were centered using their respective centers of density and rotationally aligned to a reference structure generated in a pairwise reduction scheme (see Supplementary Note~5). To rotationally align a pair of reconstructions, we used  their rotational cross-correlation \cite{Kostelec08},
\begin{align}
    \label{eq:rot_CC}
    C(\omega) = \int \mathrm{d}\mathbf{r} \rho(\mathbf{r})\, \mathbf{R}_\omega \left[\rho'(\mathbf{r}) \right], 
\end{align}
where $\rho(\mathbf{r})$ and $\rho'(\mathbf{r})$ are the corresponding 3D electron density distributions, and $\mathbf{R}_\omega$ defines a rotation by $\omega \in \mathrm{SO}(3)$.

The quality of the reconstructions was assessed using three different metrics: the phase retrieval transfer function (PRTF), the Fourier shell correlation (FSC), and an additional 3D confidence metric defined by the ratio of the mean to the standard deviation (see Supplementary Notes~6 and 7 for details).
In the present framework, the reconstructions are determined in a spherical coordinate system defined by $(q,\theta,\phi)$ in reciprocal space. The xFrame software package \cite{Berberich2024} uses uniform sampling points in $\phi$ and Gau\ss-Legendre nodes in $\theta$, yielding the following discretized expression for the FSC metric: 
\begin{equation}
    \label{eq:FSC_spherical_disctrete}
        \textrm{FSC}(q) = \frac{\sum_i \sum_j w_j\sin(\theta_j)\widehat{\rho}_1(q,\theta_j,\phi_i) \widehat{\rho}_2(q,\theta_j,\phi_i)^* }{\sqrt{\left(\sum_i \sum_j w_j \sin(\theta_j)\ |\widehat{\rho}_1(q,\theta_j,\phi_i)|^2 \right)\left( \sum_{i} \sum_j w_j \sin(\theta_j)|\widehat{\rho}_2(q,\theta_j,\phi_i)|^2 \right)}},
\end{equation}
where $\widehat{\rho}_1$ and $\widehat{\rho}_2$ are the scattering amplitudes determined from two reconstructions, $w_j$ are the Gau\ss-Legendre quadrature weights, and $\sin(\theta_j)$ accounts for the varying weights of sampling points (see Supplementary Note~6). 
This formulation of FSC appears to be equivalent to the following expression:
\begin{equation}
        \label{eq:FSC_spherical_harm}
        \textrm{FSC}(q) = \frac{\left(\widehat{\rho}_1(q,\theta,\phi) \widehat{\rho}_2(q,\theta,\phi)^*\right)^0_0 }{\sqrt{\left(|\widehat{\rho}_1(q,\theta,\phi)|^2 \right)^0_0\left(|\widehat{\rho}_2(q,\theta,\phi)|^2 \right)^0_0 }},
\end{equation} 
where $(f)^0_0$ defines a spherical harmonic expansion coefficient of the function $f$, enclosed in round brackets, for degree $l = 0$ and order $m = 0$. 

The PRTF customized for MTIP was computed as,
\begin{equation}
        \label{eq:PRTF}
        \mathrm{PRTF}(q)= \sum_i \sum_j w_j\sin(\theta_j)\frac{|\langle\widehat{\rho}_k(q,\theta_j,\phi_i)\rangle_k|}{\sqrt{\langle I_k(q,\theta_j,\phi_i) \rangle_k}},
\end{equation}
where $\langle\cdot\rangle_k$ is the average over all reconstructions, $\widehat{\rho}_k$ represents the scattering amplitudes reconstructed after the last real-space projection step, and $I_k$ represents the intensities reconstructed after the last reciprocal-space projection step. 

The determined PRTF curves are shown in Supplementary Figs.~9a,b and 10a,b, for the amo06516 and amo86615 reconstructions, respectively.
Using the common $\frac{1}{e}$ resolution limit, we find that the reconstructions for all experimental datasets and size parts have a PRTF resolution of approximately 18\,nm.
The \textit{shtns} software \cite{Schaeffer2013} for spherical harmonic expansions was employed to compute the FSC profiles using Eq.~\eqref{eq:FSC_spherical_harm}. Considering the limited number of patterns used in each data subset (see Table~\ref{TabS:TableNumbersPatterns}), we chose to determine the FSC for reconstructions from the data subsets corresponding to similar virus sizes $D_{\textrm{icos}}$ from the amo06516 (Supplementary Figs.~9e,f) and amo86615 (Supplementary Figs.~10e,f) experiments. This metric provides a lower boundary assessment of the classically determined FSC metric, indicating resolutions in the range of $7$ to $13$\,nm for different pairs of reconstructions.    

To mitigate the effect of sharp dips in the obtained resolution metrics in the regions of the structure-factor minima, we also convolved the PRTF and FSC curves with a Gaussian of FWHM equal to the critical Shannon pixel size, which for a 70~nm virus is approximately $0.04$ nm$^{-1}$. From such smoothed PRTF profiles (see Supplementary Figs.~9c,d and 10c,d), the resolution can be estimated to lie between 11 and 13~nm in all reconstructions, whereas the smoothed FSC profiles yield about 6~nm for all reconstructions in the amo06516 experiment (see Supplementary Figs.~9g,h), and about 9~nm in the amo86615 experiment (see Supplementary Figs.~10g,h).

An additional 3D confidence metric was calculated as the ratio of the average 3D reconstruction to the 3D standard deviation (see Supplementary Note~7 and Supplementary Figs.~7 and 8). This metric, which is the inverse of the coefficient of variation, allows one to estimate the confidence of different regions in the obtained reconstructions.

\subsection{Forward-scattering modeling \label{subsec4H}}

We constructed bead (dummy-atom) models \cite{Vela2020} of icosahedrally-shaped particles that match the size (approximately 70~nm) of the PR772 virus. In our simulations, we used a spherical bead with a diameter of 1~nm and an electron density of $\rho_{\textrm{el}}=0.325~\textrm{electrons}/\text{\AA}$, which is also assumed to be the electron density $\rho_{\textrm{s}}$ of the virus capsid. In more complex models, such as the core-shell model, the beads composing the core have an electron density $\rho_{\textrm{c}}$ specified as a fraction of the capsid (shell) density $\rho_{\textrm{s}}$. We also employ an atomistic model of the hollow PR772 capsid \cite{Reddy2019}.

 Using the constructed models, we simulate X-ray diffraction under conditions similar to the experimental setup: photon energy $E=1.6\;\text{keV}$, a sample-detector distance of 581~mm, a square detector with $512\times 512$ pixels, and a pixel size of $300\times 300\;\mu\text{m}$ (four times larger compared to the experiment to speed-up the simulations). For a particle size of 70~nm, such conditions correspond to a detector oversampling of about 21~pixels/speckle and a maximum resolution $q_{\text{max}}=1\;\text{nm}^{-1}$ at the detector edge. For each constructed model of a virus particle, a set of $M=2000$ single-particle 2D patterns is simulated from the particle in uniformly distributed orientations within the SO(3) rotational group, mimicking the experimental measurement. The set of simulated patterns then undergoes a processing pipeline  as described above to obtain $\mathbf{C}_n$ or $\mathbf{B}_l$. See Supplementary Notes ~8 and 9 for detailed simulation results.
 
\subsection{Data availability}\label{subsec4I}
The single-particle X-ray diffraction data from PR772 reported in \cite{Reddy2017} and \cite{Li2020} are publicly accessible via the Coherent X-ray Imaging Data Bank \cite{Maia2012} under CXIDB IDs 58 and 156. The difference-pattern CCFs and SAXS intensities used for \textit{ab initio} 
reconstructions in xFrame are available under CXIDB ID 238.
Additional data are available from the corresponding author upon request.

\subsection{Code availability}\label{subsec4I}
The open-source software suite xFrame for determining the rotational invariants from diffraction patterns and performing \textit{ab initio} reconstructions using MTIP is available on GitHub at \href{https://github.com/European-XFEL/xFrame}{https://github.com/European-XFEL/xFrame} \cite{Berberich2024}.

\bibliography{references}

\section{Acknowledgements}
Use of the Linac Coherent Light Source (LCLS), SLAC National Accelerator Laboratory, is supported by the U.S. Department of Energy, Office of Science, Office of Basic Energy Sciences under Contract No. DE-AC02-76SF00515.
This research was supported in part through the Maxwell computational resources operated at Deutsches Elektronen-Synchrotron DESY, Hamburg, Germany.
J.A.S. acknowledges the Swedish Research Council (2023-06350), the Göran Gustafsson Foundation (2044) and the Carl Tryggers Stiftelse för Vetenskaplig Forskning (CTS 21-1427). P.L.X. thanks Joachim Herz Stiftung for a fellowship and European Research Council—Frontiers in Attosecond X-ray Science: Imaging and Spectroscopy (AXSIS) (ERC-2013-SyG 609920) and the graduate student training program of LCLS-Stanford. This research also used resources of the National Synchrotron Light Source II, a U.S. Department of Energy (DOE) Office of Science User Facility operated for the DOE Office of Science by Brookhaven National Laboratory under Contract No. DE-SC0012704.
This  work  was also supported by the National Science Foundation through the  BioXFEL Science and Technology Center grant 1231306. The work of P.F. was also supported by the Biodesign Center for Applied Structural Discovery at Arizona State University.
R.P.K. acknowledges useful communications with Charlotte Uetrecht and Filipe Maia.

\section{Author contributions}

J.B., J.A.S., B.J.D., C.N., P.L.X., I.A.V., G.J.W., R.B., P.F., A.A., A.P.M., and R.P.K. participated in the X-ray experiments and collected the data; R.P.K. performed cross-correlation analysis of the experimental data and model-guided structural analysis; T.B.B. performed \textit{ab initio} structure reconstructions; T.B.B. and R.P.K. wrote the manuscript.
All authors contributed to the discussion of the experimental and simulation results and edited the manuscript.


\section{Competing interests}
The authors declare no competing interests.

\pagebreak

\begin{figure}
\centering
\includegraphics[width=0.9\linewidth]{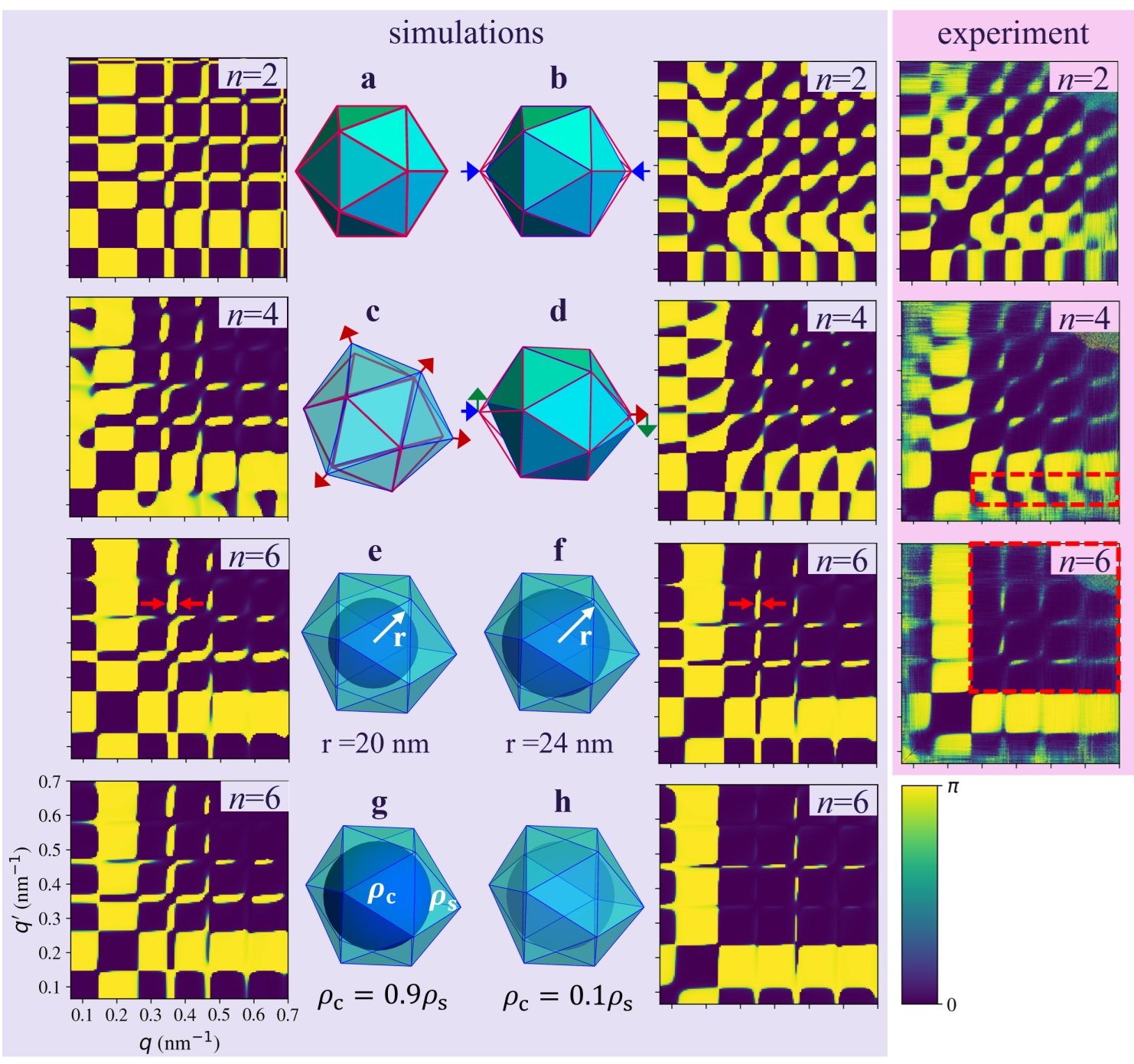}
\caption{\textbf{Model-guided structural analysis.} \textbf{a-h} Schematic illustration of the applied bead-model structures, each accompanied by a simulated 2D phase map $|\textrm{arg}[C_{n}(q, q')]|$ of the specified order $n$, positioned to the left (a,c,e,g) or to the right (b,d,f,h) of the respective structure. The phase maps illustrate changes induced by specific structural features within the corresponding model structures. The tested models include: (a) perfect solid icosahedron; (b) solid icosahedron with two opposite vertices symmetrically compressed radially by $7.5\%$; (c) solid icosahedron with asymmetric radial elongation of four vertices, each elongated by $15\%$; (d) solid icosahedron with a composite distortion, including radial compression of one vertex by $10\%$ and extension of the opposite vertex by  $5\%$, combined with a tangential shift of these two vertices in opposite directions; (e) core-shell icosahedron distorted as in (d), featuring a spherical core with radius $r=20\;\textrm{nm}$ and core electron density $\rho_{\textrm{c}}=0.7\rho_{\textrm{s}}$, where $\rho_{\textrm{s}}$ is the shell density; (f) same structure as (e), but with larger core radius $r=24\;\textrm{nm}$; (g) same structure as (f), but with higher core density $\rho_{\textrm{c}}=0.9\rho_{\textrm{s}}$; (h) same structure as (g), but with lower core density $\rho_{\textrm{c}}=0.1\rho_{\textrm{s}}$. The experimental phase maps for orders $n=2,4$, and 6, determined for subset $\mathrm{III}$ of the amo86615 experiment, are presented in the rightmost column for comparison. The axes on the bottom-left map and the color bar are the same for all phase maps $|\textrm{arg}[C_{n}(q, q')]|$ in this figure. Red arrows and dashed lines on the phase maps indicate the features discussed in the text.}
\label{fig:fig5}
\end{figure}
\end{document}



\section*{\fontsize{24}{28}\selectfont Supplementary Information}

\section*{Supplementary Note 1: Derivation of the difference-pattern cross-correlation function (CCF) \label{sec:DiffCCF}}

In this work we employ a modified form of the CCF to access $C_n(q,q')$, which helps to reduce the effect of scattering background that may contaminate the commonly used CCF $C(q,q',\Delta)$.

Within this approach we determine the $i$-th difference pattern as,
\begin{equation}
 I_{\textrm{diff}, i}(q,\phi)=I_j(q,\phi)-I_k(q,\phi),
\label{eq:Idiff}
\end{equation}
where the subscripts $j$ and $k$ denote two different ($j\neq k$) diffraction patterns.

We then determine the CCF $C_{\textrm{diff}, i}(q,q',\Delta)$ for the $i$-th difference pattern $I_{\textrm{diff}, i}(q,\phi)$ as 
\begin{equation}
 C_{\textrm{diff}, i}(q,q',\Delta)=\langle I_{\textrm{diff}, i}(q,\phi)I_{\textrm{diff}, i}(q',\phi+\Delta)\rangle_{\phi},
\label{eq:Cdiff}
\end{equation}
where $q=\lvert\bm{q}\rvert$, $q'=\lvert\bm{q'}\rvert$, and the angles $\Delta$ and $\phi$ are determined in the range $[0, 2\pi)$. 
The averaging $\langle \cdot \rangle_{\phi}$ is performed over the angular coordinate $\phi$, 
defined as $\langle I_{\textrm{diff},i}(q,\phi)\rangle_{\phi}=1/(2\pi)\int_0^{2\pi} I_{\textrm{diff},i}(q,\phi) \mathrm{d}\phi$. 
Statistical averaging is then performed over $M_\textrm{diff}$ difference-pattern CCFs $C_{\textrm{diff}, i}(q,q',\Delta)$ as follows:
\begin{equation}
 C_{\textrm{diff}}(q,q',\Delta) \equiv \langle C_{\textrm{diff}, i}(q,q',\Delta) \rangle_{i}=\frac{1}{M_\textrm{diff}}\sum_{i=1}^{M_\textrm{diff}}C_{\textrm{diff}, i}(q,q',\Delta).
\label{eq:CdiffAver}
\end{equation}
In the calculations of $I_{\textrm{diff}, i}(q,\phi)$ Eq.~\eqref{eq:Idiff} we use each available diffraction pattern only once, thus for $M$ measured diffraction patterns the summation in Eq.~\eqref{eq:CdiffAver} is performed over $M_\textrm{diff}=\left\lfloor \frac{M}{2} \right\rfloor $ unique difference patterns (here $\lfloor \cdot\rfloor$ denotes the floor operation).

It can be shown (see below), that the angular FCs of the difference CCF averaged over a set of single-particle diffraction patterns,
\begin{equation}
C_{\textrm{diff},n}(q,q')=1/(2\pi)\int_{0}^{2\pi}\langle C_{\textrm{diff}, i}(q,q',\Delta) \rangle_{i}\exp{(-in\Delta)}\mathrm{d}\Delta,
\label{eq:Cndiff}
\end{equation}
and FCs of the ``classical'' average CCF $C(q,q',\Delta)$,
\begin{equation}
C_n(q,q')=1/(2\pi)\int_{0}^{2\pi}\langle C_{j}(q,q',\Delta)\rangle_{j}\exp{(-in\Delta)}\mathrm{d}\Delta,
\label{eq:Cn}
\end{equation}
are directly related as,
\begin{equation}
 C_n(q,q')=C_{\textrm{diff},n}(q,q')/2,\quad n\neq 0.
\label{eq:CntoCdiffn}
\end{equation}
Indeed, one can expand Eq.~\eqref{eq:Cdiff} using  Eq.~\eqref{eq:Idiff} as,
\begin{eqnarray}
&C_{\textrm{diff}, i}(q,q',\Delta)=\langle [I_j(q,\phi)-I_k(q,\phi)]\cdot[I_j(q',\phi+\Delta)-I_k(q',\phi+\Delta)]\rangle_{\phi}\nonumber \\
&=\langle I_j(q,\phi)I_j(q',\phi+\Delta)\rangle_{\phi}+\langle I_k(q,\phi)I_k(q',\phi+\Delta)\rangle_{\phi}\nonumber \\
&-\langle I_j(q,\phi)I_k(q',\phi+\Delta)\rangle_{\phi}-\langle I_k(q,\phi)I_j(q',\phi+\Delta)\rangle_{\phi}\label{Eq:Cdiffexp1}.
\end{eqnarray}
Note that the first two terms on the RHS of Eq.~\eqref{Eq:Cdiffexp1} are expressed as products of intensities determined from individual images, while the last two terms are defined as cross-products of intensities from distinct diffraction patterns $(j\neq k)$. We can then express the statistically averaged difference-pattern CCF in terms of the corresponding angular CCFs as,  
\begin{eqnarray}
\langle C_{\textrm{diff},i}(q,q',\Delta)\rangle_{i} =
\langle C_{j}(q,q',\Delta) \rangle_{j}+\langle C_{k}(q,q',\Delta) \rangle_{k} -\langle C_{jk}(q,q',\Delta) \rangle_{j\neq k} - \langle C_{kj}(q,q',\Delta) \rangle_{j\neq k},
\label{Eq:Cdiffexp2}
\end{eqnarray}
where the four CCFs on the RHS of equation are defined by the respective four terms in Eq.~\eqref{Eq:Cdiffexp1}, statistical averaging defined by $\langle\cdot\rangle_{j}$ and $\langle\cdot\rangle_{k}$ is performed over two different halves of a given dataset (each containing $M/2$ diffraction patterns), while averages  
$\langle\cdot\rangle_{i}$ and $\langle\cdot\rangle_{j\neq k}$ are performed over $\left\lfloor M/2 \right\rfloor$ unique pairs of patterns ($j\neq k$).
Notice, that $\langle C_{j}(q,q',\Delta) \rangle_{j}$  and $\langle C_{k}(q,q',\Delta) \rangle_{k}$ are statistically equivalent to the CCF $C(q,q',\Delta)$, where averaging is performed over individual diffraction patterns. 
We can then reformulate Eq.~\eqref{Eq:Cdiffexp2} in terms of the  Fourier components as,
\begin{equation}
C_{\textrm{diff},n}(q,q') = 2C_{\textrm{intra},n}(q,q') - C_{\textrm{inter},n}(q,q'),
\label{Eq:Cndiffexp1}
\end{equation}
where the FCs $C_{\textrm{diff},n}(q,q')$ and $C_{\textrm{intra},n}(q,q')\equiv C_n(q,q')$ are defined in Eqs.~\eqref{eq:Cndiff} and \eqref{eq:Cn}, respectively, and the FCs of the combined inter-pattern CCFs are determined as
\begin{eqnarray}
C_{\textrm{inter},n}(q,q') =1/(2\pi)\int_{0}^{2\pi}[\langle C_{jk}(q,q',\Delta) \rangle_{j\neq k} + \langle C_{kj}(q,q',\Delta) \rangle_{j\neq k}]\exp{(-in\Delta)}\mathrm{d}\Delta.
\label{Eq:Cndiffexp2}
\end{eqnarray}
The form of Eq.~\eqref{Eq:Cndiffexp1} is very close to the constructed ``difference Fourier spectrum'' in Ref.~\cite{Kurta2017}, however here it naturally arises because of performing the correlation analysis of difference patterns (Eq.~\eqref{eq:Idiff}). The basic idea behind this equation, is that in the absence of correlations between different diffraction patterns, the Fourier spectrum of the difference-pattern CCF, $C_{\textrm{diff},n}(q,q')$, coincides with the Fourier spectrum $C_{n}(q,q')$ (for $n\neq 0$) of the classical CCF, up to a scaling constant of 2. Such conditions arise when the distribution of particle orientations in the measured ensemble is uniform, and correlations in the detector background are absent. Assuming a uniform orientational probability distribution is reasonable for aerosolized particles without significant shape anisotropy.
However, in the presence of unwanted correlations $C_{jk}$ between different diffraction patterns, they will be accumulated in $C_{\textrm{inter},n}(q,q')$ and subtracted from 
$C_{\textrm{intra},n}(q,q')$, so that $C_{\textrm{diff},n}(q,q')$ becomes a corrected estimation of $C_{n}(q,q')$. 
Applications of such correction technique to the X-ray scattering data have shown substantial improvement of the experimentally-extracted correlation functions and their Fourier spectra \cite{Kurta2014, Kurta2017, Kurta2019}.

Recall that the FCs of the angular CCFs can be expressed using the angular FCs of the corresponding scattered intensities \cite{Altarelli2010}. Thus, one can write for $C_{jk}(q,q',\Delta)$,
\begin{eqnarray}
&1/(2\pi)\int_{0}^{2\pi}C_{jk}(q,q',\Delta)\exp{(-in\Delta)}\mathrm{d}\Delta\nonumber \\
&=1/(2\pi)\int_{0}^{2\pi}\langle I_j(q,\phi)I_k(q',\phi+\Delta)\rangle_{\phi}\exp{(-in\Delta)}\mathrm{d}\Delta=I_{j,n}(q)\cdot I^{\ast}_{k,n}(q'),
\label{eq:FTCCFintens}
\end{eqnarray}
where $I_{j,n}(q)$ and $I_{k,n}(q')$ are the angular FCs of intensities 
$I_j(q,\phi)$ and $I_k(q',\phi)$, respectively, 
\begin{equation}
I_{j,n}(q) =1/(2\pi)\int_{0}^{2\pi}I_j(q,\phi)\exp{(-in\phi)}\mathrm{d}\phi,
\quad I_{k,n}(q') =1/(2\pi)\int_{0}^{2\pi}I_k(q',\phi)\exp{(-in\phi)}\mathrm{d}\phi,
\label{eq:FTintens}
\end{equation}
and $\ast$ denotes complex conjugation.
Using Eqs.~\eqref{eq:FTCCFintens} and \eqref{eq:FTintens}, we can rewrite Eq.~\eqref{Eq:Cndiffexp2} as
\begin{eqnarray}
C_{\textrm{inter},n}(q,q') &=&\langle I_{j,n}(q)\cdot I^{\ast}_{k,n}(q') \rangle_{j\neq k} + \langle I_{k,n}(q)\cdot I^{\ast}_{j,n}(q') \rangle_{j\neq k}\nonumber\\
&=&\langle I_{j,n}(q)\rangle_{j} \langle I^{\ast}_{k,n}(q') \rangle_{k} + \langle I_{k,n}(q)\rangle_{k} \langle I^{\ast}_{j,n}(q') \rangle_{j},\label{Eq:Cndiffexp3}
\end{eqnarray}
where, in the last step, we factorized the average of products into the products of averages, which is possible because each pair of intensities in the product, $I_j(q,\phi)$ and $I_k(q,\phi)$, are mutually independent, since they are measured from independent sample realizations. 
Moreover, since diffraction patterns are measured from particles in random orientations, the higher order terms ($n>0$) in Eq.~\eqref{Eq:Cndiffexp3} vanish (assuming a uniform distribution of particle orientations), and only the zero-order FC survives after averaging. As a consequence, the two terms
in Eq.~\eqref{Eq:Cndiffexp3} become identical in statistical sense, and we can finally rewrite Eq.~\eqref{Eq:Cndiffexp1} as
\begin{equation}
C_{\textrm{diff},n}(q,q') = 2[C_{n}(q,q') - \delta_{n,0}I_{\text{SAXS}}(q)I_{\text{SAXS}}(q')],
\label{Eq:Cndiffexp4}
\end{equation}
where $\delta_{n,0}$ is the Kronecker delta, and we denoted $I_{\text{SAXS}}(q)\simeq \langle I_{j,0}(q)\rangle_{j}\simeq \langle I_{k,0}(q)\rangle_{k}$, and $I_{\text{SAXS}}(q')\simeq \langle I^{\ast}_{k,0}(q') \rangle_{k}\simeq \langle I^{\ast}_{j,0}(q') \rangle_{j}$. 
As one can see, for $n\neq0$ we arrive at Eq.~\eqref{eq:CntoCdiffn}. The 0-th order component $C_{\textrm{diff},0}(q,q')$ is reduced by the product $I_{\text{SAXS}}(q)I_{\text{SAXS}}(q')$ of small-angle X-ray scattering (SAXS) intensities. 
Notice, that Eq.~\eqref{Eq:Cndiffexp4} also arises in the correlation analysis of difference images in the pump-probe X-ray experiments on molecular solutions, where it defines time-independent ground-state molecular correlations \cite{Kurta2023}.

\section*{Supplementary Note 2: Details of handling missing data in diffraction patterns in CCF calculations \label{sec:CCFMask}}

In the correlation analysis of experimental data, certain parts of the diffraction patterns recorded by a 2D detector—e.g., gaps between tiles of a modular detector, dead pixels, etc.—may not contain useful information and therefore need to be excluded from the analysis (masked). Considering each image on a polar grid, where the angular grid points, $\Delta_t$ and $\phi_t$, are given by $\Delta_t=\phi_t=t2\pi/N_{\phi}$, $N_\phi$ is the number of angular grid points, and $q_{p}$ are the radial grid points, Eq.~\eqref{eq:CdiffAver} 
can be expanded on this grid in the following form \cite{Berberich2024},
\begin{align}
  \label{eq:CdiffAverPrac}
  C_{\textrm{diff}}(q_p,q'_p,\Delta_t) = \frac{1}{M_\textrm{diff}}\sum_{i=1}^{M_\textrm{diff}} \left[\frac{\sum_{f=0}^{N_{\phi}-1}  I_{\textrm{diff},i}(q_p,\phi_f) W_{i}(q_p,\phi_f) I_{\textrm{diff},i}(q'_p,\Delta_t + \phi_f)W_{i}(q'_p,\Delta_t + \phi_f)}{\sum_{f=0}^{N_{\phi}-1} W_{i}(q_p,\phi_f) W_{i}(q'_p,\Delta_t + \phi_f)} \right],
\end{align}
where $W_{i}(q_p,\phi_f)=W_{j}(q_p,\phi_f)\cdot W_{k}(q_p,\phi_f) $ is a binary mask for the $i$-th difference image, which is determined as a product of the binary masks defined for $j$-th and $k$-th patterns forming the $i$-th difference pattern (see Eq.~\eqref{eq:Idiff}). The mask has the value of $0$ for all sampling points $(q_p,\phi_f)$ for which image data should be excluded from the analysis, and the value of $1$ otherwise. 
It is, therefore, possible to rewrite Eq.~\eqref{eq:CdiffAverPrac} as,
\begin{align}
  \label{eq:CdiffAverPrac1}
  C_{\textrm{diff}}(q_p,q'_p,\Delta_t) = \frac{1}{M_\textrm{diff}}\sum_{i=1}^{M_\textrm{diff}} 
  \frac{C_{\textrm{diff\_masked},i}(q_p,q'_p,\Delta_t)}{C_{\textrm{mask},i}(q_p,q'_p,\Delta_t)},
\end{align}
where $C_{\textrm{diff\_masked},i}(q_p,q'_p,\Delta_t)=\langle I_{\textrm{diff},i}(q_p,\phi_f) W_{i}(q_p,\phi_f) I_{\textrm{diff},i}(q'_p,\Delta_t + \phi_f)W_{i}(q'_p,\Delta_t + \phi_f)\rangle_{\phi_f}$ is the CCF of the $i$-th masked difference pattern, and $C_{\textrm{mask},i}(q_p,q'_p,\Delta_t)=\langle W_{i}(q_p,\phi_f) W_{i}(q'_p,\Delta_t + \phi_f) \rangle_{\phi_f}$ is the CCF of the $i$-th mask. Thus, in practical applications the CCF of the masked image needs to be divided by the CCF of the corresponding binary mask. Such normalization properly accounts for the missing (masked) data by taking into account the correct number of terms in the angular sum over $\phi_f$ for each coordinate triplet $(q_p,q'_p,\Delta_t)$. If the applied mask is identical for all images in the average, that is $W_{i}(q_p,\phi_f)\equiv W(q_p,\phi_f)$, the CCF of the mask can be computed only once, and Eq.~\eqref{eq:CdiffAverPrac1} simplifies to Eq.~(5) in the main text. 

\section*{Supplementary Note 3: Recovery of rotational invariants in noisy regions \label{sec:RecovInvar}}

High-$q$ regions of the extracted invariants are typically the most affected by noise. Here, we describe a method that allows one to initially determine $V_l$ in the limited range $q,q'\leq Q$ and later extend $\bm{V_l}$ to cover the entire available momentum transfer range.
We found that this procedure has only a small effect on the computed $\bm{V_l}$, when using the Ruiz equilibration scheme, nevertheless it might prove useful in other regularization schemes.

Consider the ideal noise-free case in which $\bm{B_l}=\vtrd{}\vtrd{}^\dagger$. Splitting the momentum transfer points into low-$q$ and high-$q$ regions introduces a splitting of the invariant $\bm{B_l}$ via\footnote{Note that in matrix notation the $q'$ values increase from the upper-left corner downward, whereas in image representations of $\bm{B_l}$ (see, e.g., Fig.~\ref{fig:bl_regularization}) the  $q'$ axis is flipped and the origin lies in the lower-left corner.}
\NiceMatrixOptions{xdots={line-style = ->}}
\begin{equation}
  \label{eq:invariant_q_splitting}
  \bm{B_l} \ = \ \quad \  \begin{bNiceArray}{ccc|cc}[columns-width=0.45cm,margin,first-row,first-col,extra-left-margin=-2pt,extra-right-margin=-2pt]
    &\Hdotsfor{5}^{\text{\normalsize$q$}}\\
    \Vdots_{\rotatebox{90}{\text{$q'$}}}&\Block{3-3}{\bm{B_l}^1} &&&\Block{3-2}{\bm{A_l}} \\[0.2cm]
    &&&&&\\[0.2cm]
    &&&&&\\[0.2cm]
    \hline
    &\Block{2-3}{\bm{A_l}^\dagger}&&& \Block{2-2}{\bm{B_l}^2}\\[0.1cm]
    &&&&&
  \end{bNiceArray} \,,
\end{equation}
as well as a splitting of $\vtrd{}$ in its $q$-dependent dimension via $\vtrd{} = [\vtrd{}^1,\vtrd{}^2]$, where $\vtrd{}^1$ corresponds to the low-q region of $\vtrd{}$. Using this we can reformulate $\bm{B_l}$ as follows:
\begin{equation}
  \label{eq:incariant_decomposition_q_splitting}
  \bm{B_l} =\vtrd{}\vtrd{}^T= \begin{bNiceArray}{cc}[columns-width=0.3cm,margin,baseline=5]
    \Block{3-2}{\vtrd{}^1}\\[0.2cm]
    &\\[0.2cm]
    &\\[0.2cm]
    \hline 
    \Block{2-2}{\vtrd{}^2}\\[0.2cm]
    &\\[0.2cm]
  \end{bNiceArray}
  \begin{bNiceArray}{w{c}{1.6cm}|w{c}{1cm}}[margin,baseline=2]
    \Block{2-1}{{\vtrd{}^1}^\dagger}&\Block{2-1}{{\vtrd{}^2}^\dagger}\\[0.2cm]
    &
  \end{bNiceArray}
  = \begin{bNiceArray}{ccc|cc}[columns-width=0.45cm,margin,baseline=5]
    \Block{3-3}{\vtrd{}^1 {\vtrd{}^1}^\dagger} &&&\Block{3-2}{\vtrd{}^1 {\vtrd{}^2}^\dagger} \\[0.2cm]
    &&&&\\[0.2cm]
    &&&&\\[0.2cm]
    \hline
    \Block{2-3}{\vtrd{}^2 {\vtrd{}^1}^\dagger}&&& \Block{2-2}{\vtrd{}^2 {\vtrd{}^2}^\dagger}\\[0.1cm]
    &&&&
  \end{bNiceArray}\ .
\end{equation}
One can always assume that the matrix $\vtrd{}^1$ has full rank, since it was created from scaled versions of distinct eigenvectors.  As long as the number of momentum transfer points in the low-$q$ area is larger than $2l+1$ one can assume that the matrix $\vtrd{}^1$ has linearly independent columns. In this case $\vtrd{}^1$ has a left-inverse, i.e there is a matrix ${\vtrd{}^1}^{-1}$ such that ${\vtrd{}^1}^{-1}{\vtrd{}^1}=\text{\textbf{id}}$ is the identity matrix. Combining this with Eqs.~\eqref{eq:invariant_q_splitting} and \eqref{eq:incariant_decomposition_q_splitting} one can see that the decomposition matrix $\vtrd{}^2$ in the high-$q$ area may be reconstructed from the values of $\bm{A_l}$ together with the decomposition matrix $\vtrd{}^1$ in the low-$q$ area, that is
\begin{equation}
  \label{eq:decomposition_extension}
  {\vtrd{}^1}^{-1}\bm{A_l} = {\vtrd{}^1}^{-1} \vtrd{}^1{\vtrd{}^2}^\dagger = {\vtrd{}^2}^\dagger.
\end{equation}

Using Eq.~\eqref{eq:decomposition_extension} one can extend an already known $\bm{V_l^1}$ for $q,q'\leq Q$ to the entire $q$-range without using the particular nosy region in $\bm{B_l}$ where both $q$ and $q'$ are grater than $Q$. An example of this procedure is illustrated in Supplementary Fig.~\ref{fig:regularization_masking}.

\section*{Supplementary Note 4: Details of the correlation analysis of experimental diffraction data \label{sec:InvarExp}}

The plots of $\textrm{Re}[C_{n}(q, q')]$, $|\text{arg}[C_n(q,q')]|$ and $\textrm{Re}[B_{l}(q, q')]$, determined for data part II of the amo06516 experiment (see Fig.~\ref{FigS:SizeDistributionHistograms}e), are shown in Fig.~\ref{FigS:amo06516visOptions} up to $q=q'=1~\text{nm}^{-1}$. Different data representations highlight various features, which constitute a complex reciprocal-space fingerprint of the entire 3D structure of the PR772 virus. For the small-angle geometry implemented in the experiments \cite{Reddy2017, Li2020} analyzed here, the accessible correlations are contained in the harmonic orders/degrees $n,l=2,4,6,8,10$ and 12, while all other harmonics have vanishing values and can be neglected. 

It is convenient to use $C_{n}(q, q')$ to directly compare the results of correlation analysis for different datasets and experiments. The phases $|\text{arg}[C_n(q,q')]|$ allow to judge about the statistical convergence of the correlation data. In theory, $C_n(q,q')$ [as well as $B_{l}(q, q')$] are real-valued signed quantities, therefore for ideal, statistically converged data the wrapped phases $|\text{arg}[C_n(q,q')]|$ must take the values of 0 or $\pi$, which correspond to the positive or negative sign of $\textrm{Re}[C_{n}(q, q')]$, respectively (see also Supplementary Note~8 for simulation results).

The 2D maps of $\textrm{Re}[C_{n}(q, q')]$ determined for five selected data parts from the amo06516 experiment (see Fig.~\ref{FigS:SizeDistributionHistograms}e) are shown in Fig.~\ref{FigS:amo06516Cnallparts} for a reduced region of reciprocal space, up to $q=q'=0.7~\text{nm}^{-1}$, to simplify comparison with other experiments. Similar plots for five selected data parts from the amo86615 experiment are shown in Fig.~\ref{FigS:amo86615Cnallparts}, and for the amo11416 experiment in Fig.~\ref{FigS:amo11416Cnallparts}. Note that, due to improvements in the data processing pipeline, the usable data range in the amo86615 experiment has been extended in this work by $\sim 30\%$ up to $0.7\,\text{nm}^{-1}$, compared to the previous work \cite{Kurta2017}.
Visual inspection of the $\textrm{Re}[C_{n}(q, q')]$ plots from various data parts of individual experiments and from different experiments (Figs.~\ref{FigS:amo06516Cnallparts}-\ref{FigS:amo11416Cnallparts}) indicates overall good reproducibility of the results. Clearly, the results for data parts I and V typically deviate from those obtained for data parts at the centers of the size distributions, due to limited statistics on the wings of the distributions (see Supplementary Fig.~\ref{FigS:SizeDistributionHistograms}d-f and Table~1 in the main text), as well as a larger heterogeneity in the corresponding ensembles of particles. At the same time, even in the amo11416 experiment, where the number of patterns considered in the calculations of the CCFs is only about 50 or fewer in each data part, the distribution of speckles (or fringes) emerging on the noisy maps of $\textrm{Re}[C_{n}(q, q')]$ already reveals features similar to those observed in the other two experiments with higher SNR of the determined CCFs.

At the same time, it is also possible to find the differences in Figs.~\ref{FigS:amo06516Cnallparts}-\ref{FigS:amo11416Cnallparts}. Some differences arise due to distinct amounts of patterns considered in the analysis of different data parts and experiments (see Table~1 in the main text), as well as missing data at specific detector locations. This, in particular, results in fluctuations of SNR as a function of $(q, q')$, even within a particular correlation map (i.e. harmonic order). Such effect can be identified from the 2D maps of the phases $|\textrm{arg}[C_{n}(q, q')]|$, plotted in Figs.~\ref{FigS:amo06516ArgCnallparts}, \ref{FigS:amo86615ArgCnallparts} and \ref{FigS:amo11416ArgCnallparts} for the amo06516, amo86615, and amo11416 experiments, correspondingly. Deviations of the phases from the theoretical values of $0$ or $\pi$ (see Figs.~\ref{FigS:SimCapsidandRegIcos}b,e and Supplementary Note 8) indicate that the corresponding values $C_{n}(q, q')$ did not converge. One can readily see, that the higher order FCs $(n=8,10,12)$, have much lower SNR compared to lower orders $(n=2,4,6)$. However even for lower order harmonics the convergence rate is not uniform, and SNR of correlation maps typically degrades at higher $q$, where the scattering is much weaker. Also, the phases are noisy in the regions where $\textrm{Re}[C_{n}(q, q')]$ change the sign, as well as in the regions where the maps contain very narrow speckles (see, for instance, FCs for $n=4$ and $n=6$ at higher q's in Fig.~\ref{FigS:amo86615ArgCnallparts}). 

In Fig.~\ref{FigS:CnComparisonExp} the results for three selected data parts from different experiments are plotted together for ease of comparison. One can readily observe the similarity between the correlation maps from the three independent experiments, suggesting the reproducibility of the studied ensembles of virus particles and the robustness of such SPI measurements.
We verified that our data reduction and correlation analysis pipeline, despite of an arbitrary choice of several parameters (e.g., fitting ranges or thresholds), is robust with respect to moderate variation of these parameters, leading to a rather minor changes in the experimentally determined invariants.

\section*{Supplementary Note 5: Procedure for orientational alignment of 3D reconstructions \label{sec:xframerecon}}

To avoid potential bias from choosing a single reference reconstruction for rotational alignment,  we generated the reference structure through a pairwise reduction procedure. This procedure consists of the following steps:
\begin{enumerate}
\item Select random pairs from a set of 114 real-space reconstructions (densities).
\item \label{enum:start_step} Rotationally align each pair of densities.
\item Compute the average and variance for each density pair using the formula (2.1) from Ref.~\cite{Chan82}, taking the total number of involved individual reconstructions into account.
\item Define new pairs of reconstructions by combining the $2i$-th with the $(2i+1)$-th density averages. If the number of averages is odd, skip the 2nd entry in a pair, and transfer the 1st entry to the next reduction step.
\item Stop if only one reconstructions is left, otherwise continue with step \ref{enum:start_step}.
\end{enumerate}
Fig.~\ref{fig:pairwise_alignment} schematically illustrates this reduction procedure for a set of 10 reconstructions.
On the final step, all individual reconstructions are aligned with the reference one computed by the above procedure, and their average and standard deviation are computed. This average serves as the final reconstruction result, while both 3D average and standard deviation are used to compute additional reconstruction quality metric (see Supplementary Note~7).\\
To rotationally align a pair of reconstructions, $\rho$ and $\rho'$, we used the pairwise rotational cross-correlation $C(\omega)$ defined in Eq.~(10) in the main text, which was efficiently computed using the Fourier transform on the space of rotations \cite{Kostelec08}.
The cross-correlation $C(\omega)$ is maximal for the rotation $\omega$ that  optimally aligns $\rho$ with $\rho'$.

\section*{Supplementary Note 6: Determination of the FSC and PRTF in the spherical coordinate system \label{sec:fscdetails}}

Given two reconstructions $\rho_1$ and $\rho_2$ based on different parts of the scattering data, the FSC metric in Cartesian coordinates for voxels of uniform volume is defined by \cite{Heel2005}
\begin{equation}
        \label{eq:FSC_original}
        \textrm{FSC}(q) = \frac{\sum_{\bm{q_i}\in q} \widehat{\rho}_1(\bm{q_i}) \widehat{\rho}_2(\bm{q_i})^* }{\sqrt{\sum_{\bm{q_i}\in q} |\widehat{\rho}_1(\bm{q_i})|^2\sum_{\bm{q_i}\in q} |\widehat{\rho}_2(\bm{q_i})|^2 }},
\end{equation}
where $\widehat{\rho}_1$ and  $\widehat{\rho}_2$ are the scattering amplitudes corresponding to the reconstructions $\rho_1$ and $\rho_2$, respectively.
The native coordinate system for the MTIP reconstructions is, however, spherical coordinates. In particular, this means that a simple sum over the angular sampling points to compute the FSC would unequally weight different regions on the sphere for a fixed momentum transfer magnitude $q$.
To establish the connection, one can recognize that the the sums in Eq.~\eqref{eq:FSC_original} are discrete approximations of integrals over spherical surfaces, i.e.,
\begin{equation}
        \label{eq:FSC_continouse}
        \textrm{FSC}(q) = \frac{\int_{||\bm{q}||= q} \mathrm{d}\bm{q}\ \widehat{\rho}_1(\bm{q}) \widehat{\rho}_2(\bm{q})^* }{\sqrt{\int_{||\bm{q}||= q} \mathrm{d}\bm{q}\ |\widehat{\rho}_1(\bm{q})|^2 \int_{||\bm{q}||= q} \mathrm{d}\bm{q}\ |\widehat{\rho}_2(\bm{q})|^2 }}.
\end{equation}
Using the volume element in spherical coordinates, $\mathrm{d}\bm{q} = q^2\sin(\theta)\mathrm{d}q\,\mathrm{d}\theta\,\mathrm{d}\phi$, we find the expression for the FSC in spherical coordinates to be 
\begin{equation}
        \label{eq:FSC_spherical}
        \textrm{FSC}(q) = \frac{\int_{0}^{2\pi} \mathrm{d}\phi \int_0^{\pi} \mathrm{d}\theta \ \sin(\theta)\widehat{\rho}_1(q,\theta,\phi) \widehat{\rho}_2(q,\theta,\phi)^* }{\sqrt{\left(\int_{0}^{2\pi} \mathrm{d}\phi \int_0^{\pi} \mathrm{d}\theta \ \sin(\theta)\ |\widehat{\rho}_1(q,\theta,\phi)|^2 \right)\left( \int_{0}^{2\pi} \mathrm{d}\phi \int_0^{\pi} \mathrm{d}\theta \ \sin(\theta)\ |\widehat{\rho}_2(q,\theta,\phi)|^2 \right) }}.
\end{equation}
Finally, the discretization of the integrals in Eq.~\eqref{eq:FSC_spherical} depends on the chosen sampling points in spherical coordinates. The xFrame software package \cite{Berberich2024} uses uniform sampling points in $\phi$ and Gau\ss-Legendre nodes in $\theta$, which leads to Eq.~(12) in the main text.

Similar to Eq.~(\ref{eq:FSC_spherical}), the PRTF customized for MTIP \cite{Kurta2017} can be expressed in spherical coordinates as,
\begin{equation}
        \label{eq:PRTFspher}
        \mathrm{PRTF}(q)= \int_{0}^{2\pi} \mathrm{d}\phi \int_0^{\pi} \mathrm{d}\theta \sin(\theta)\frac{|\langle \widehat{\rho}_k(q,\theta,\phi)\rangle_k|}{\sqrt{\langle I_k(q,\theta,\phi) \rangle_k}},
\end{equation}
where $\langle\cdot\rangle_k$ is the average over all reconstructions, $\widehat{\rho}_k$ represents the scattering amplitudes reconstructed after the last real-space projection step, and $I_k$ represents the intensities reconstructed after the last reciprocal-space projection step. Again, considering the sampling of spherical coordinates implemented in xFrame we obtain Eq.~(13) in the main text.

The determined PRTF curves are shown in Figs.~9a,b and Figs.~10a,b, for the amo06516 and amo86615 reconstructions, respectively.
Since the number of scattering patterns per size-part in the experimental reconstructions is quite low (ranging from 387 to 1928, see Table~1 in the main text and Supplementary Fig.~\ref{FigS:SizeDistributionHistograms}), it is rather impractical to further divide individual size-part into two halves to determine the FCS metric. Instead, we computed the FSC metric for pairs of reconstructions from neighboring size-parts in each of the experimental datasets. This serves as a lower boundary of the ``true'' FSC metric, since there are systematic differences between different size parts (see below).
The FSC curves for the amo06516 dataset are shown in Figs.~\ref{fig:resolution_06516}e,f, and their counterparts for the amo86615 dataset are presented in Figs.~\ref{fig:resolution_86615}e,f.
In the case of FSC, the $\frac{1}{2}$ bit resolution \cite{Heel2005}\footnote{Note that we use a spherical grid in which the number of points in each angular shell is constant, this causes the $\frac{1}{2}$ bit threshold to be constant as a function of the momentum transfer $q$.} varies between the experiments and different subsets.
For the amo06516 experiment, the FSC metric estimates the resolution of "I vs II" to be 8\,nm, 9\,nm for "II vs III", and 7\,nm for "III vs IV". In the amo86615 experiment, both "II vs III" and  "III vs IV" have the FSC resolution of 9\,nm, while for "IV vs V" we obtain 13\,nm. 

Note, however, that for all reconstructions the resolution limit is violated only at specific locations of reciprocal space, where the scattering form-factor of the particle (or SAXS intensity) has interference minima, and the measured intensities are typically characterized by lower signal to noise ratio. In the FSC metric these dips are additionally broadened due to the particle size differences between different data parts which are compared. The reconstructed values of the particle structure factor in such regions may be less accurate, leading to apparently lower resolution estimations using PRTF \cite{Lundholm2018, Ayyer2021}. In the present case, similar effects can be also observed for FSC, since the 3D distribution of scattered intensity is also unknown and reconstructed simultaneously with the real-space structure in MTIP.  
To mitigate these effects, we also convolved the FSC and PRTF curves with a Gaussian of FWHM equal to the critical Shannon pixel size, and estimated the resolution from these smoothed profiles (see Figs.~\ref{fig:resolution_06516}c,d,g,h and Figs.~\ref{fig:resolution_86615}c,d,g,h).

\section*{Supplementary Note 7: Confidence metric for three-dimensional reconstructions \label{sec:3ddev}}

As mentioned in Supplementary Note~5, we computed the 3D standard deviation (STD) for a set of 114 aligned reconstructions for each experimental subset of data. By dividing the 3D averaged electron density by this STD, we obtain a 3D confidence metric that accounts for variability in the reconstructed densities. This metric equals unity in a given 3D voxel if the density fluctuations of the individual reconstructions match the magnitude of the final averaged density value in that voxel. Consequently, the higher this metric, the more confidently the structure can be interpreted in the corresponding region of the averaged reconstruction.
Supplementary Figs.~\ref{fig:06516_mean_over_var} (amo06516) and \ref{fig:86615_mean_over_var} (amo86615) show this confidence metric for the same density cuts that are presented in Figs.~3 and 4 of the main text, respectively.
In the case of amo86615, Fig.~\ref{fig:86615_mean_over_var} displays a relatively uniform confidence level throughout the virus capsid, where the mean is more than 5 times higher than the STD, with an exception of a small region at the very center of the capsid. The higher confidence values of data part III compared to the other size parts also indicates that a larger number of diffraction patterns $M$ directly leads to reconstructions of better quality.
For amo06516, Fig.~\ref{fig:06516_mean_over_var} shows that the confidence values are not uniformly distributed over the virus volume. Interestingly, that high density areas in the average reconstructions (see Fig.~3 in the main text) overlap with the high-confidence areas of the metric. This reinforces our findings on the presence of high and low density regions inside of the viruses in the amo06516 experiment. The cuts highlighted in black in Fig.~\ref{fig:06516_mean_over_var} also reveal higher-confidence regions extending from the top-right vertex of the capsid. While this feature is clearly visible only for size part III in the averaged reconstruction, this confidence metric hints at its presence in all other data subsets of the amo06516 experiment as well.

\section*{Supplementary Note 8: Detailed model‑guided analysis of reciprocal‑space structural fingerprints \label{sec:Fingerprints}}

The experimental correlation maps in Figs.~\ref{FigS:amo06516Cnallparts} to \ref{FigS:amo11416ArgCnallparts} contain a complex fingerprint (signature) of the entire 3D structure of the PR772 virus. Here, we analyze how distinct real-space structural attributes of the virus are revealed by this reciprocal-space fingerprint. 

We start with simulations of the empty PR772 capsid atomic structure determined at $2.3\,\text{\AA}$ by cryo-EM technique [Protein Data Bank (PDB) entry 6Q5U] \cite{Reddy2019}.
The simulated correlation maps presented in Figs.~\ref{FigS:SimCapsidandRegIcos}a-c provide several insights.
First of all, the results of simulations for the empty PR772 capsid strongly deviate from the experimental observations (compare, for instance, Fig.~\ref{FigS:SimCapsidandRegIcos}a and Fig.~\ref{FigS:amo86615Cnallparts}), except of the harmonic order/degree $n=l=10$. The observed deviations can arise due to several reasons: the experimentally studied particles are not just empty capsids, but filled particles, and they do not necessary have perfect icosahedral symmetry \cite{Kurta2017, Rose2018, Assalauova2020}. Note that the capsid structure reported in Ref.~\cite{Reddy2019} was refined by applying icosahedral symmetry. In this case, $B_6(q,q')$ (for $l=6$) defines the predominant low-order contribution, indicating the icosahedral symmetry of the object \cite{Cohan1958}. As a consequence, due to the relation in Eq.~(4) in the main text, the maps of $C_{n}(q, q')$ of orders $n=2,4,6$ look similar (Fig.~\ref{FigS:SimCapsidandRegIcos}a), being dominated by the morphology of $B_6(q,q')$ (see Fig.~\ref{FigS:SimCapsidandRegIcos}c).   
Clearly, the experimental $B_l(q,q')$ reveal a distinct picture (see Fig.~\ref{FigS:amo06516visOptions}, bottom row), where also the invariants of lower degrees $l=2$ and 4 have magnitudes comparable to $B_6(q,q')$, that also leads to a distinct morphology of $C_{n}(q, q')$ of orders $n=2,4$ and 6 (see Fig.~\ref{FigS:amo06516visOptions}, bottom and middle rows). Thus, our experimental results (Figs.~\ref{FigS:amo06516Cnallparts}-\ref{FigS:CnComparisonExp}) consistently show deviations of the shapes of the studied virus particles from a perfect icosahedral symmetry in all three considered here experiments.
To confirm, that the main discrepancy between the experimental and simulated results stems from the assumption of perfect icosahedral symmetry, we also performed simulations for a bead model of a perfect solid (uniform-density) icosahedral particle (see Figs.~\ref{FigS:SimCapsidandRegIcos}d-f). While these results reflect the structural difference between the empty capsid and the filled particle, our arguments about the dominating effect of symmetry on the correlation maps are supported in this case.

The next step is to identify the types of deviations from perfect icosahedral symmetry that could explain the experimental observations. We begin by introducing simple distortions into the bead model of a perfect solid icosahedron, such as particle compression or elongation. In Fig.~\ref{FigS:ReCnSimDistort2Vert} the results are shown for symmetric radial compression/elongation of two opposite vertices of the perfect icosahedron, along one of the five-fold symmetry axes. As one can see, the FCs $C_n(q,q')$ are very sensitive to the structural distortions, which change the morphology of the correlation maps of different harmonics $n$ already at very small distortions of $\pm 2.5\%$ (compare with Fig.~\ref{FigS:SimCapsidandRegIcos}e). Most prominently, at specific magnitudes of distortions, for instance $-7.5\%$ or $+12.5\%$, the FC of the order $n=2$ reproduces the morphology of the corresponding experimental correlation map very well (compare Figs.~\ref{FigS:ReCnSimDistort2Vert}d,h with Figs.~\ref{FigS:CnComparisonExp} at $n=2$). This supports the conclusion that PR772 particles studied in our experiments do not possess perfect icosahedral symmetry. 
We also found that other types of radial distortions — for instance, asymmetric one-, two-, three- and four-vertex extension — can lead to morphologies of the $n=2$ FCs matching the experimental observations (compare maps for $n=2$ in Figs.~\ref{FigS:ArgCnAsymDistort}a-d and Figs.~\ref{FigS:amo86615ArgCnallparts}). Also $n=8$ harmonic reveals morphologies close to the experimental ones. The leftmost column in Figs.~\ref{FigS:ArgCnAsymDistort}a-d
illustrates the type and magnitude of the distortion with respect to a perfect icosahedron (shown in violet), with $20\%$ elongation of one vertex (Fig.~\ref{FigS:ArgCnAsymDistort}a) along one of the five-fold axes, as well as elongation of two adjacent vertices (Fig.~\ref{FigS:ArgCnAsymDistort}b), three vertices (Fig.~\ref{FigS:ArgCnAsymDistort}c) and four vertices (Fig.~\ref{FigS:ArgCnAsymDistort}d) by $15\%$ along the respective rotational axes. 
These results suggest that the experimentally observed morphology of the $n=2$ component is a generic indicator of deviations of a certain magnitude from a perfect icosahedral shape.

The agreement with the experimental data improves when more complex distortions are considered.
Figs.~\ref{FigS:ArgCnAsymDistort}e-h illustrate the results for several bead models 
involving a tangential component of distortion in addition to the radial one. 
These four models represent: radial elongation of two opposite vertices by $5\%$  (Fig.~\ref{FigS:ArgCnAsymDistort}e), radial compression of two opposite vertices by $5\%$ (Fig.~\ref{FigS:ArgCnAsymDistort}f), radial compression of one vertex by $5\%$ and elongation of the opposite vertex by $5\%$ (Fig.~\ref{FigS:ArgCnAsymDistort}g), radial compression of one vertex by $10\%$ and elongation of the opposite vertex by $5\%$ (Fig.~\ref{FigS:ArgCnAsymDistort}h), combined in each case with a tangential shift of the two vertices by approximately $20\%$ in  mutually opposite directions. 
Such composite distortions, involving compression, elongation and skew leads to enhanced agreement of the simulated $n=4$ and 6 harmonics with the experimental data (see Figs.~\ref{FigS:amo06516ArgCnallparts} to \ref{FigS:amo11416ArgCnallparts}). Specifically, such composite distortions allow to correctly reproduce the connectivity between speckles in the low-resolution part of the $n=4$ maps, indicated by the red dashed rectangular shapes in  Fig.~\ref{FigS:ArgCnAsymDistort}h and Part III in Fig.~\ref{FigS:amo86615ArgCnallparts}.
These observations suggest a rather complex and asymmetric character of distortions in the studied ensembles of PR772 viruses.

On the next step, we implement a core-shell model of the virus, which is closer to the realistic structure of PR772, consisting of a thin icosahedrally-shaped protein capsid (shell) that encapsulates internal content (core) of typically lower density $\rho_{\textrm{c}}$, as compared to capsid density $\rho_{\textrm{s}}$.
In Figs.~\ref{FigS:ArgCnAsymDistortInternSpher}a-h we show the results for a set of such core-shell models, which represent the same shape distortion as in Figs.~\ref{FigS:ArgCnAsymDistort}a-h, respectively, and contain a spherical vesicle (core) of the radius $r=24\;\textrm{nm}$, uniformly filled with a density $\rho_{\textrm{c}}$ of $70\%$ of the capsid density $\rho_{\textrm{s}}$. As compared to the models of solid icosahedral particle (Figs.~\ref{FigS:ArgCnAsymDistort}a-h), introduction of the core leads, in particular, to further improvement of the $n=4$ and $n=6$ FCs. Specifically, the morphology of the bright narrow features encompassed within the white dashed square areas in Fig.~\ref{FigS:amo86615ArgCnallparts} for Part III at $n=4$ and $n=6$, are quite well reproduced in Figs.~\ref{FigS:ArgCnAsymDistortInternSpher}f-h. 
To test the sensitivity of these regions of the correlation maps to the parameters of the core-shell model, we performed simulations for a set of models with different sizes of the core (Fig.~\ref{FigS:ArgCnSpherSizes}). In this set of simulations we applied the same type of shape distortion as in Fig.~\ref{FigS:ArgCnAsymDistort}g, and varied the radius of the core in the range from $14\;\textrm{nm}$ (Fig.~\ref{FigS:ArgCnSpherSizes}a) to $25\;\textrm{nm}$ (Fig.~\ref{FigS:ArgCnSpherSizes}h), while keeping the vesicle density $\rho_{\textrm{c}}$ fixed to $70\%$ of the capsid density $\rho_{\textrm{s}}$ in all simulations. The results clearly indicate, that the agreement of the $n=4$ and $n=6$ phase maps in the specified above locations improves at larger core sizes, where the capsid thickness approaches the expected values.
It is quite remarkable that, despite the limited resolution of the present experiments, the reciprocal-space analysis of correlation maps appears sensitive to structural features—such as the presence of a thin capsid—that cannot be easily resolved in the real-space reconstructions.

To explore more the sensitivity of the core-shell models, we performed extended simulations of the core-shell model represented in Fig.~\ref{FigS:ArgCnSpherSizes}g, by varying the density of the core in the range from $10\%$ (Fig.~\ref{FigS:ArgCnSpherDensity}a) to $100\%$ of the capsid density $\rho_{\textrm{s}}$ (Fig.~\ref{FigS:ArgCnSpherDensity}j), while keeping all other model parameters fixed. Again, one can see a notable effect on the $n=6$ component, where the thickness and location of the features changes upon filling-in the core (compare the locations highlighted with white dashed square areas in Fig.~\ref{FigS:amo86615ArgCnallparts} at $n=6$ for Part III).
Thus, variations in the filling and density of the internal content of the virus, at other fixed structural parameters, can influence the morphology of the specified parts of the correlation maps.
This may explain, why the corresponding areas of the maps in the amo06516 experiment (see the regions encompassed within the white dashed square areas in Fig.~\ref{FigS:amo06516ArgCnallparts} for Part III at $n=4$ and $n=6$) look almost featureless, while it is not the case in the amo86615 experiment. The xFrame reconstructions from the amo06516 experiment show partially filled, nonuniform internal structure of the virus, in contrast to more uniformly filled structures reconstructed from the amo86615 experiment.
For a partially filled virus $(20-50\%)$ the speckle structure at $n=6$ notably varies, which may lead to vanishing features in the specified regions of the correlations in the average FCs.
Another feature clearly visible in Figs.~\ref{FigS:ArgCnSpherDensity}, is the decrease of the height $h$ and increase of the width $w$ of the speckles denoted in Fig.~\ref{FigS:ArgCnSpherDensity}a and Fig.~\ref{FigS:ArgCnSpherDensity}j at $n=2$, when the core density decreases from $100\%$ to $10\%$ of $\rho_{\textrm{s}}$. The vertical dashed line through the $n=2$ plots in Fig.~\ref{FigS:ArgCnSpherDensity} shows the decrease of the height $h$ towards lower core density. The same effect can be identified in the experimental data (see Fig.~\ref{FigS:CnComparisonExp}), where similar speckle analysis shows 
decreased values of $h$, and increased values of $w$ in the amo06516 experiment, as compared to the amo86615 experiment. 
Therefore, both real-space reconstructions and, reciprocal space analysis of rotational invariants, consistently show differences in the internal filling of the viruses in the amo06516 and amo86615 experiments. 
Notice, that such density variation is not observed between different data parts in individual experiments (see vertical red dashed lines at $n=2$ in Fig.~\ref{FigS:amo06516ArgCnallparts} and Fig.~\ref{FigS:amo86615ArgCnallparts}).
This indicates, that PR772 batches studied in the two abovementioned experiments contain viruses in different states.

\section*{Supplementary Note 9: Particle sphericity and the scaling relationship between SAXS intensity and higher-order invariants \label{sec:SetsComparison}}

The qualitative analysis of individual harmonics, $C_n(q,q')$ or $B_l(q,q')$, allows to associate various morphologies on the 2D correlation maps with specific structural features of the 3D viruses.
A more comprehensive analysis should also include quantitative relationship between the separate harmonics ($n$ or $l$) and, in particular, the SAXS intensity $I_{\text{SAXS}}(q)$. 
It is crucial to maintain this relationship when performing the {\it ab initio} reconstructions using  xFrame or when making quantitative comparisons of different experimental datasets and simulation results.
To achieve this, it is necessary to normalize the correlation data consistently. 

Note that, owing to the specifics of the experimental geometry or the data-processing settings, scattering data from different experiments (or simulations) may differ in overall scattering strength and sampling, placing $I_{\text{SAXS}}(q)$ and $C_n(q,q')$ on arbitrary scales.
For example, the number of angular grid points $N_\phi$ can be chosen optimally for each particular experiment - depending on detector size, resolution and extent of scattering in reciprocal space - to accelerate computation while using the maximum available information. 
The application of standard numerical libraries for DFT to determine the angular FCs $C_n(q,q')$ (see Methods in the main text) may introduce additional scaling of $C_n(q,q')$ by $N_\phi$, depending on the used convention. For instance, the numpy function \texttt{fft.fft}, which is utilized in this work to compute $C_n(q,q')$, produces unscaled values by default. In this case, the values of $C_n(q,q')$ should be rescaled by $1/N_\phi$ to represent their actual contribution to the CCF. Also, as it is shown in Supplementary Note~1, when employing the difference image CCF $ C_{\textrm{diff}}(q,q',\Delta)$, its FCs $C_{\textrm{diff},n}(q,q')$ must be divided by 2 to yield the single-particle FCs $C_n(q,q')$.
When rescaling the $I_{\text{SAXS}}(q)$ profile for comparison with another experiment or simulation, it is crucial to apply appropriate rescaling also to $C_n(q,q')$, to preserve consistency between $I_{\text{SAXS}}(q)$ and $C_n(q,q')$.
Clearly, multiplying $I_{\text{SAXS}}(q)$ by a constant factor $A$ necessitates that $C_n(q,q')$ be multiplied by the squared factor $A^2$, to maintain this consistency \cite{Kurta2019}.

An example of this rescaling procedure is demonstrated in Fig.~\ref{FigS:ScalingSAXSExample}, where we consider two correlation datasets (denoted as \texttt{sim1} and \texttt{sim2}).  Each dataset consist of SAXS intensity, $I_{\text{SAXS}}(q)$, and FCs, $C_n(q,q')$, of orders $n=2$ and 4, calculated from the same set of simulated diffraction patterns generated for the virus particle model shown in Fig.~\ref{FigS:ArgCnAsymDistortInternSpher}h.
In Fig.~\ref{FigS:ScalingSAXSExample}, all curves corresponding to \texttt{sim1} are represented by solid lines, while dashed curves correspond to \texttt{sim2}.
The difference between these two datasets is that they are defined on polar grids with different angular sampling $N_{\phi}$. Additionally, diffraction patterns were normalized by the corresponding intensities $\langle I(q) \rangle_{q}$, averaged over the range $q=(0.1, 0.2)\;\text{nm}^{-1}$, during the calculation of the second dataset.
The SAXS intensities $I_{\text{SAXS}}(q)$ and one-dimensional profiles of the amplitudes of the FCs $|C_n(q,q')|$ at a fixed $q'=0.19\;\text{nm}^{-1}$, corresponding to these two datasets, are shown Fig.~\ref{FigS:ScalingSAXSExample}a. 
As can be seen, even the relative positions of the $I_{\text{SAXS}}(q)$ and $|C_n(q,q')|$ curves differ between the two datasets.
After dividing $C_n(q,q')$ by the respective values of $N_{\phi}$, as well as dividing by a factor of 2 (since the difference-pattern CCF was utilized in both calculations), the relative positions of the curves change (see Fig.~\ref{FigS:ScalingSAXSExample}b). 
Notice, that the datasets in Fig.~\ref{FigS:ScalingSAXSExample}b are both internally consistent and can be used for fingerprint analysis or {\it ab initio} reconstructions, however additional rescaling is required to compare them quantitatively.
Finally, after normalizing each SAXS profile in Fig.~\ref{FigS:ScalingSAXSExample}b by the respective values of $A=\langle I_{\text{SAXS}}(q) \rangle_{q}$, where averaging is performed over the range  $0.2\;\text{nm}^{-1}\leq q\leq 0.5\;\text{nm}^{-1}$, and dividing the FCs $C_n(q,q')$ by $A^2$, the two datasets coincide exactly (see Fig.~\ref{FigS:ScalingSAXSExample}c), indicating that they correspond to the exactly same structure. Note, that there is an offset between the magnitude of the SAXS intensity $I_{\text{SAXS}}(q)$ and $|C_n(q,q')|$, denoted by two-sided arrows in Fig.~\ref{FigS:ScalingSAXSExample}c. 
The magnitude of this offset, in particular, defines the relative strengths of the higher-order FCs (or higher-order rotational invariants) as compared to the zero-order invariant, directly related to the SAXS intensity. For instance, the magnitude of this offset can be related to the sphericity of a scattering object (see Fig.~\ref{FigS:SphericityAnalysis}). Clearly, an ideal spherical object (with sphericity equal to 1) can be completely described by SAXS intensity, the zero-order invariant, while all higher-order invariants vanish. For objects with lower sphericity, the higher-order invariants will increase in magnitude upon increasing the deviations from a spherical shape, as it is clearly visible in Figs.~\ref{FigS:SphericityAnalysis}a-f. This means, that the offset between SAXS and higher order invariants (indicated by two-sided arrows in Fig.~\ref{FigS:ScalingSAXSExample}c) will increase upon increasing the sphericity of a particle. At the same time, the experimental amo86615 data (see Figs.~\ref{FigS:SphericityAnalysis}g-i) do not show such a pronounced offset variation between data parts corresponding to different particle sizes, indicating the absence of effects related to changes in sphericity.

In Fig.~\ref{FigS:ScalingSAXSExpSim}, we also compare the rescaled results of the amo06516 and amo86615 experiments for two selected parts of data with the results of simulations for the distorted core-shell structure (shown in Fig.~\ref{FigS:ArgCnSpherDensity}h), where the curves of different color correspond to distinct datasets. Here we focus on the SAXS intensities $I_{\text{SAXS}}(q)$, and one-dimensional sections of FCs $|C_2(q,q')|$ of the order $n=2$ at three distinct fixed values of $q'$. The plots in Figs.~\ref{FigS:ScalingSAXSExpSim}a-c consistently indicate several features. The rescaled experimental SAXS profiles, as well as the magnitudes of $|C_2(q,q')|$ are in good agreement for the two selected data parts, suggesting substantial similarity of the studied particles. The differences mostly present at lower $q$ (see Fig.~\ref{FigS:ScalingSAXSExpSim}a), as it was also discussed in the previous section. A similar situation is observed for other data parts and for higher orders $n>0$ of the FCs (not presented here).
The results of simulations for the distorted core-shell structure (denoted by red curves in Fig.~\ref{FigS:ScalingSAXSExpSim}) align very well with the experimental results. The positions, shapes and magnitudes of the distinct peaks in the experimental $|C_2(q,q')|$ curves at $q'=0.3\;\text{nm}^{-1}$ and $q'=0.5\;\text{nm}^{-1}$ are closely reproduced in the simulations, although there is no perfect match across the entire $q$ range, particularly at low $q'$ (see Fig.~\ref{FigS:ScalingSAXSExpSim}a). 
The absence of the offset between the experimental and simulated curves indicates that there no notable effects related to changes in particle sphericity.


\begin{figure}
\centering
\includegraphics[width=\textwidth]{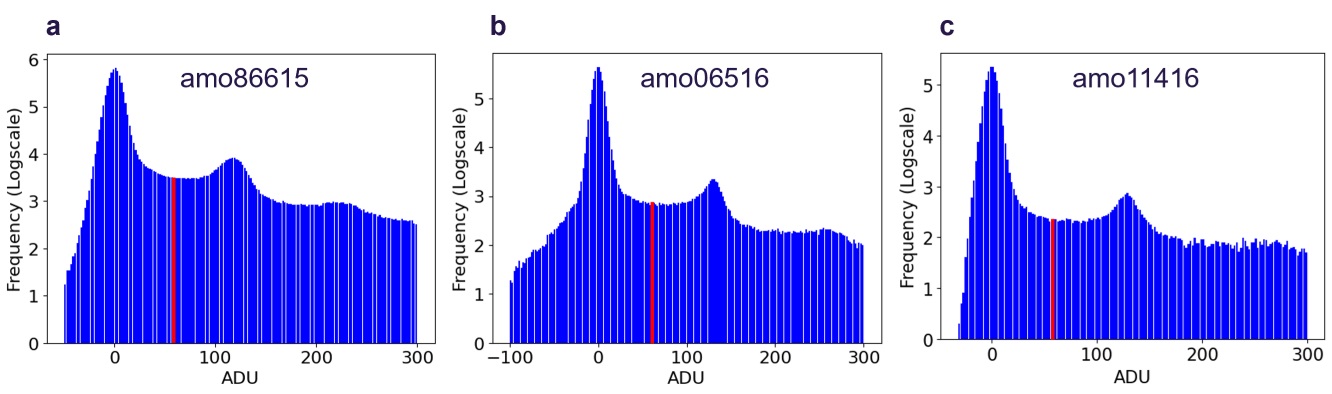}
\caption{\textbf{Experimental photon distribution histograms.} The histograms are presented for the amo86615 (a), amo06516 (b), and amo11416 (c) experiments. Each histogram was determined from 100 single-particle patterns by counting all unmasked pixels on each diffraction pattern. The histograms extend up to 300 ADU, reflecting the noisy character of the measured patterns, with a dominant zero-photon peak and much weaker one-photon and two-photon peaks. The red vertical line on each histogram defines an ADU threshold set at $I_{\textrm{min}}=60$~ADU, approximately midway between the zero-photon and one-photon peaks (all pixels with values less than $I_{\textrm{min}}$ were set to zero).}
\label{FigS:ADUThresholds}
\end{figure}

\begin{figure}
\centering
\includegraphics[width=\textwidth]{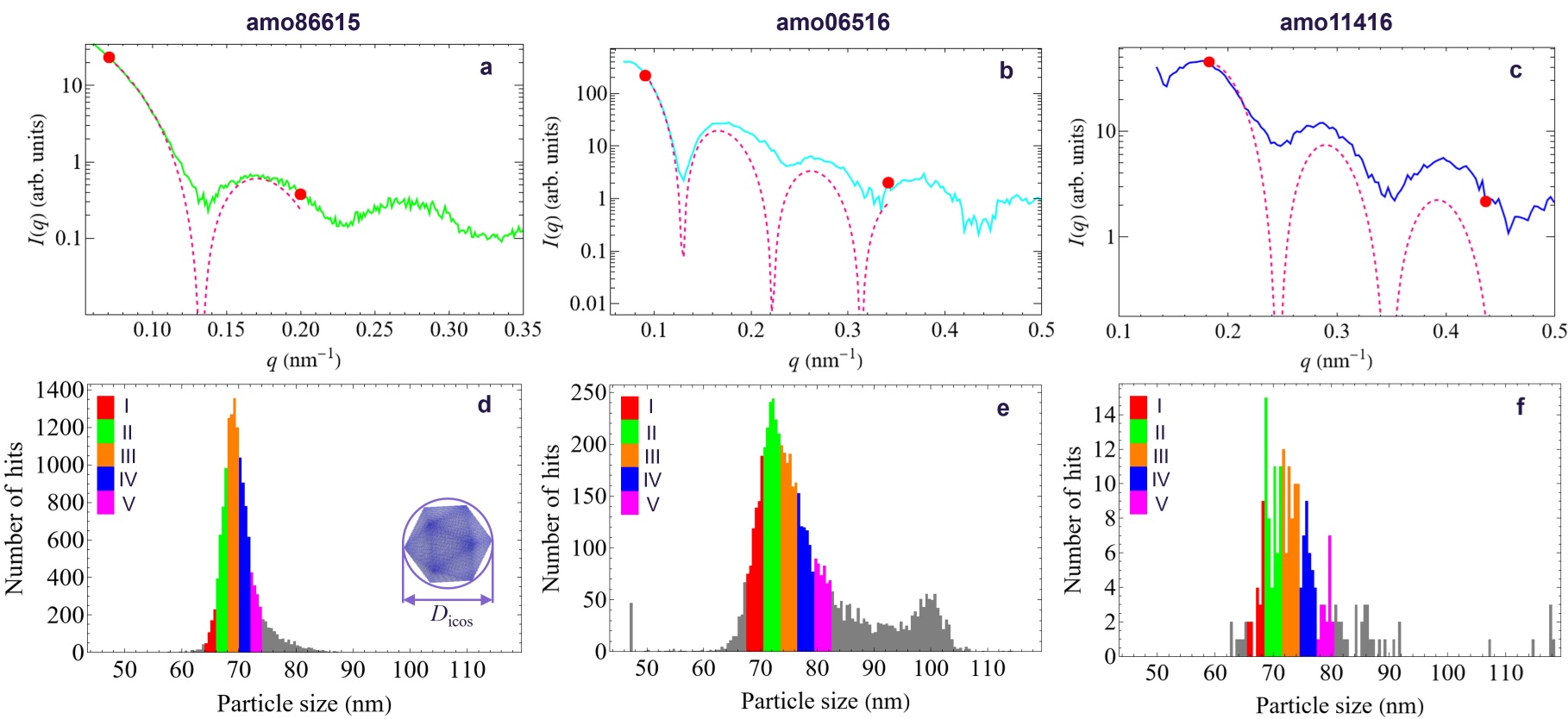}
\caption{\textbf{Size determination of the PR772 particles.} (a-c) Examples of the radial intensity profiles $I(q)$ determined from the representative amo86615 (a), amo06516 (b), and amo11416 (c) experimental single-particle hits.
Solid lines correspond to the experimental data $I(q)$, while dashed lines show the results of fitting $I(q)$ with a form factor of a spherical particle, $I_{\textrm{spher}}(q)$. Red markers indicate the fitting ranges for each experiment (see Methods). (d-f) Size distribution histograms of the PR772 viruses, corresponding to the selected high-intensity single-particle hits in the amo86615 (d), amo06516 (e), and amo11416 (f) experiments. The definition of the icosahedron size $D_{\text{icos}}$ used in this work is schematically shown in the inset of (d). In each experiment, five distinct parts of the data (I-V), corresponding to different average particle sizes, are denoted by red, green, orange, blue, and magenta colors in the histograms (d-f), were analyzed individually. Each of the five selected data parts in (d) corresponds to a particle polydispersity of 2~nm (i.e., a 2~nm-wide size bin), while the data parts in (e) and (f) are defined with 3~nm bins.}
\label{FigS:SizeDistributionHistograms}
\end{figure}

\begin{figure}
\centering
\includegraphics[width=\textwidth]{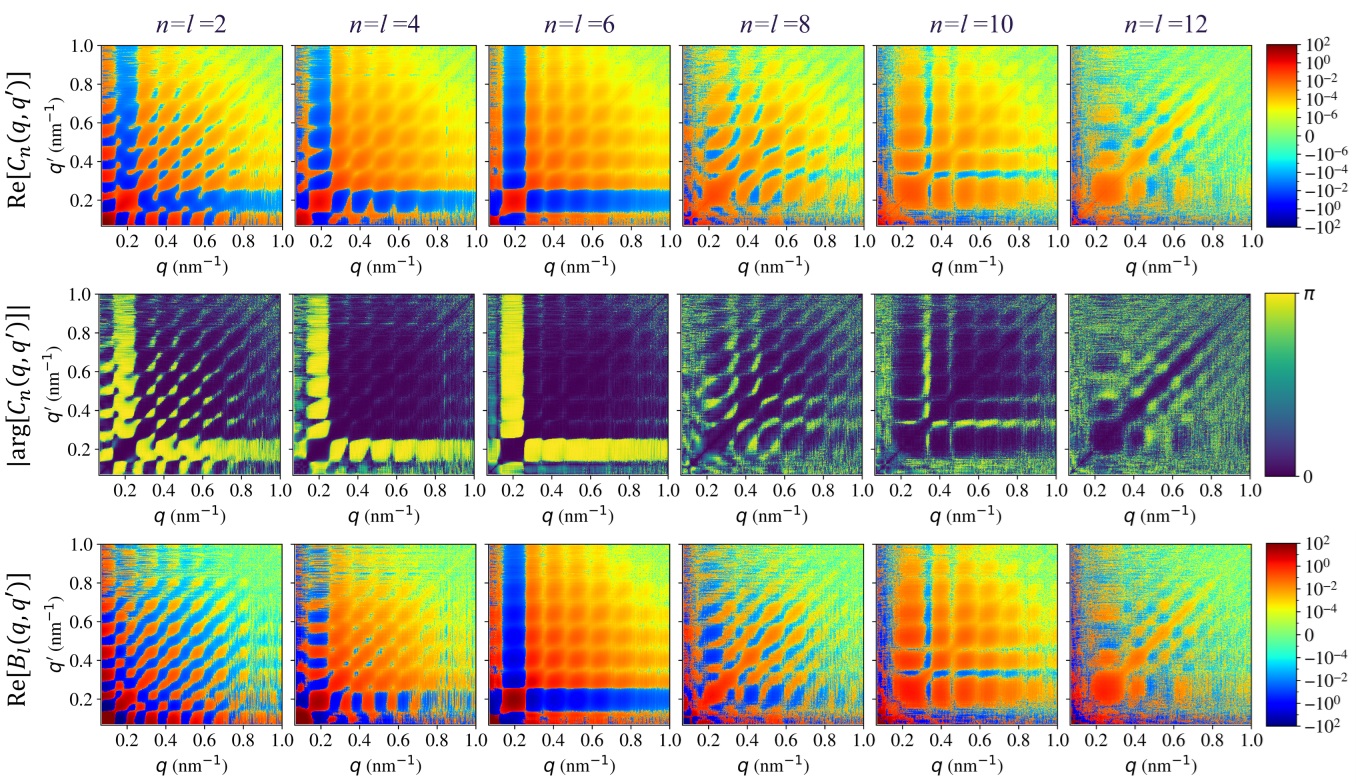}
\caption{\textbf{Distinct representations of the rotational invariants.} The plots generated from data part II of the amo06516 experiment (see Fig.~\ref{FigS:SizeDistributionHistograms}e) show: the real part of the Fourier components $\textrm{Re}[C_{n}(q, q')]$ (top row), their corresponding phases $|\textrm{arg}[C_{n}(q, q')]|$ wrapped into the region $(0,\pi)$ radians (middle row), and the real part of the 3D single-particle invariants $\textrm{Re}[B_{l}(q, q')]$ (bottom row). Each column corresponds to the specified order $n$ of $C_{n}$ or degree $l$ of $B_l$ ($n,l=2,4,6,8,10$ and 12). The values of $C_{n}$ and $B_l$ are presented in arbitrary units and plotted on a symmetrical logarithmic scale.}
\label{FigS:amo06516visOptions}
\end{figure}

\begin{figure}
\centering
\includegraphics[width=0.8\textwidth]{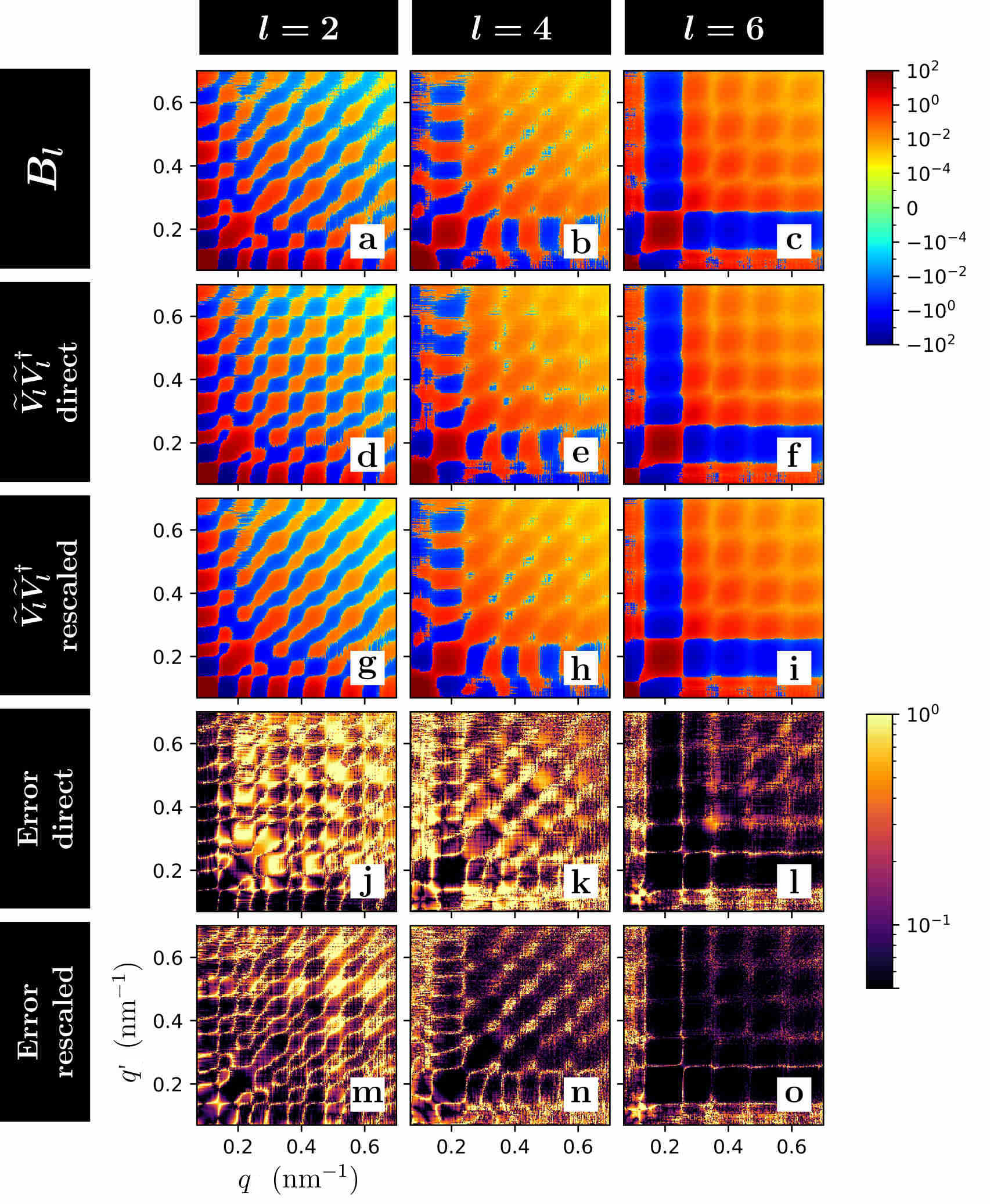}
\caption{\textbf{Regularisation of the rotational invariants.} (a-c) Directly extracted $\bm{B_l}$ for $l=2,4,6$ using Eq.~(4) from the main text. (d-f) Matrix products $\vtrd \vtrd^\dagger$, where $\vtrd$ is directly computed using the $2l+1$ highest eigenvalues of $\bm{B_l}$. (g-i) Matrix products $\vtrd \vtrd^\dagger$, where $\vtrd$ is computed using using Ruiz equalization (see Methods). The relative error $|\vtrd \vtrd^\dagger-\bm{B_l}|/|\bm{B_l}|$for both versions of $\vtrd$ is presented in (j-l) and (m-o),respectively. It can be seen that direct calculation causes $\vtrd \vtrd^\dagger$ to diverge significantly from $\bm{B_l}$ for all $q,q'>0.2$~nm$^{-1}$. The proposed regularization approach results in error values that are more evenly distributed across the argument range, yielding a significantly better agreement between the phases (signs) of $\bm{B_l}$ and its decomposition $\vtrd \vtrd^\dagger$. The data in this figure correspond to part II of the experiment amo06516 (see Fig.~\ref{FigS:SizeDistributionHistograms}e).}
\label{fig:bl_regularization}
\end{figure}

\begin{figure}
\centering
\includegraphics[width=0.9\textwidth]{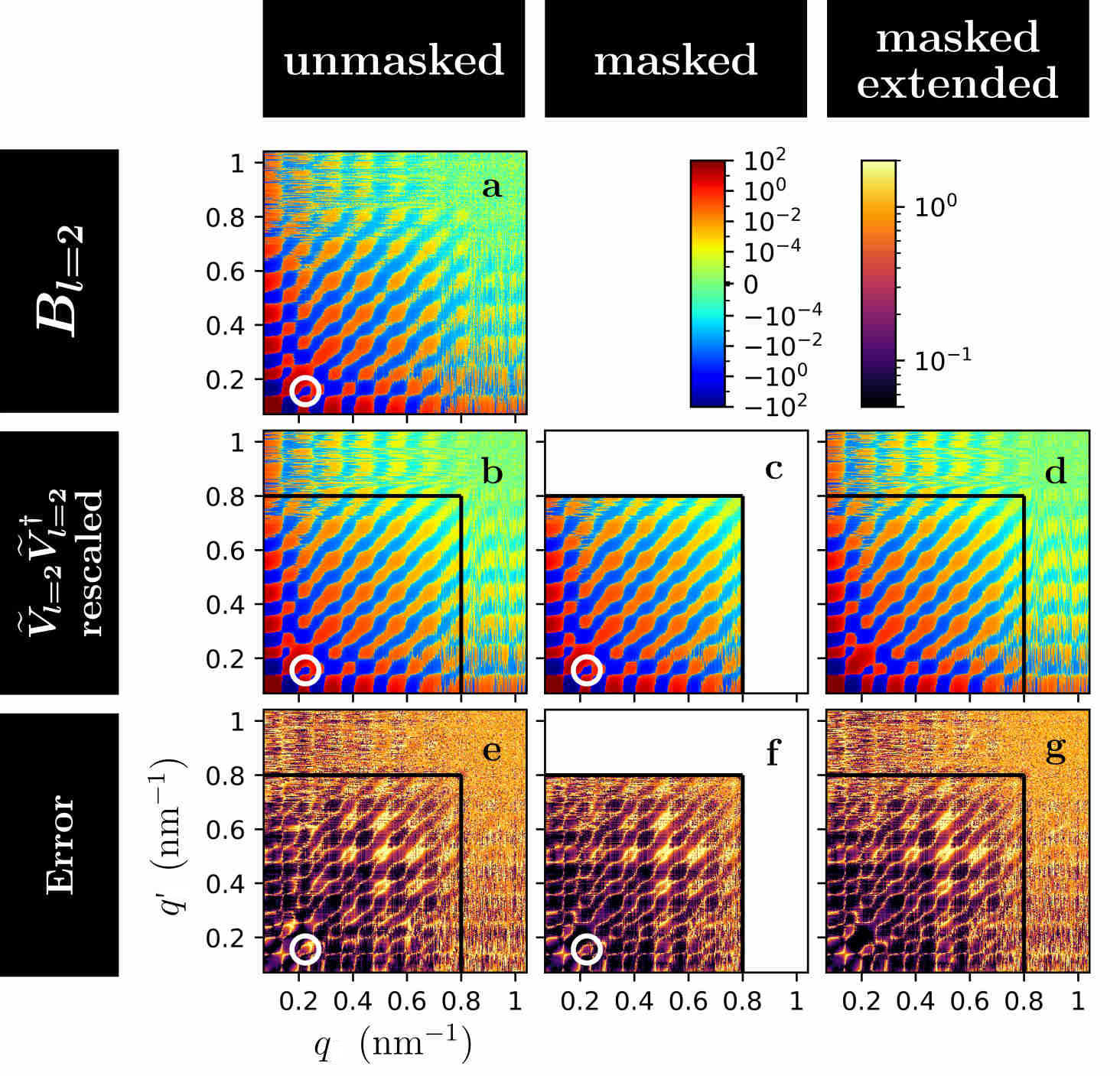}
\caption{\textbf{Comparison of regularization schemes for rotational invariants.} The data in this figure correspond to part II of the experiment amo06516 (see Fig.~\ref{FigS:SizeDistributionHistograms}e). (a) Directly determined rotational invariant $\bm{B_l}$ of degree $l=2$. Regularized invariants without masking the high-$q$ region (b), with masking (c), and after applying the recovery scheme described in Supplementary Note~3 to the masked invariant in (c). (e-g) Relative difference between the regularized invariants (b-d) and the unregularized invariant shown in (a). The white circles in (a,b,c,e) and (f) mark areas in which the ``masked'' and ``mask extended'' regularizations exhibit lower errors than the directly computed decomposition $\vtrd \vtrd^\dagger$. This shows that masking the high-$q$ regions can positively effect the values of invariants in the low-$q$ region after applying regularization.}
\label{fig:regularization_masking}
\end{figure}

\begin{figure}
\centering
\includegraphics[width=0.9\textwidth]{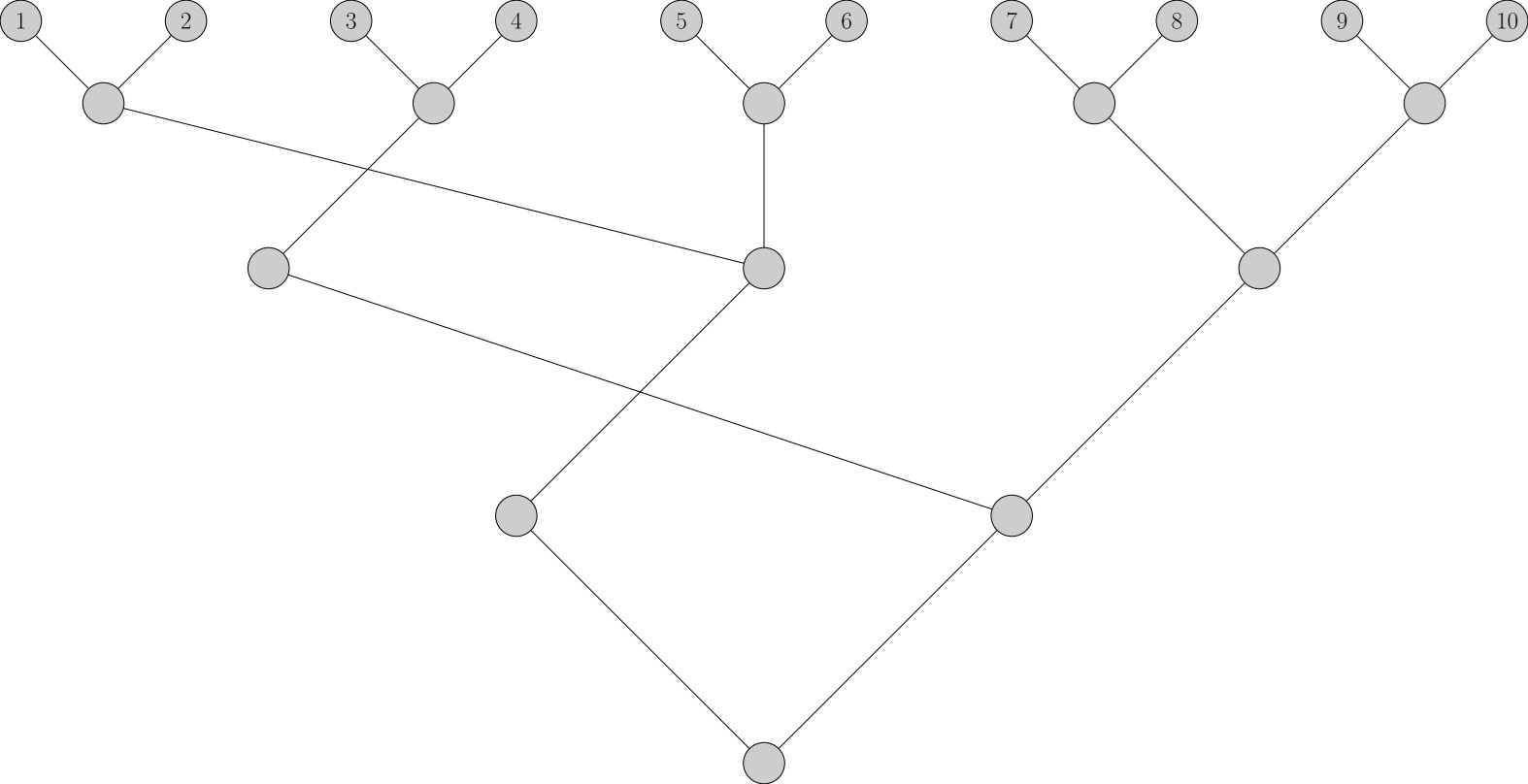}
\caption{\textbf{Pairwise reduction procedure for obtaining a reference reconstruction.} This example illustrates the creation of the 3D reference from the aligned 10 individual reconstructions labeled 1 to 10. In this case, the reduction takes four steps, with the number of involved reconstructions at each step reducing as $10\rightarrow 5 \rightarrow 3 \rightarrow 2 \rightarrow 1$. Whenever two nodes are connected to one node at a lower level, this indicates that the lower node represents the weighted average/variance of the two upper nodes. Note that in steps 2 and 3, there is an odd number of structures, which in our algorithm implies that the second node is simply passed as the first entry into the next step without averaging. Swapping the position of the leftover node from position 2 to position 1 prevents a single density from passing through all iterative steps, except for the final one, without being combined with other structures.}
\label{fig:pairwise_alignment}
\end{figure}

\begin{figure}
\centering
\includegraphics[width=0.8\textwidth]{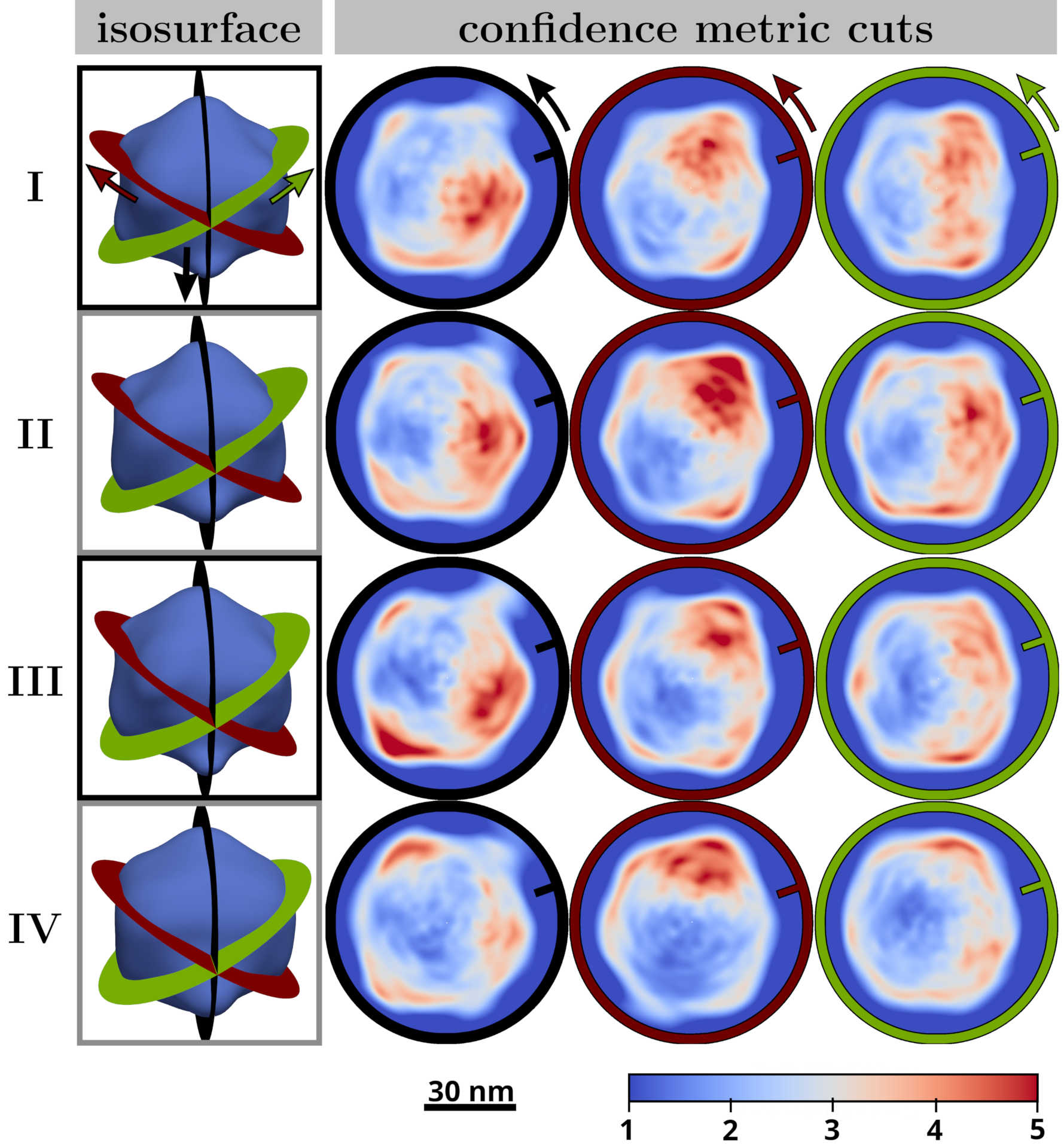}
\caption{\textbf{3D confidence metric for the amo06516 reconstructions.} This figure supplements Fig.~3 in the main text, where the reconstruction results for the amo06516 experiment are presented. Here, the isosurfaces are reproduced from Fig.~3 for convenience, while the  cuts display the mean value divided by the standard deviation, determined over 114 individual aligned reconstructions. High values of this metric indicate high consistency in densities across different reconstructions. Note that the density extension at one of the icosahedral vertices, visible in the black cut for data part III in Fig.~3 of the main text, is also visible here in all other data parts (I to IV).}
\label{fig:06516_mean_over_var}
\end{figure}

\begin{figure}
\centering
\includegraphics[width=0.8\textwidth]{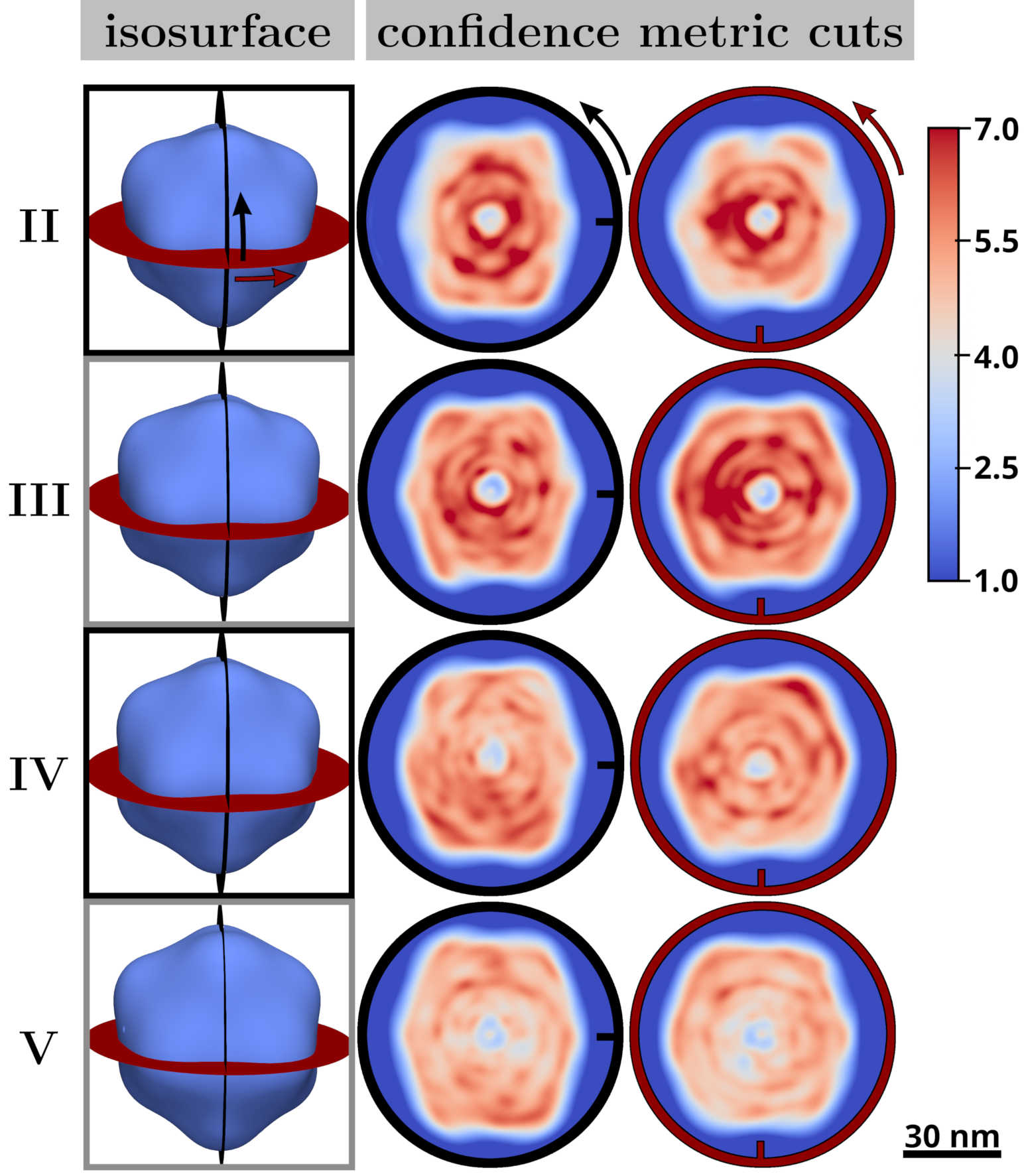}
\caption{\textbf{3D confidence metric for the amo86615 reconstructions.} This figure supplements Fig.~4 in the main text, where the reconstruction results for the amo86615 experiment are presented. Similar to Fig.~\ref{fig:06516_mean_over_var}, the cuts display the mean value divided by the standard deviation, determined over 114 individual aligned reconstructions. Note that the color scale used here ranges from 1 to 7 and differs from Fig.~\ref{fig:06516_mean_over_var}.}
\label{fig:86615_mean_over_var}
\end{figure}

\begin{figure}
\centering
\includegraphics[width=0.9\textwidth]{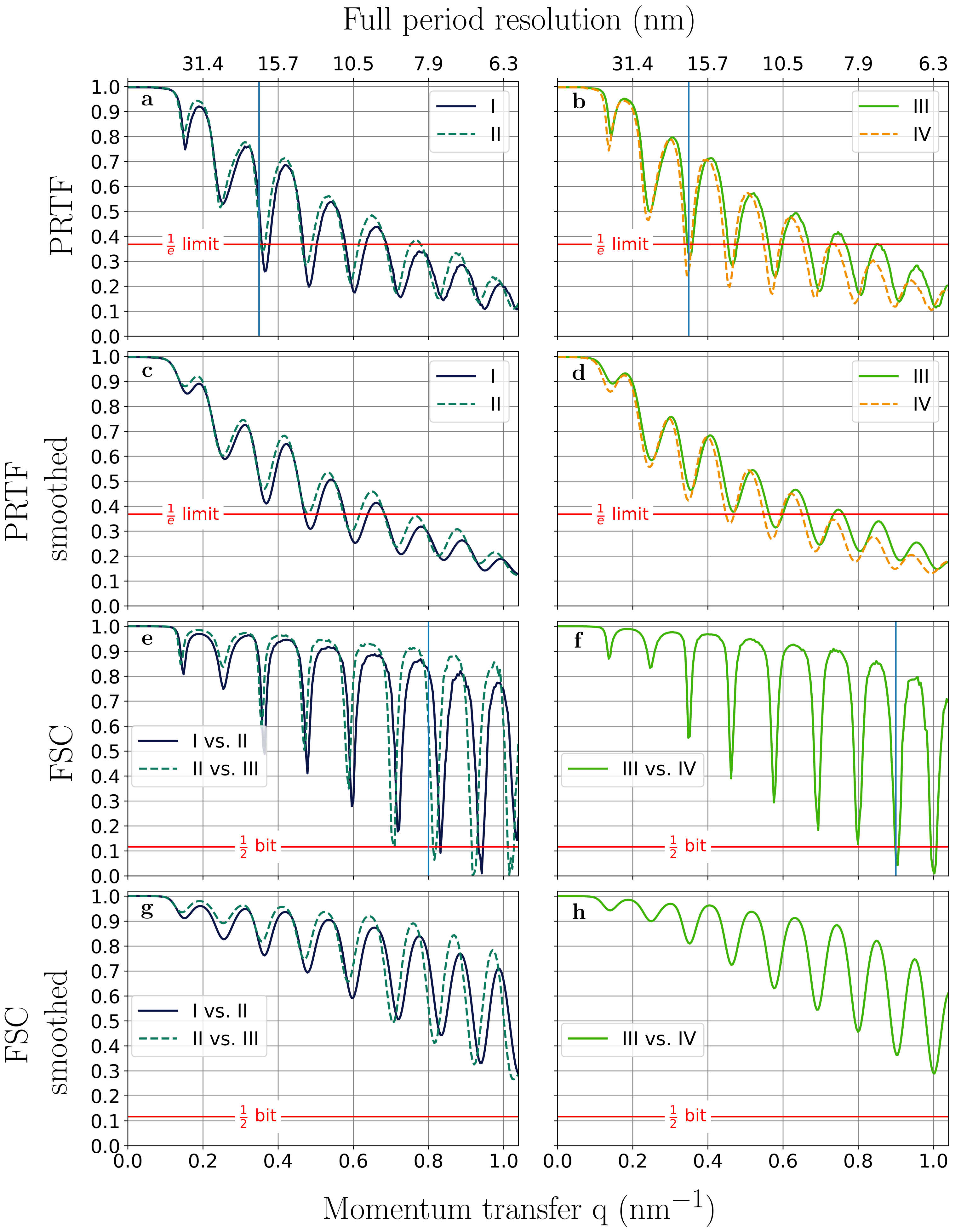}
\caption{\textbf{Resolution metrics for the amo06516 reconstructions.} (a,b) PRTF curves determined using Eq.~(13) from the main text for the different data parts I-IV, and (c,d) the same curves after smoothing (convolving) with a Gaussian of $0.04$nm$^{-1}$ FWHM (corresponding to the width of the critical Shannon pixel for a 70~nm virus).
The red horizontal lines in (a-d) indicate the $1/e$ resolution threshold. (e,f) FSC curves measuring the similarity between reconstructions from the different data parts, and (g,h) the same curves after smoothing with a Gaussian of FWHM=$0.04$nm$^{-1}$. The red horizontal lines in (e-h) indicate the $1/2$-bit resolution threshold, which is constant since the reconstructions were computed on a uniform spherical grid.}
\label{fig:resolution_06516}
\end{figure}

\begin{figure}
\centering
\includegraphics[width=0.9\textwidth]{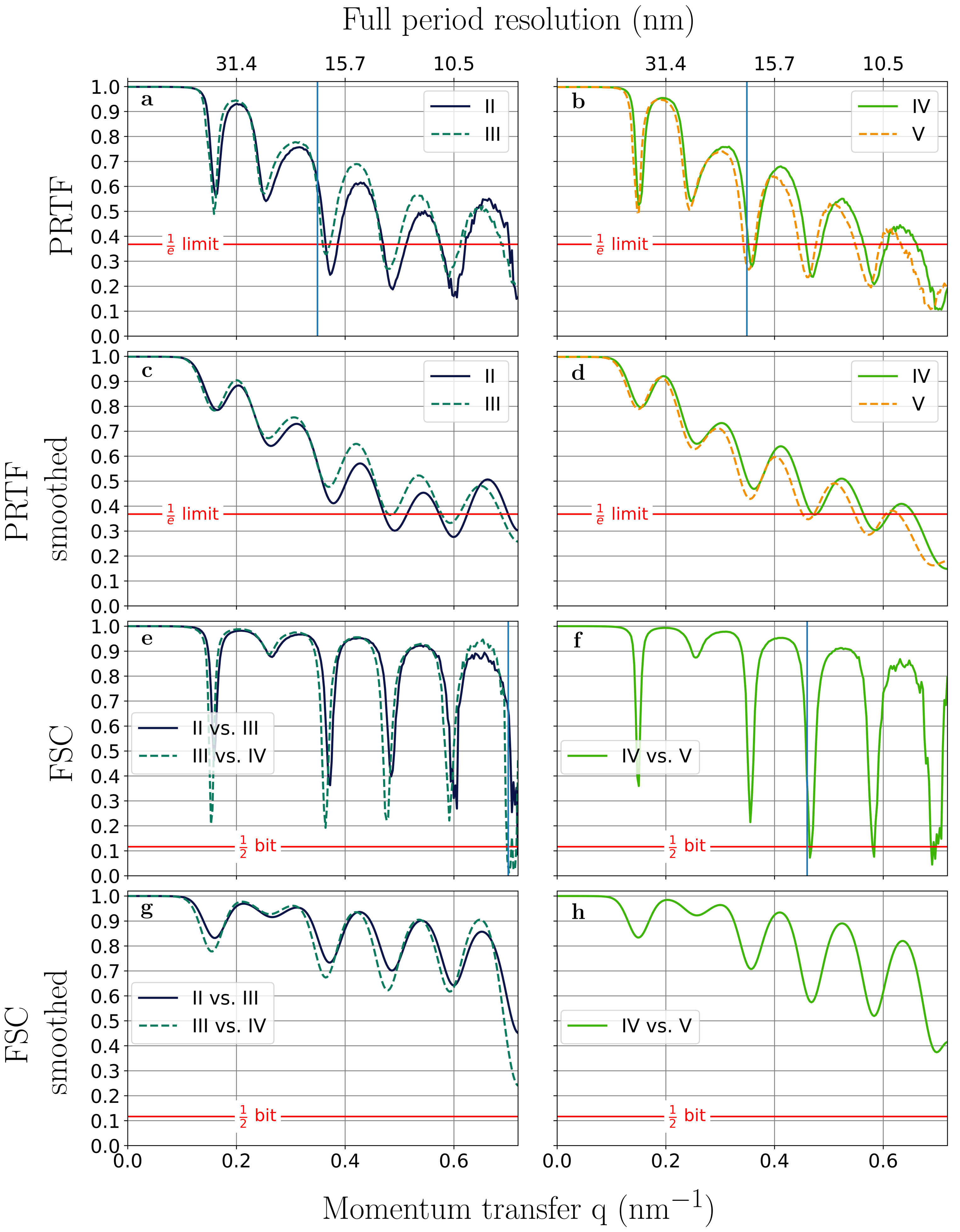}
\caption{\textbf{Resolution metrics for the amo86615 reconstructions.} (a,b) PRTF curves for the different data parts I-IV, and (c,d) the same curves after convolving with a Gaussian of $0.04$nm$^{-1}$ FWHM. (e,f) FSC curves measuring the similarity between reconstructions from the different data parts, and (g,h) the same curves after smoothing with a Gaussian of FWHM=$0.04$nm$^{-1}$. The notation and definitions of the resolution threshold are the same as in Supplementary Fig.~\ref{fig:resolution_06516}.}
\label{fig:resolution_86615}
\end{figure}

\begin{figure}
\centering
\includegraphics[width=\textwidth]{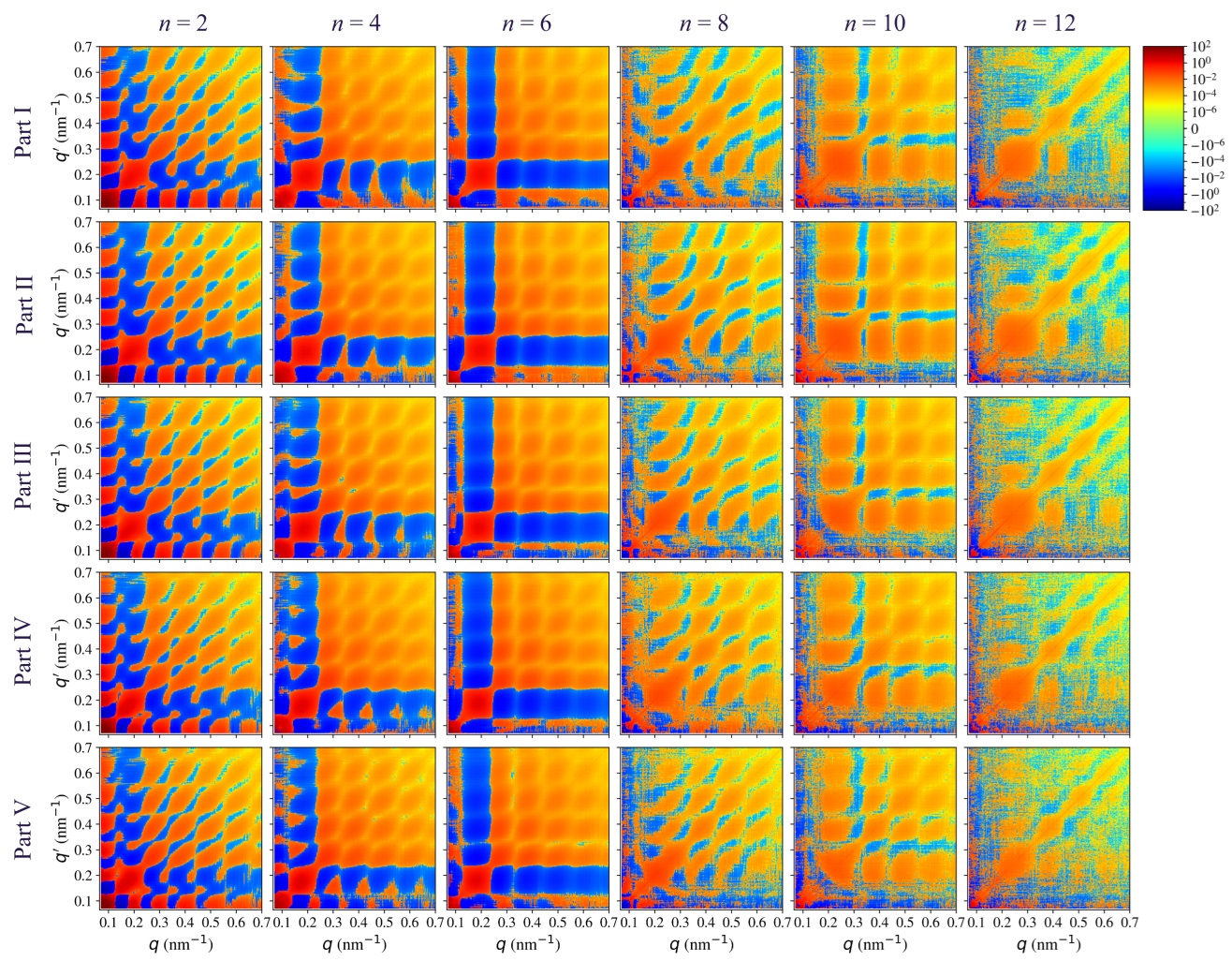}
\caption{\textbf{The real parts of the Fourier components $\textrm{Re}[C_{n}(q, q')]$ determined for the amo06516 experiment}. Different columns correspond to the specified orders $n$ of $C_{n}$, while rows correspond to different data parts I to V (see Supplementary Fig.~\ref{FigS:SizeDistributionHistograms}e). The values of $C_{n}$ are presented in arbitrary units and plotted on a symmetrical logarithmic scale.}
\label{FigS:amo06516Cnallparts}
\end{figure}

\begin{figure}
\centering
\includegraphics[width=\textwidth]{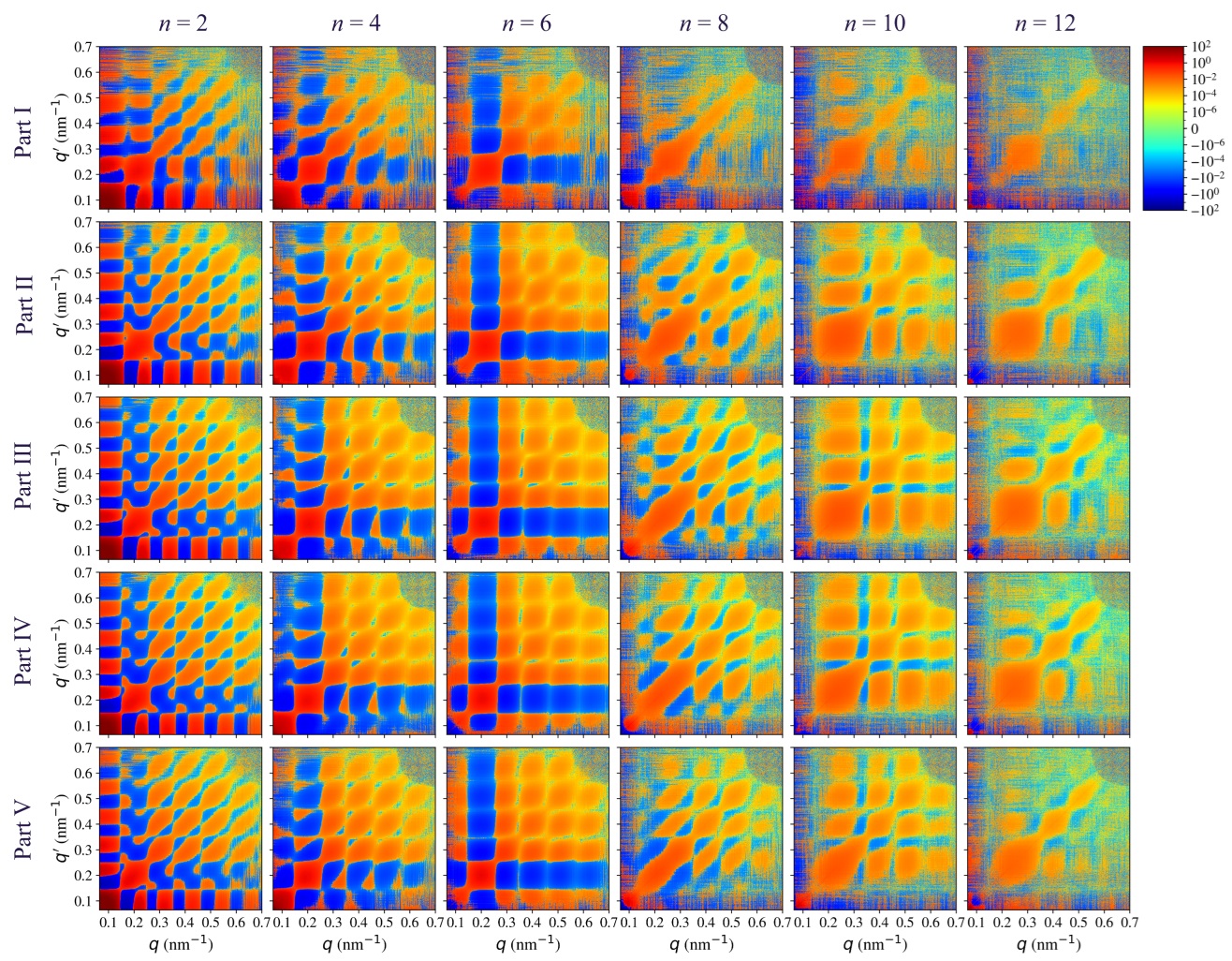}
\caption{\textbf{The real parts of the Fourier components $\textrm{Re}[C_{n}(q, q')]$ determined for the amo86615 experiment}. Different columns correspond to the specified orders $n$ of $C_{n}$, while rows correspond to different data parts I to V (see Fig.~\ref{FigS:SizeDistributionHistograms}d).
The shaded area in the top right corner of each 2D map is an artifact of the CCF calculation, arising from significant portions of missing data at the edges of the detector.
The values of $C_{n}$ are presented in arbitrary units and plotted on a symmetrical logarithmic scale.}
\label{FigS:amo86615Cnallparts}
\end{figure}

\begin{figure}
\centering
\includegraphics[width=\textwidth]{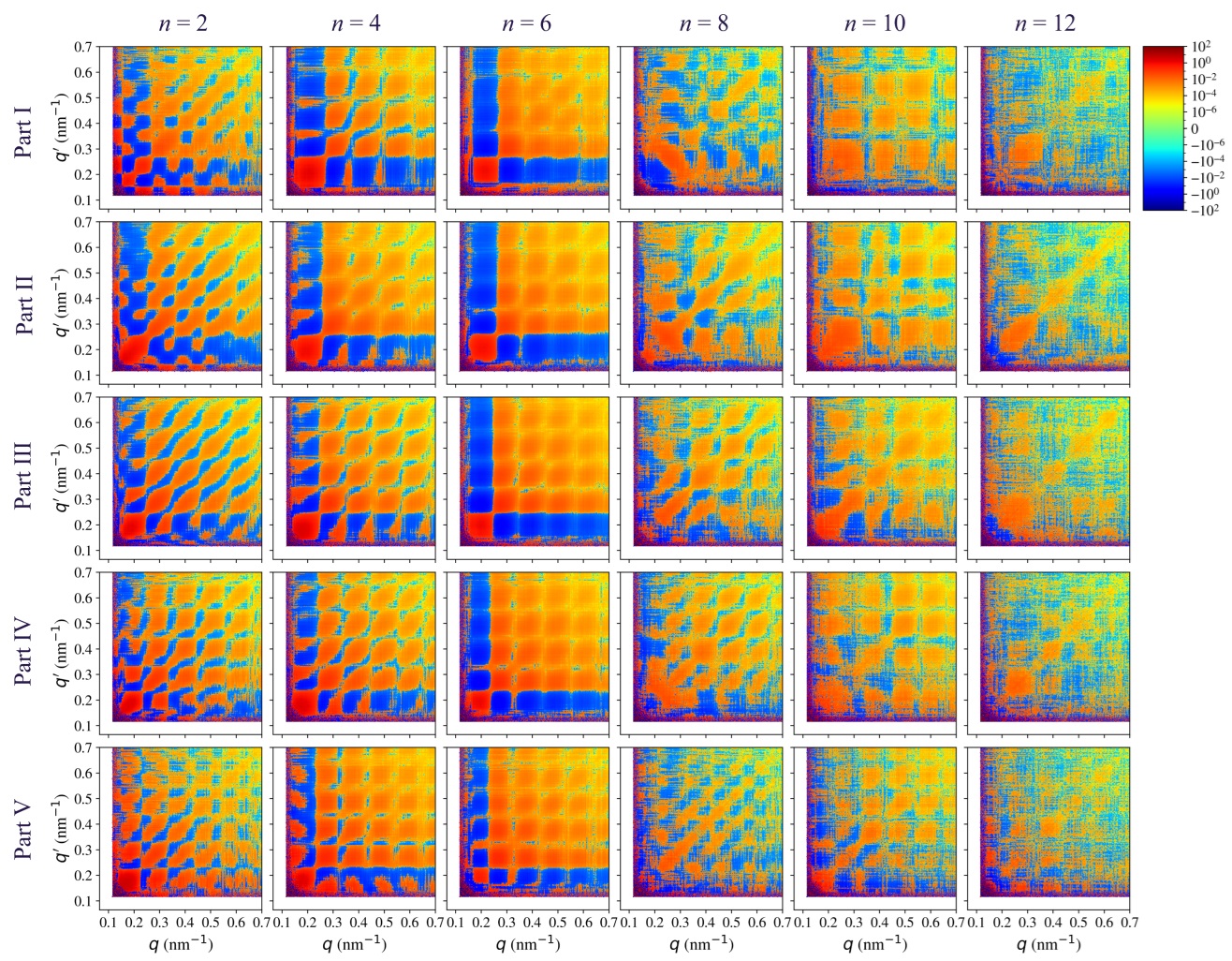}
\caption{\textbf{The real parts of the Fourier components $\textrm{Re}[C_{n}(q, q')]$ determined for the amo11416 experiment}. Different columns correspond to the specified orders $n$ of $C_{n}$, while rows correspond to different data parts I to V (see Fig.~\ref{FigS:SizeDistributionHistograms}f). The values of $C_{n}$ are presented in arbitrary units and plotted on a symmetrical logarithmic scale.}
\label{FigS:amo11416Cnallparts}
\end{figure}

\begin{figure}
\centering
\includegraphics[width=\textwidth]{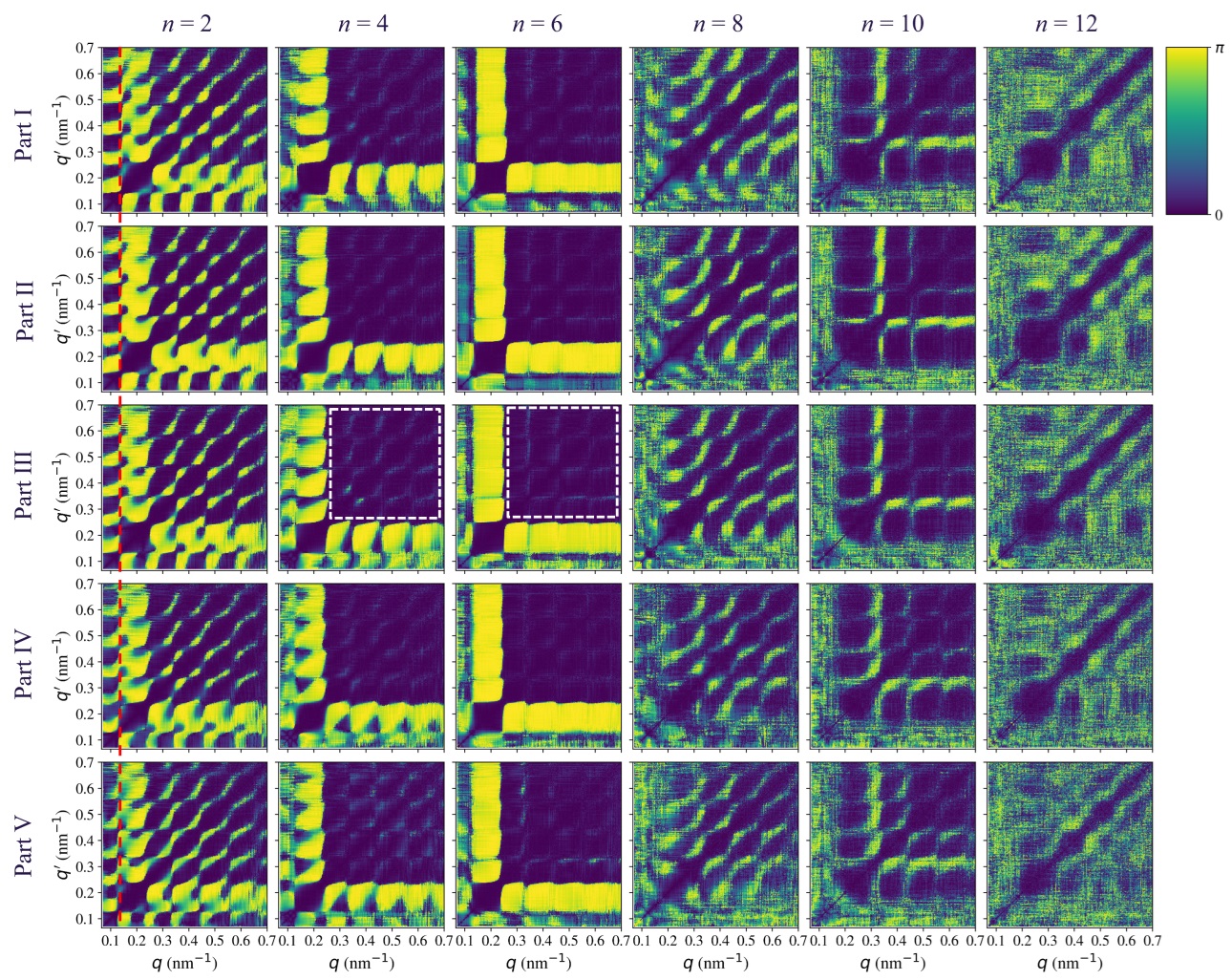}
\caption{\textbf{The phases $|\textrm{arg}[C_{n}(q, q')]|$ determined for the amo06516 experiment}. Different columns correspond to the specified orders $n$ of $C_{n}$, while rows correspond to different data parts I to V (see Supplementary Fig.~\ref{FigS:SizeDistributionHistograms}e). The phases are presented in radians. The dashed lines highlight distinct features discussed in Supplementary Note~8.}
\label{FigS:amo06516ArgCnallparts}
\end{figure}

\begin{figure}
\centering
\includegraphics[width=\textwidth]{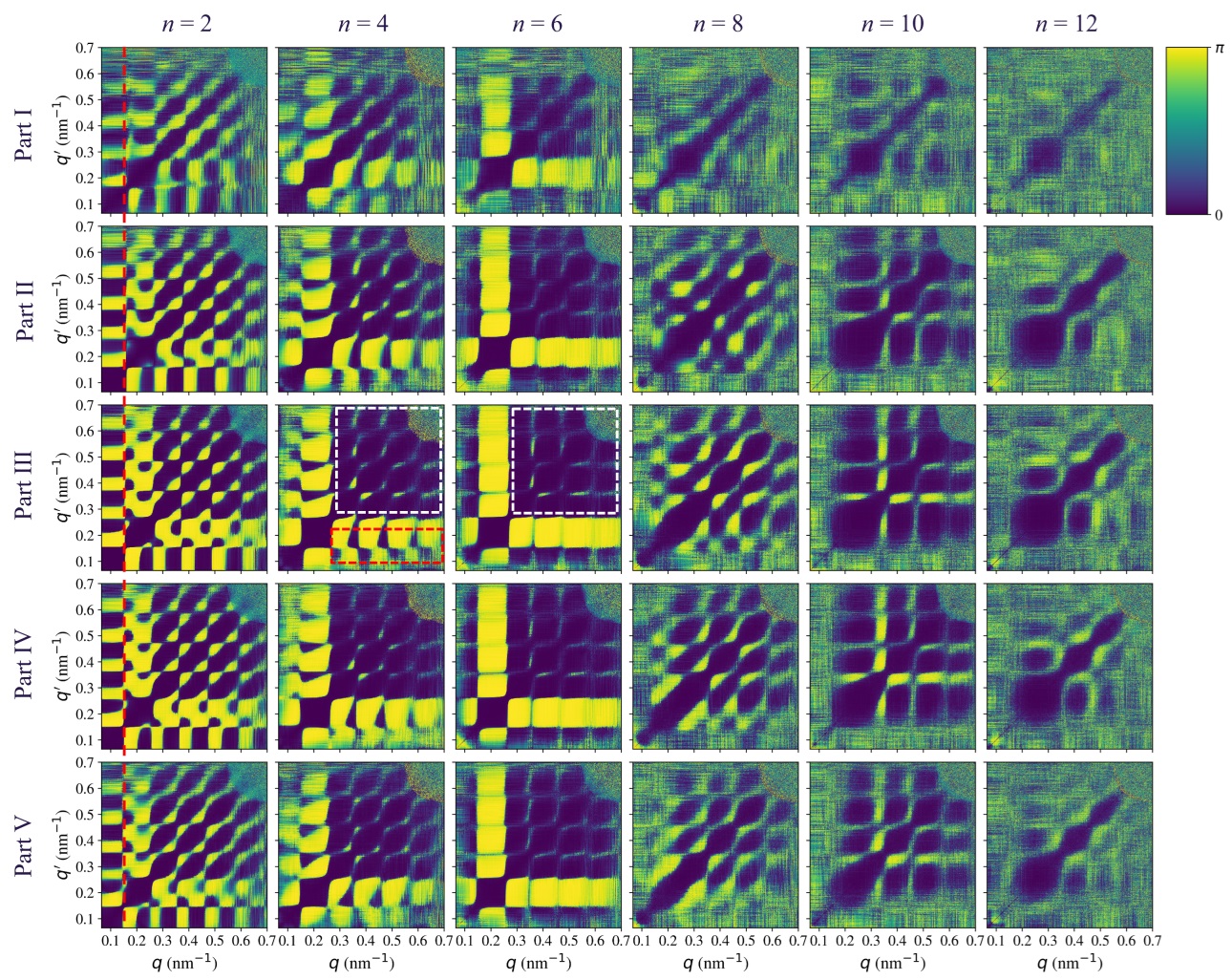}
\caption{\textbf{The phases $|\textrm{arg}[C_{n}(q, q')]|$ determined for the amo86615 experiment}. Different columns correspond to the specified orders $n$ of $C_{n}$, while rows correspond to different data parts I to V (see Fig.~\ref{FigS:SizeDistributionHistograms}d). 
The phases are presented in radians. The dashed lines highlight distinct features discussed in Supplementary Note~8.}
\label{FigS:amo86615ArgCnallparts}
\end{figure}

\begin{figure}
\centering
\includegraphics[width=\textwidth]{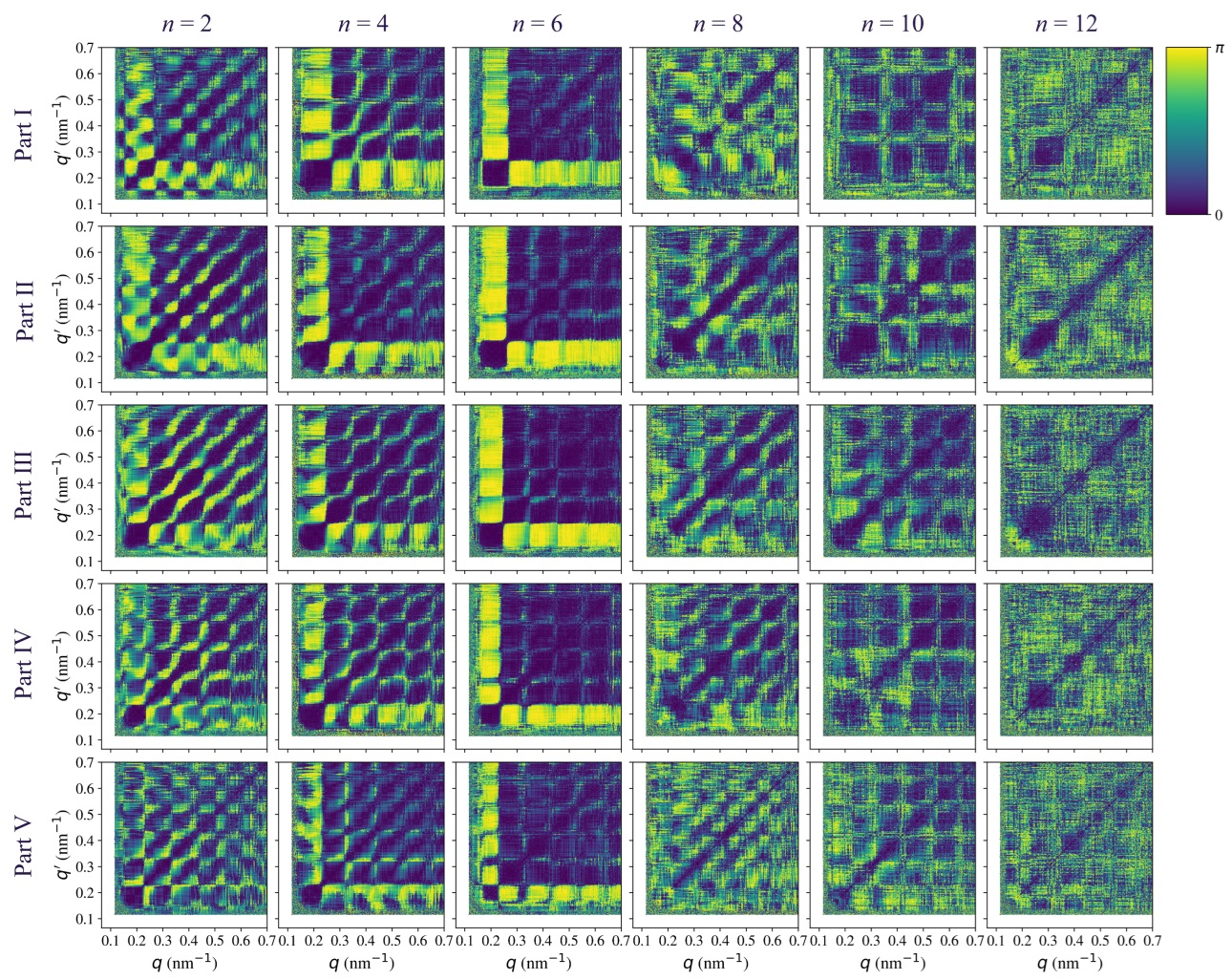}
\caption{\textbf{The phases $|\textrm{arg}[C_{n}(q, q')]|$ determined for the amo11416 experiment}. Different columns correspond to the specified orders $n$ of $C_{n}$, while rows correspond to different data parts I to V (see Fig.~\ref{FigS:SizeDistributionHistograms}f). The phases are presented in radians.}
\label{FigS:amo11416ArgCnallparts}
\end{figure}

\begin{figure}
\centering
\includegraphics[width=\textwidth]{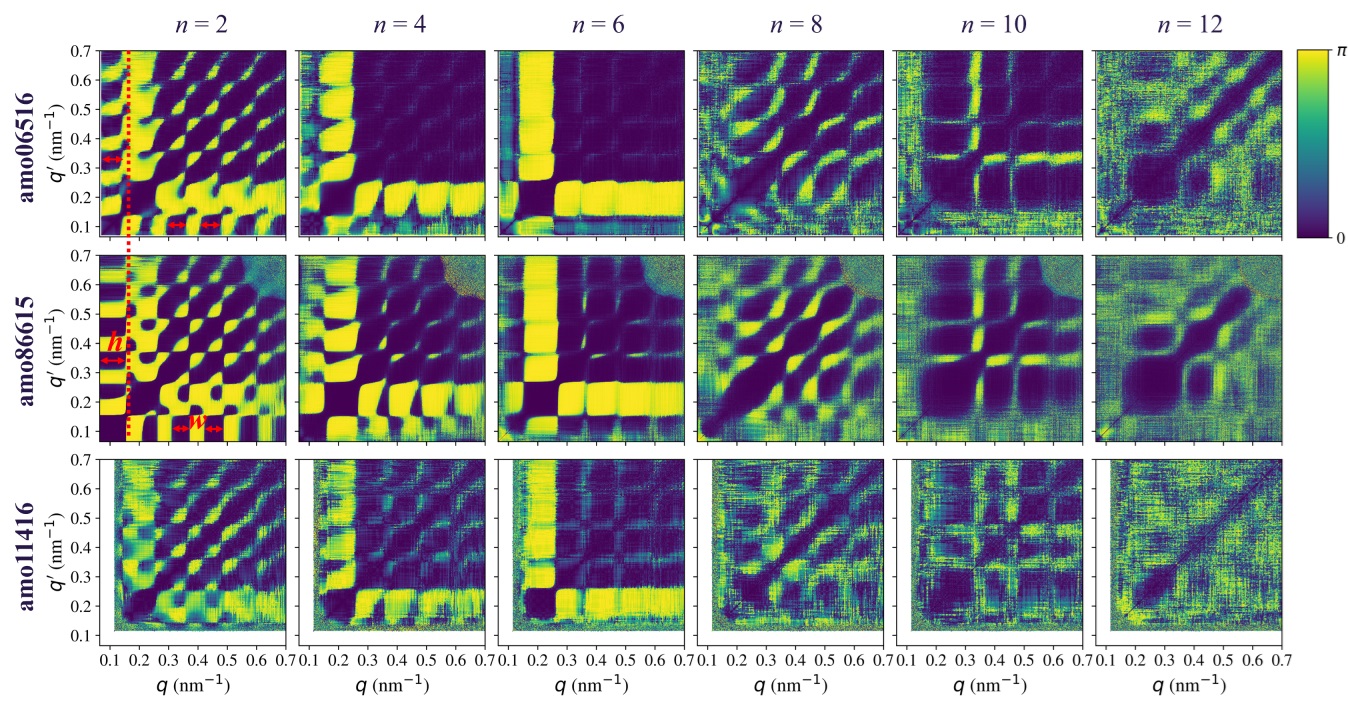}
\caption{\textbf{Comparison of the phases $|\textrm{arg}[C_{n}(q, q')]|$ across the three experiments.} The phases are presented for data part II of the amo06516 experiment (top row), part III of the amo86615 experiment (middle row), and  part II of the amo11416 experiment (bottom row). Different columns correspond to the specified orders $n$ of $C_{n}$, and the phases are presented in radians. The red dashed line at $n=2$, along with the arrows defining the height $h$ and width $w$ of the selected speckles in the amo06516 and amo86615 experiments, highlights features discussed in Supplementary Note~8. The plots shown here duplicate the corresponding plots in Figs.~\ref{FigS:amo06516Cnallparts}-\ref{FigS:amo11416Cnallparts} and are reproduced here for ease of comparison.}
\label{FigS:CnComparisonExp}
\end{figure}

\begin{figure}
\centering
\includegraphics[width=\textwidth]{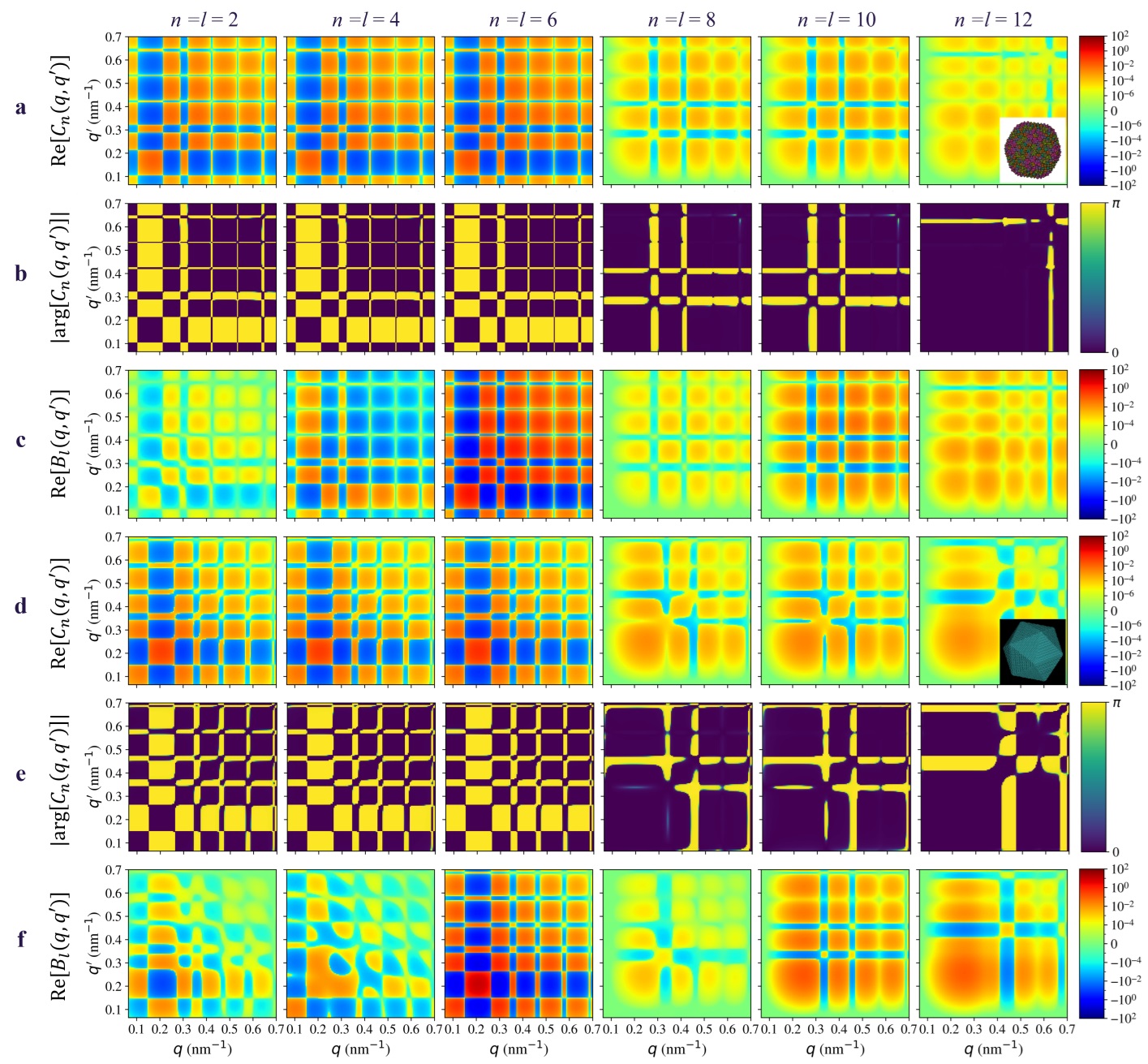}
\caption{\textbf{Cross-correlation analysis for particles with perfect icosahedral symmetry.} The results are presented for an atomistic model of the empty PR772 capsid [rows (a) to (c)], and a bead model of perfect solid icosahedral particle of 70~nm in size [rows (d) to (f)]: (a,d) the real parts of the Fourier components $\textrm{Re}[C_{n}(q, q')]$, (b,e) their corresponding phases $|\textrm{arg}[C_{n}(q, q')]|$, and (c,f) the real parts of the invariants $\textrm{Re}[B_{l}(q, q')]$. Each column corresponds to the specified order $n$ of $C_{n}$, or degree $l$ of $B_l$ ($n,l=2,4,6,8,10$ and 12). The values of $C_{n}$ and $B_l$ are specified in arbitrary units and plotted on a symmetrical logarithmic scale, and the phases of $C_{n}$ are presented in radians.
The structures of the empty PR772 capsid and of the ideal solid icosahedron are shown in the insets in (a) and (d), respectively.}
\label{FigS:SimCapsidandRegIcos}
\end{figure}

\begin{figure}
\centering
\includegraphics[width=0.85\textwidth]{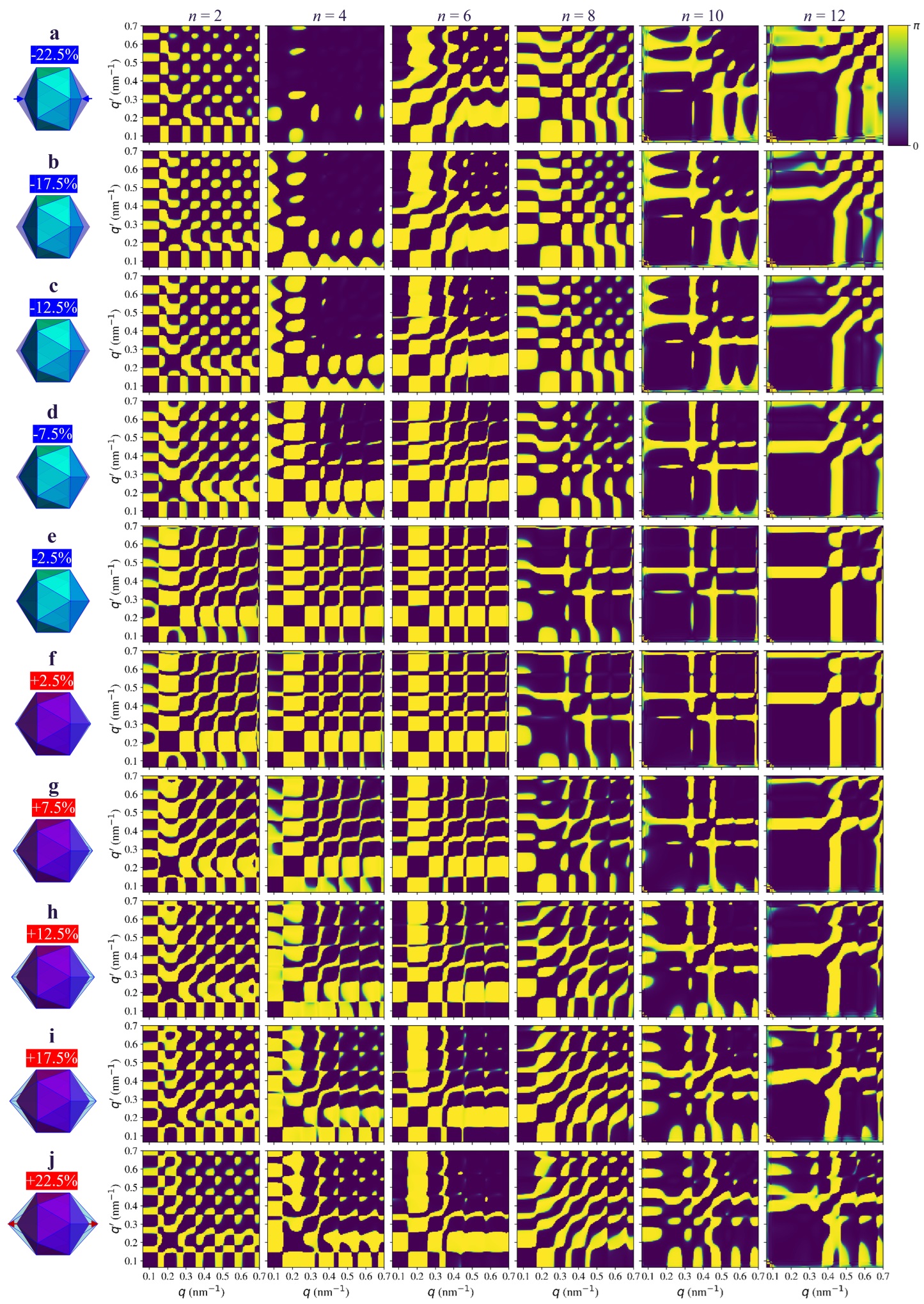}
\caption{\textbf{Cross-correlation analysis for icosahedral particles with uniaxial distortions.} The phases $|\textrm{arg}[C_{n}(q, q')]|$ are simulated for a bead model of a solid icosahedral particle with a radial distortion of two opposite vertices of different magnitudes. Different columns correspond to the specified orders $n$ of $C_{n}$, while the leftmost column schematically illustrates the amount of distortion relative to the perfect icosahedron (shown in violet). The values of compression (a-e) and elongation (f-j) specify how much each vertex has been displaced either towards (``-'') the center of the icosahedron or outwards (``+''), respectively. The phases $|\textrm{arg}[C_{n}(q, q')]|$ are specified in radians.}
\label{FigS:ReCnSimDistort2Vert}
\end{figure}

\begin{figure}
\centering
\includegraphics[width=0.85\textwidth]{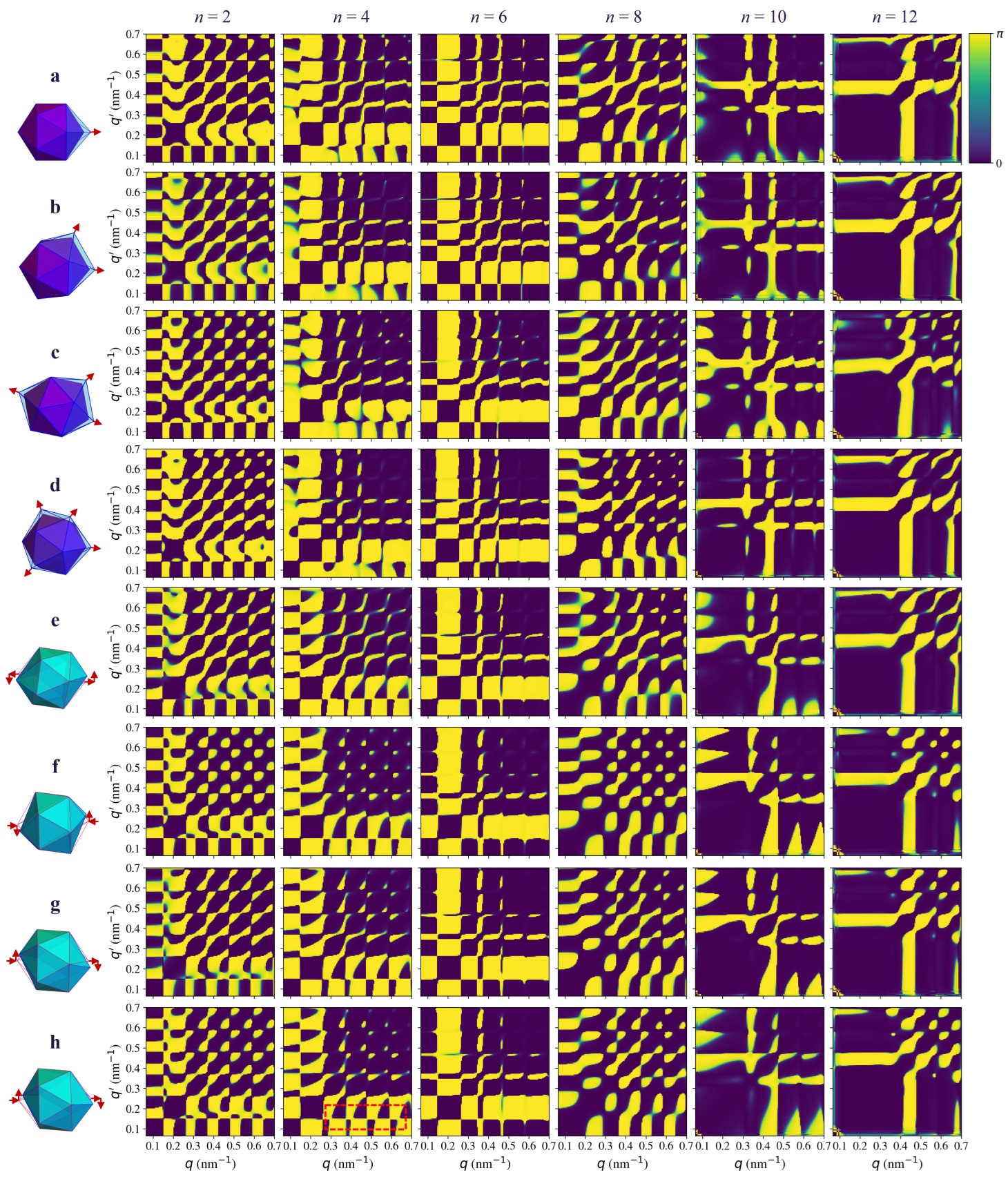}
\caption{\textbf{Cross-correlation analysis for icosahedral particles with composite distortions.} The phases $|\textrm{arg}[C_{n}(q, q')]|$ are simulated for a bead model of a solid icosahedral particle with (a-d) distinct types of asymmetric radial distortions of (a) one, (b) two, (c) three, and (d) four vertices, as well as with (e-h) composite radial and tangential distortions of two opposite vertices. Different columns correspond to the orders $n=2, 4, 6, 8, 10$ and 12 of $C_{n}$, while the leftmost column schematically illustrates the particle distortion with respect to the perfect icosahedral shape, with the latter shown as a solid violet icosahedron in (a-d), and red icosahedral wireframe in (e-h). The phases $|\textrm{arg}[C_{n}(q, q')]|$ are presented in radians. The dashed contour in (h) at $n=4$ highlights the features discussed in Supplementary Note~8.}
\label{FigS:ArgCnAsymDistort}
\end{figure}

\begin{figure}
\centering
\includegraphics[width=0.85\textwidth]{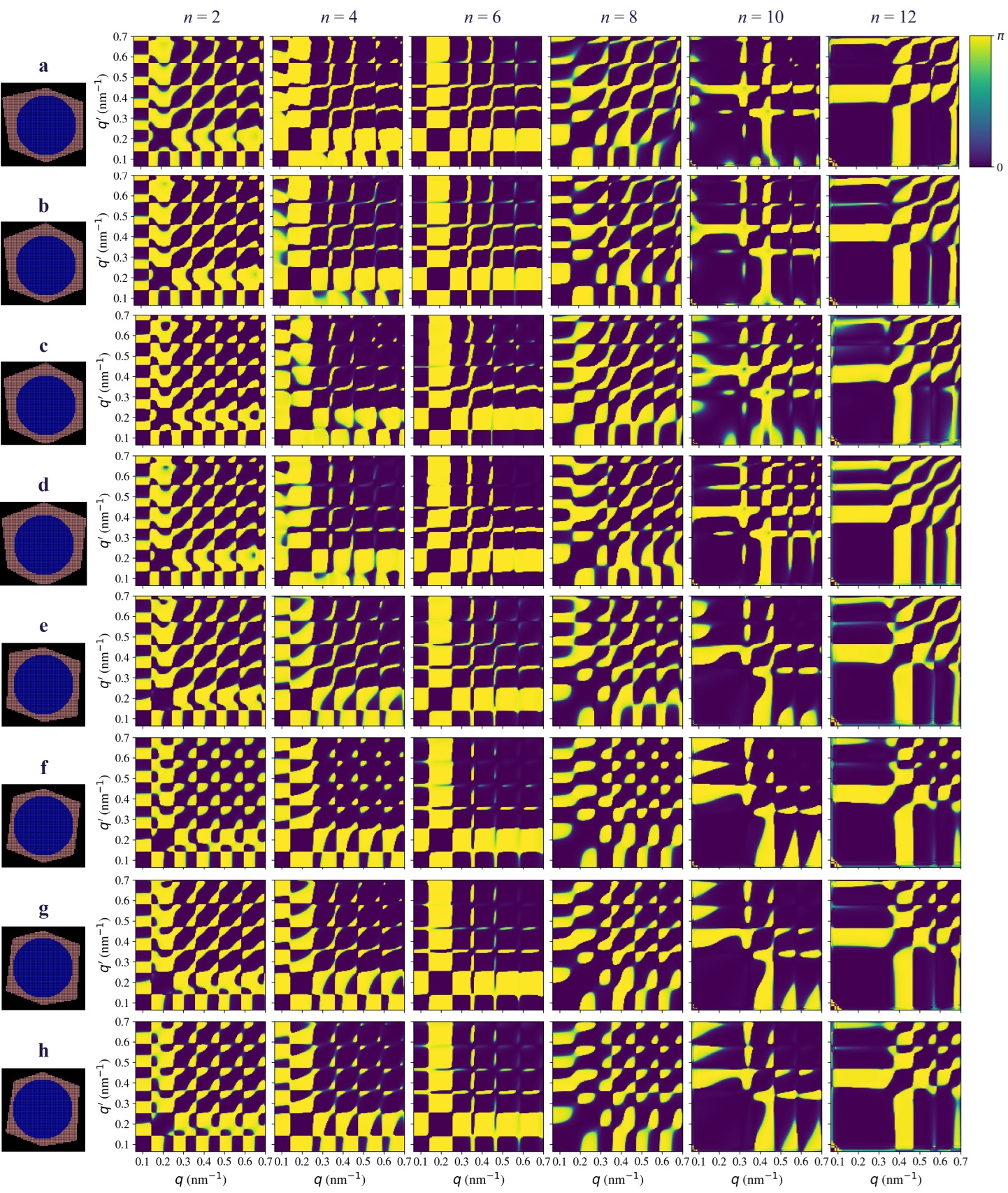}
\caption{\textbf{Cross-correlation analysis for core-shell icosahedral particles with composite distortions.} The phases $|\textrm{arg}[C_{n}(q, q')]|$ are simulated for a bead model of a core-shell icosahedral particle, with the same distortions as in Figs.~\ref{FigS:ArgCnAsymDistort}a-h, respectively, and additionally containing an internal spherical vesicle (core) of a radius  $r=24\;\textrm{nm}$ and density $\rho_{\textrm{c}}$ of 
$70\%$ of the capsid (shell) density $\rho_{\textrm{s}}$. Different columns correspond to the orders $n=2, 4, 6, 8, 10$ and 12 of $C_{n}$, while the leftmost column shows the central cuts through the corresponding bead models, with the spherical region of lower density shown in blue. The phases $|\textrm{arg}[C_{n}(q, q')]|$ are presented in radians.}
\label{FigS:ArgCnAsymDistortInternSpher}
\end{figure}

\begin{figure}
\centering
\includegraphics[width=0.85\textwidth]{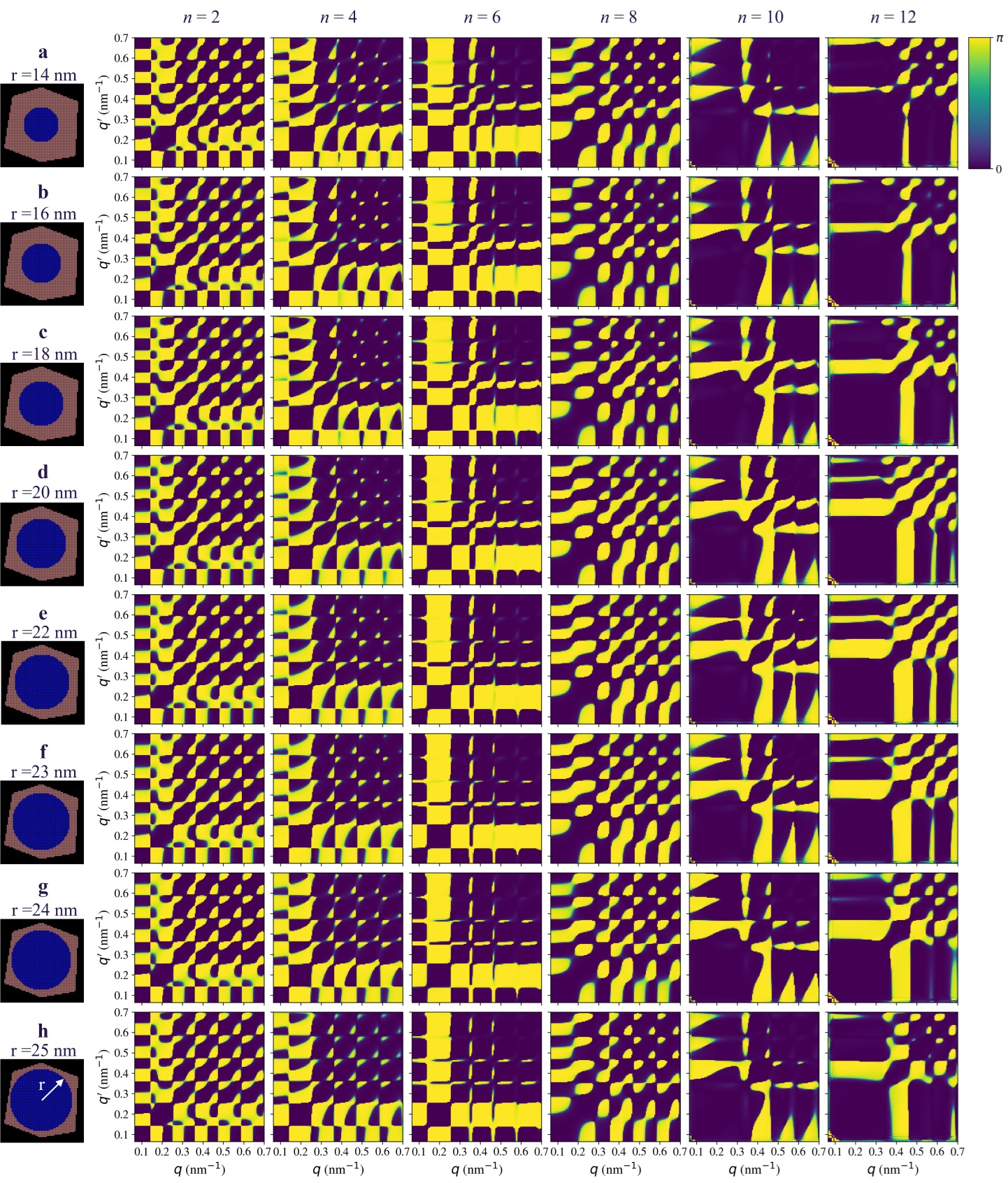}
\caption{\textbf{Cross-correlation analysis for distorted core-shell icosahedral particles with distinct core sizes.} The phases $|\textrm{arg}[C_{n}(q, q')]|$ are simulated for a bead model of a core-shell icosahedral particle with the same distortion as in Fig.~\ref{FigS:ArgCnAsymDistort}g, and additionally containing an internal spherical vesicle with a radius in the range from (a) $r=14\;\textrm{nm}$ to (h) $r=25\;\textrm{nm}$, and density of 
$70\%$ of the capsid density $\rho_{\textrm{s}}$. Different columns correspond to the orders $n=2, 4, 6, 8, 10$ and 12 of $C_{n}$, while the leftmost column shows the central cuts through the corresponding bead models, with the spherical region of lower density shown in blue. The phases $|\textrm{arg}[C_{n}(q, q')]|$ are presented in radians.}
\label{FigS:ArgCnSpherSizes}
\end{figure}

\begin{figure}
\centering
\includegraphics[width=0.85\textwidth]{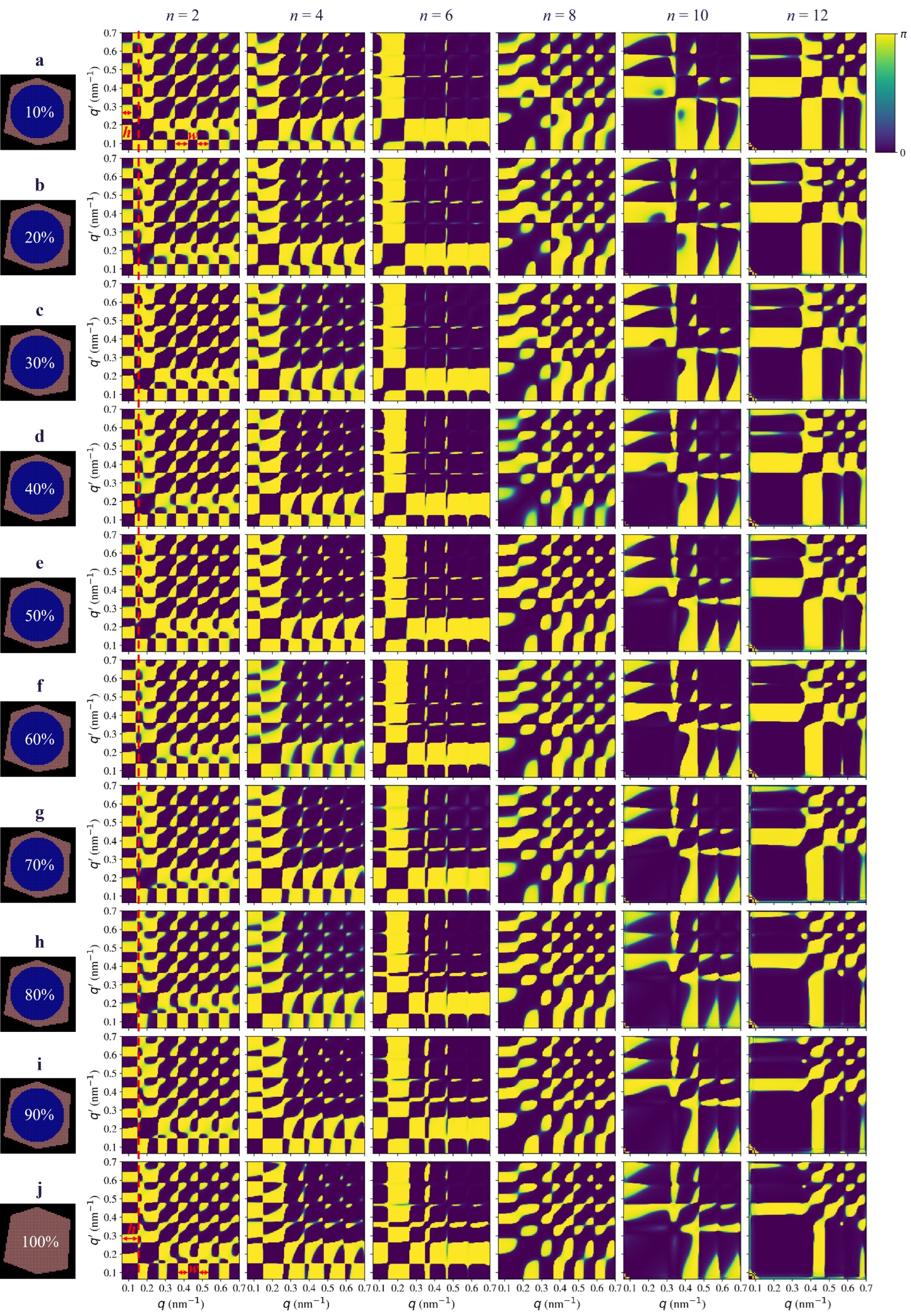}
\caption{\textbf{Cross-correlation analysis for distorted core-shell icosahedral particles with distinct core densities.} The phases $|\textrm{arg}[C_{n}(q, q')]|$ are simulated for a bead model of a core-shell icosahedral particle with the same distortion and vesicle size ($r=24\;\textrm{nm}$) as in Fig.~\ref{FigS:ArgCnSpherSizes}g, and variable density $\rho_{\textrm{c}}$ of the vesicle content, ranging from (a) $10\%$ to (j) $100\%$ of the capsid density $\rho_{\textrm{s}}$. Different columns correspond to the orders $n=2, 4, 6, 8, 10$ and 12 of $C_{n}$, while the leftmost column shows the central cuts through the corresponding bead models, with the specified values of the vesicle density. The phases $|\textrm{arg}[C_{n}(q, q')]|$ are presented in radians.  The red dashed line at $n=2$, along with the red arrows defining the height $h$ and width $w$ of the selected speckles in (a) and (j), highlight features discussed in Supplementary Note~8 and main text.}
\label{FigS:ArgCnSpherDensity}
\end{figure}

\begin{figure}
\centering
\includegraphics[width=0.8\textwidth]{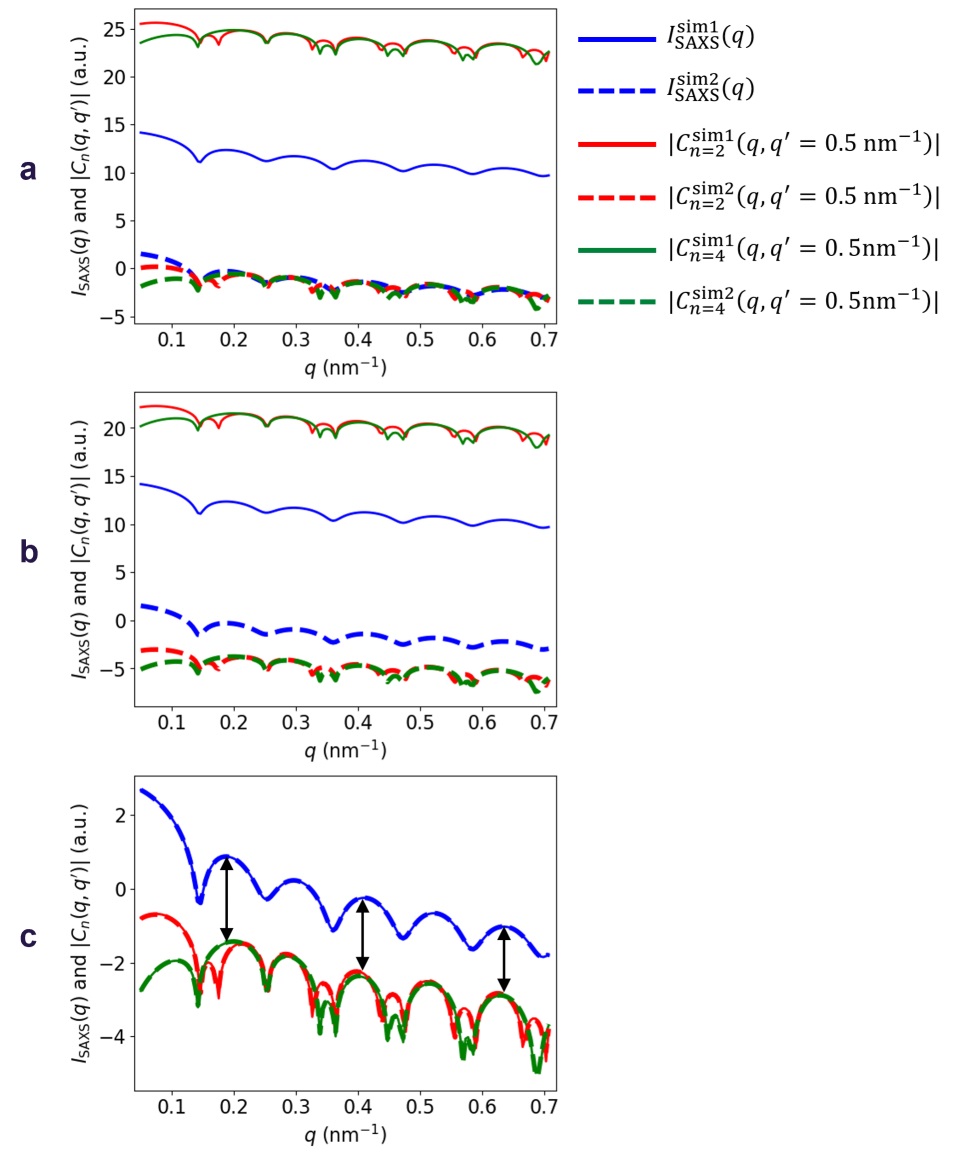}
\caption{\textbf{Scaling relationship between SAXS intensity and higher-order invariants.} This illustration shows the rescaling of SAXS intensity, $I_{\text{SAXS}}(q)$, and FCs, $C_n(q,q')$, for orders $n=2$ and 4 at a fixed $q'=0.5\;\text{nm}^{-1}$, determined from the same simulated structure (shown in Fig.~\ref{FigS:ArgCnAsymDistortInternSpher}h), using different processing parameters (denoted as \texttt{sim1} and \texttt{sim2}). In \texttt{sim1}, the diffraction patterns were considered on a polar grid with $N_{\phi}^{\text{sim1}}=1100$ azimuthal points, while in \texttt{sim2}, $N_{\phi}^{\text{sim2}}=800$ was used. Additionally, each diffraction pattern in \texttt{sim2} was normalized using the $\langle I(q) \rangle_{q}$ intensities averaged over the range of $q=(0.1, 0.2)\;\text{nm}^{-1}$. (a) Results of the calculation without any rescaling; (b) after rescaling by a factor of 2 and the respective $N_\phi$ values; and (c) after rescaling the SAXS profiles by the corresponding values of $A=\langle I_{\text{SAXS}}(q) \rangle_{q}$, with  $0.2\;\text{nm}^{-1}\leq q\leq 0.5\;\text{nm}^{-1}$, and the FCs $C_n(q,q')$ by $A^2$ (see Supplementary Note~9 for details). After rescaling, all solid and dashed curves of the same color coincide precisely in (c), as expected. Black two-sided arrows in (c) indicate the offset between $I_{\text{SAXS}}(q)$ and $|C_n(q,q')|$. The data are presented on a logarithmic scale.}
\label{FigS:ScalingSAXSExample}
\end{figure}

\begin{figure}
\centering
\includegraphics[width=0.9\textwidth]{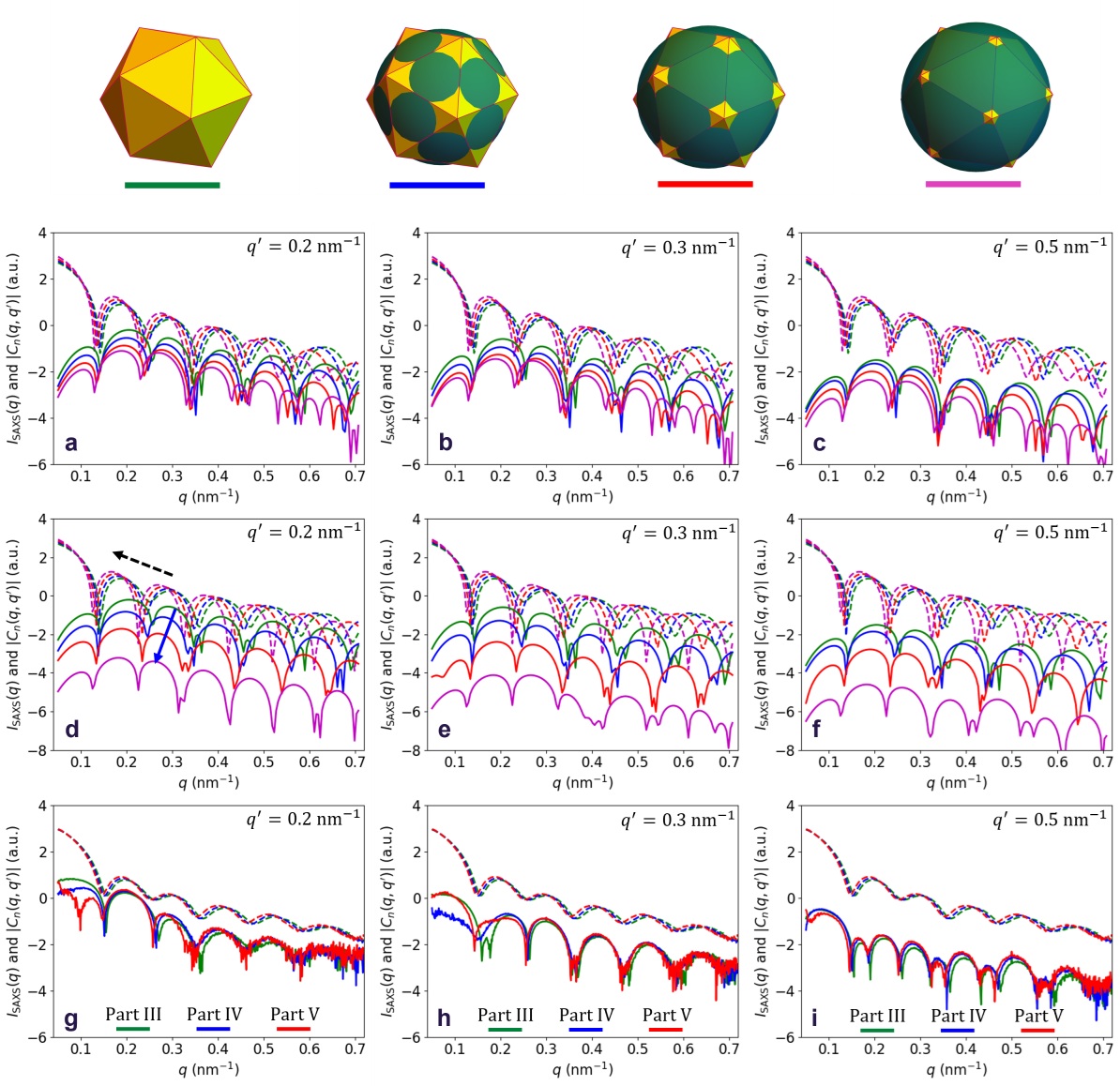}
\caption{\textbf{Effect of particle sphericity on the correlation data.} (a-f) Rescaled SAXS intensities (solid lines), $I_{\text{SAXS}}(q)$, and magnitudes of FCs (dashed lines), $|C_2(q,q')|$ for $n=2$,
determined for four simulated icosahedral core-shell virus structures with distinct degree of sphericity, illustrated in the top of the figure. The colors of the curves in (a-f) correspond to the colors of the lines
underscoring the structures on top of the figure. The change in sphericity is simulated by placing additional material on top of the perfect icosahedral particle, starting from a perfect icosahedron (leftmost structure)
and ending with a spherical shell that almost completely covers the particle (rightmost structure). The density of the additional material is $70\%$ of the capsid density in (a-c) and $100\%$ of the capsid density in (d-f).
(g-i) SAXS intensities and FCs  $|C_2(q,q')|$ determined for parts III, IV, and V of the amo86615 experiment (see Fig.~\ref{FigS:amo86615Cnallparts}). 
The results are presented for slices through $C_2(q,q')$ at (a,d,g) $q'=0.2\;\text{nm}^{-1}$, (b,e,h) $q'=0.3\;\text{nm}^{-1}$, and (c,f,i) $q'=0.5\;\text{nm}^{-1}$. 
The black dashed arrow and blue arrow in (d) indicate the shift of the peaks (and dips) in the SAXS profiles toward lower $q$, as well as the decrease in the magnitudes $|C_2(q,q')|$
upon increasing the sphericity of the simulated virus particle. This effect is more pronounced for a higher density of added material (d-f). 
The data are presented on a logarithmic scale.}
\label{FigS:SphericityAnalysis}
\end{figure}

\begin{figure}
\centering
\includegraphics[width=0.8\textwidth]{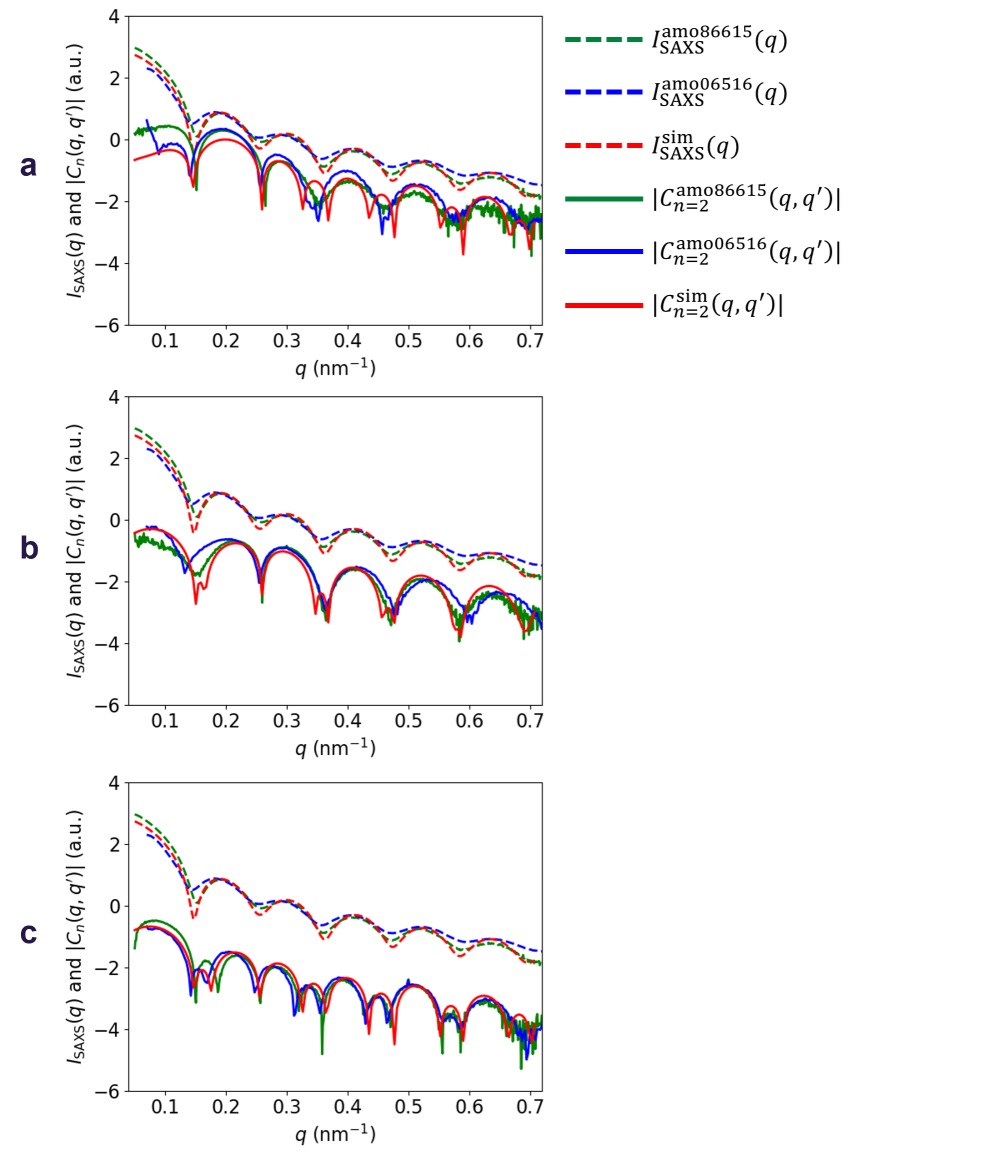}
\caption{\textbf{Comparison of experimental SAXS intensities and higher-order invariants.} Rescaled SAXS intensity, $I_{\text{SAXS}}(q)$, and FCs, $C_2(q,q')$, determined for two selected experimental datasets, Part IV of the amo86615 experiment and Part II from the amo06516 experiment, as well as a simulated dataset for the structure shown in Fig.~\ref{FigS:ArgCnSpherDensity}h. The results are presented for slices through $C_2(q,q')$ at (a) $q'=0.2\;\text{nm}^{-1}$, (b) $q'=0.3\;\text{nm}^{-1}$, and (c) $q'=0.5\;\text{nm}^{-1}$. The data are presented on a logarithmic scale.}
\label{FigS:ScalingSAXSExpSim}
\end{figure}

\FloatBarrier

\renewcommand{\refname}{Supplementary References}
\bibliography{references_supplementary}